\documentclass[english]{article}

\pdfoutput=1

\usepackage{lmodern}
\usepackage[T1]{fontenc}
\usepackage[latin9]{inputenc}
\usepackage{amsmath}
\usepackage{amssymb}
\usepackage{mathrsfs}
\usepackage{bm}
\usepackage{graphicx}
\usepackage{esint}
\usepackage{xcolor}
\usepackage{babel}
\usepackage{fullpage}
\usepackage{hyperref}
\usepackage{caption}
\usepackage{subcaption}
\usepackage{float}
\usepackage{cite}
\usepackage{comment}
\usepackage{datetime}
\usepackage{cancel}
\allowdisplaybreaks[4]
\definecolor{blus}{cmyk}{1,1,0,0.6}
\definecolor{verdes}{cmyk}{0.99,0,0.59,0.65}
\definecolor{rossos}{cmyk}{0,1,1,0.55}
\definecolor{redy}{cmyk}{0,1,1,0.7}
\definecolor{greeny}{cmyk}{0.99,0,0.59,0.98}
\definecolor{green-go}{cmyk}{0.79,0,0.59,0.5}

\hypersetup{colorlinks,bookmarksopen,bookmarksnumbered,
citecolor=blue,
linkcolor=black,
pdfstartview=green,
urlcolor=purple}

\def\be{\begin{equation}}
\def\ee{\end{equation}}

\def\bea{\begin{eqnarray}}
\def\eea{\end{eqnarray}}

\def\hhref#1{\href{http://arxiv.org/abs/#1}{arXiv:#1}} 

\numberwithin{equation}{section}

\begin{document}

\begin{titlepage}

\rule{0pt}{1.5cm}


\centerline{
\Large \bf  
On shape dependence of holographic entanglement entropy
}

\rule{0pt}{.0cm}

\centerline{
\Large \bf  
in AdS$_4$/CFT$_3$
}

\vspace{1cm}

 \centerline{\large
Piermarco Fonda$^{a,}$\footnote[1]{piermarco.fonda@sissa.it},  
Domenico Seminara$^{b,}$\footnote[2]{seminara@fi.infn.it}
 and Erik Tonni$^{a,}$\footnote[3]{erik.tonni@sissa.it}
}

\vspace{1.4cm}

\centerline{$^a${\it  
SISSA and INFN, via Bonomea 265, 34136, Trieste, Italy }}
\vspace{.3cm}
\centerline{$^b${\it 
Dipartimento di Fisica, Universit\'a di Firenze and INFN, via G. Sansone 1, 50019, 
Sesto Fiorentino, Italy}}

\vspace{2truecm}

\centerline{\bf Abstract}

We study the finite term of the holographic entanglement entropy of finite domains with smooth shapes and for four dimensional gravitational backgrounds. 
Analytic expressions depending on the unit vectors normal to the minimal area surface are obtained for both stationary and time dependent spacetimes. 
The special cases of AdS$_4$, asymptotically AdS$_4$ black holes, domain wall geometries and Vaidya-AdS backgrounds have been analysed explicitly. 
When the bulk spacetime is AdS$_4$, the finite term is the Willmore energy of the minimal area surface viewed as a submanifold of the three dimensional 
flat Euclidean space.
For the static spacetimes, some numerical checks involving spatial regions delimited by ellipses and non convex domains have been performed.
In the case of AdS$_4$, the infinite wedge has been also considered, recovering the known analytic formula for the coefficient of the logarithmic divergence.

\vspace{.5truecm}

\end{titlepage}

\tableofcontents

\newpage

\section{Introduction}

Entanglement entropy of extended quantum systems has attracted a lot of interest during the last decades and its importance is firmly established within different areas of theoretical physics like condensed matter, quantum information and quantum gravity \cite{ent-revs}.

Given a quantum system in a state characterised by the density matrix $\rho$ and whose Hilbert space can be written as $\mathcal{H}= \mathcal{H}_A \otimes \mathcal{H}_B$, the reduced density $\rho_A$ matrix associated to $\mathcal{H}_A $ is obtained by taking the partial trace over $\mathcal{H}_B $, namely $\rho_A = \textrm{Tr}_B \rho$. The entanglement entropy is the von Neumann entropy of  $\rho_A$, i.e. $S_A = - \textrm{Tr}_A (\rho_A \log \rho_A)$.  
When $\rho$ is a pure state, the entanglement entropy is a good measure of the entanglement associated to the bipartition of the Hilbert space and $S_A =S_B$.
One of the most important properties of this quantity with respect to other measures of entanglement is the strong subadditivity \cite{Lieb-Ruskai}.
Here we will consider only geometric bipartitions, i.e. cases where $A$ is a spatial part of the whole system and $B$ is its complement (notice that $A$ can be the union of many disjoint regions).
For a quantum field theory in a $d$ dimensional spacetime, the spatial domain $A$ is $(d-1)$ dimensional and the hypersurface $\partial A = \partial B$ separating $A$ and $B$ is $(d-2)$ dimensional.

Among the quantum field theories, conformal field theories (CFTs) are the ones for which the entanglement entropy has been mostly studied.
In general, $S_A$ can be written as a series expansion in terms of the ultraviolet cutoff $\varepsilon \to 0$ and the leading term is $S_A \propto \textrm{Area}(\partial A)/\varepsilon^{d-2}  + \dots$, where the dots denote subleading terms. 
This behaviour is known as the area law of the entanglement entropy \cite{Sorkin:2014kta, Bombelli:1986rw, Srednicki:1993im}.
For two dimensional CFTs on the infinite line at zero temperature, when $A$ is an interval of length $\ell$ the famous formula $S_A = (c/3) \log (\ell/\varepsilon) + \textrm{const}$ holds, where $c$ is the central charge of the theory \cite{Callan:1994py, Holzhey:1994we, Vidal:2002rm, Calabrese:2004eu} (see \cite{Calabrese:2009qy} for a review).
In this manuscript we will employ the holographic prescription of \cite{Ryu:2006bv, Ryu:2006ef} to compute the entanglement entropy for quantum field theories with a gravity dual (see \cite{Nishioka:2009un} for a review).

Extending the definition of the central charge also to non critical models, Zamolodchikov proved that the central charge decreases along a renormalization group (RG) flow going from the ultraviolet to the infrared fixed point \cite{Zamolodchikov:1986gt}. 
This result can be derived also from the strong subadditivity of the entanglement entropy \cite{Casini:2004bw}.
In higher dimensions, important results have been obtained for spherical domains \cite{Casini:2011kv}.
In particular, in $2+1$ dimensions, it has been found that the constant term occurring in the $\varepsilon \to 0$ expansion of $S_A$ for a disk decreases along an RG flow ($F$ theorem) 
\cite{Myers:2010xs, Myers:2010tj, Jafferis:2011zi, Klebanov:2011gs, Liu:2012eea, Casini:2012ei}.
Thus, in three spacetime dimensions this quantity plays a role similar to the central charge $c$ in two dimensions.

In the context of quantum gravity, a remarkable progress in the comprehension of entanglement has been done through the AdS/CFT correspondence. 
An important result is the holographic formula to compute the entanglement entropy of a $d$ dimensional CFT having a gravitational holographic dual characterised by an asymptotically AdS$_{d+1}$ background.
For static backgrounds it is given by \cite{Ryu:2006bv, Ryu:2006ef}
\be
\label{RT formula intro}
S_A = \frac{\mathcal{A}_A}{4G_N}\,,
\ee
where $G_N$ is the $(d+1)$ dimensional gravitational Newton constant and $\mathcal{A}_A\equiv \mathcal{A}[\hat{\gamma}_\varepsilon ]$ is the area of the $(d-1)$ dimensional (codimension two) hypersurface $\hat{\gamma}_\varepsilon $ obtained from $\partial A$ as follows. 
Given the hypersurface $\partial A$ on some constant time slice of the CFT living at the boundary of the asymptotically AdS$_{d+1}$ background, one must consider all the spatial hypersurfaces $\gamma_A$ in the bulk such that $\partial \gamma_A =\partial A$.
Among these hypersurfaces, we have to find the one having minimal area, which will be denoted by $\hat{\gamma}_A$.
Since these hypersurfaces reach the boundary of the asymptotically AdS$_{d+1}$ spacetime, which is located at $z=0$ in some convenient system of coordinates, their area is infinite. 
The regularization of this divergence is done by restricting to $z\geqslant \varepsilon>0$, where $\varepsilon$ is a small quantity which coincides with the ultraviolet cutoff of the dual CFT, according to the AdS/CFT dictionary.
Denoting by $\hat{\gamma}_\varepsilon$ the restriction of $\hat{\gamma}_A$ to $z\geqslant \varepsilon$, its area $ \mathcal{A}[\hat{\gamma}_\varepsilon ]$ can be written as a series expansion for $\varepsilon \to 0$ and the terms of this expansion can be compared with the ones occurring in the expansion of $S_A$ computed through CFT techniques.
This prescription has been derived through a generalization of the usual black hole entropy in \cite{Lewkowycz:2013nqa}.
The covariant generalisation of (\ref{RT formula intro}), which allows to deal with time dependent gravitational backgrounds, has been found in \cite{Hubeny:2007xt}. In this case the formula is formally identical to (\ref{RT formula intro}) but $\mathcal{A}_A$ is evaluated by extremizing the area functional without forcing the spatial hypersurfaces $\gamma_A$ to live on some constant time slice.

The formula (\ref{RT formula intro}) has passed many consistency checks (e.g. it satisfies the strong subadditivity property \cite{Headrick:2007km}) and nowadays it is a well established piece of information within the holographic dictionary.
When the dual CFT is at finite temperature, the dual gravitational background is an asymptotically AdS$_{d+1}$ black hole and (\ref{RT formula intro}) provides the corresponding holographic entanglement entropy.
Let us remind that the entanglement entropy is not a measure of entanglement when the whole system is in a mixed state.
It is important to remark that (\ref{RT formula intro}) holds for those regimes of the CFT parameters which are described by classical gravity through the AdS/CFT correspondence. 
The corrections coming from quantum effects have been discussed in \cite{Faulkner:2013ana}.

The minimal area surface entering in the holographic formula (\ref{RT formula intro}) for the entanglement entropy is difficult to find analytically for domains $A$ which are not highly symmetric because typically a partial differential equation must be solved. 
Numerical methods can be employed, but for non trivial domains finding a convenient parameterisation of the surface is already a non trivial task. 
The shape dependence of some subleading terms in the expansion of $\mathcal{A}_A$ as $\varepsilon \to 0$  have been studied in various papers 
\cite{Solodukhin:2008dh, Hubeny:2012ry, Klebanov:2012yf, Myers:2013lva, Fursaev:2013fta, Astaneh:2014uba, Allais:2014ata, Fonda:2014cca}.

The holographic formula (\ref{RT formula intro}) can be employed also when $A=\cup_i A_i$ is the union of two or more disjoint spatial domains $A_i$.
In these cases, one can construct combinations of entanglement entropies which are finite as $\varepsilon\to 0$: the simplest case is the mutual information $I_{A_1, A_2} \equiv S_{A_1} + S_{A_2} - S_{A_1 \cup A_2} $ when $A=A_1 \cup A_2$. 
For two dimensional CFTs, the mutual information or its generalizations to more than two intervals encode all the CFT data of the model  \cite{Caraglio:2008pk, Furukawa:2008uk, Casini:2009vk, Calabrese:2009ez, Alba:2009ek, Headrick:2010zt, Calabrese:2010he, Coser:2013qda, DeNobili:2015dla}.
Some results for the mutual information are also known in $2+1$ dimensions from the quantum field theory point of view, where the analysis is more difficult because of the non local nature of $\partial A$ \cite{Cardy:2013nua, Shiba:2012np, Casini:2008wt, Schnitzer:2014zva, Agon:2015twa, Casini:2015woa}. 
As for the holographic analysis for disjoint domains through (\ref{RT formula intro}), the main feature to deal with is the occurrence of two or more local extrema of the area functional \cite{Gross:1998gk, Drukker:2005cu, Hirata:2006jx, Headrick:2010zt, Tonni:2010pv, Nakaguchi:2014pha}. Thus, the holographic mutual information is zero when the two regions are distant enough (see \cite{Fonda:2014cca} for the transition curves of domains $A_1$ and $A_2$ which are not disks).

The covariant prescription of \cite{Hubeny:2007xt} has been employed to study the behaviour of the holographic entanglement entropy during a thermalization process.
The  simplest holographic models are provided by  the Vaidya-AdS backgrounds \cite{Vaidya:1951zz, Bonnor:1970zz}, which have been largely studied during recent years \cite{AbajoArrastia:2010yt, Albash:2010mv, Balasubramanian:2010ce, Balasubramanian:2011at, Allais:2011ys, Callan:2012ip, Caceres:2012em, Hubeny:2013hz, Liu:2013iza, Keranen:2011xs, Alishahiha:2014cwa, Fonda:2014ula}.

In this paper we will consider only asymptotically AdS$_4$ bulk spacetimes whose boundary is the three dimensional Minkowski spacetime.
Given a finite domain $A$ delimited by a finite and smooth boundary $\partial A$ (entangling curve), the expansion of the area of the surface $\hat{\gamma}_\varepsilon$ entering in the holographic entanglement entropy (\ref{RT formula intro}) reads
\be
\label{hee ads4 intro}
\mathcal{A}[\hat{\gamma}_\varepsilon]
= 
\frac{P_A}{\varepsilon} 
- F_A
+ o(1)\,,
\ee
where $P_A = \textrm{length}(\partial A)$ is the perimeter of the spatial region $A$ (we set the AdS radius to one).
In order to find the $O(1)$ term $F_A$, the whole surface $\hat{\gamma}_A$ is needed.
Exact analytic expressions of the $F_A$ are known only for few cases which are highly symmetric like the disk \cite{Berenstein:1998ij, Gross:1998gk} and the annulus for AdS$_4$ \cite{Drukker:2005cu, Hirata:2006jx, Dekel:2013kwa}. 
Among the infinite domains, namely the ones elongated in one particular direction, the strip has been studied because its symmetry makes it the simplest case to address from the analytical point of view \cite{Aharony:1999ti, Ryu:2006bv, Ryu:2006ef}. In \cite{Fonda:2014cca} the interpolation between the disk and the elongated strip through various domains has been considered.

In this paper, we derive closed expressions for $F_A$  in terms of the unit vectors normal to $\hat{\gamma}_A$ for both static and time dependent backgrounds.
When the bulk spacetime is AdS$_4$, our formula for $F_A$ becomes the Willmore energy \cite{Thomsen, blaschke, willmorebound,willmorebook} of the minimal area surface $\hat{\gamma}_A$ viewed as a submanifold of $\mathbb{R}^3$, recovering the result of \cite{Babich:1992mc, AM}.
The formulas for some static backgrounds are checked numerically for regions delimited by ellipses and also for non convex domains, while for the Vaidya-AdS spacetime only disks are employed as benchmark of our results.
The numerical analysis for generic entangling curves have been performed by employing {\it Surface Evolver} \cite{evolverpaper, evolverlink}.
We will not consider spacetimes which are asymptotically global AdS. In these cases the homology constraint in the holographic prescription (\ref{RT formula intro}) plays a crucial role \cite{Headrick:2007km, Hubeny:2013gta, Hubeny:2013gba, Headrick:2013zda}.

The paper is organized as follows. 
In \S\ref{sec static} we find $F_A$ for generic static backgrounds which are conformally related to asymptotically flat spacetimes and then specialise the formula to the explicit examples given by AdS$_4$, asymptotically AdS$_4$ black holes \cite{Romans:1991nq, London:1995ib, Chamblin:1999tk} and domain wall geometries  \cite{Freedman:1999gp, Girardello:1998pd, Girardello:1999bd, Myers:2010tj, Myers:2012ed, Liu:2013una}. The latter spacetimes are simple holographic models dual to RG flows in the boundary theory. 
In \S\ref{sec time-dep} we extend the analysis to the time dependent spacetimes, considering then the Vaidya-AdS backgrounds as special case.  
In \S\ref{sec examples} some particular domains are discussed for the above backgrounds, in order to recover the known results for disks and strips and extend them through the formulas found in the previous sections.
Spatial regions delimited by ellipses and also a non convex domain are considered. 
When the bulk geometry is AdS$_4$, we also consider the infinite wedge \cite{Drukker:1999zq}, which includes also a logarithmic divergence as $\varepsilon \to 0$ (see \cite{Bueno:2015rda, Bueno:2015xda} for recent developments about entangling curves with corners).
Some consequences for the holographic mutual information are addressed in \S\ref{sec HMI} and concluding remarks are given in \S\ref{sec conclusions}.
In the appendices \ref{sec app normal vector},  \ref{sec app higher dims}, \ref{app helfrich},  \ref{sec app wedge} and \ref{sec app annulus} we have collected technical details and some further discussions related to issues occurred in the main text.


\section{Static backgrounds}
\label{sec static}

In this section we derive a formula for $F_A$ in (\ref{hee ads4 intro}) for static backgrounds which are conformally related to asymptotically flat spacetimes whose boundary is the four dimensional Minkowski space. The discussion for the general case is given in \S\ref{sec general case}, while in \S\ref{sec static backs} we specify the result to some explicit backgrounds: AdS$_4$, asymptotically AdS$_4$ black holes and domain wall geometries.

\subsection{General case}
\label{sec general case}

Let us consider the three dimensional Euclidean space $\mathcal{M}_3$ obtained by taking a constant time slice of a static asymptotically AdS$_4$ background, namely
\be
\label{ansatz t=cost}
ds^2 \big|_{t=\textrm{const}} 
= g_{\mu\nu} \,dx^\mu dx^\nu \,.
\ee
Given a two dimensional surface $\gamma$ embedded into $\mathcal{M}_3$, let us denote by $n_\mu$ the spacelike unit vector normal to $\gamma$ and by $h_{\mu\nu} = g_{\mu\nu} - n_\mu n_\nu $ the metric induced on $\gamma$ (first fundamental form).
The trace of the induced metric is $h_{\mu\nu} g^{\mu\nu} =  h_{\mu\nu} h^{\mu\nu} = 2$ and the tensor $h_\mu^{\;\;\,\nu}$ allows to project all the other tensors on $\gamma$.
The extrinsic curvature (second fundamental form) of $\gamma$ embedded in $\mathcal{M}_3$ is defined as
\be
\label{extrinsic curv def}
K_{\mu\nu} =  h_\mu^{\;\;\alpha}h_\nu^{\;\;\beta}\,  \nabla_\alpha n_\beta\,,
\ee
where $\nabla_\alpha$ is the torsionless covariant derivative compatible with $g_{\mu\nu}$.
We find it convenient to introduce also the following traceless tensor constructed through the extrinsic curvature 
\be
\label{traceless Kmunu def}
\mathcal{K}_{\mu\nu}
= K_{\mu\nu}-\frac{\textrm{Tr}K}{2} \,h_{\mu\nu}\,.
\ee

An important identity to employ in our analysis is the following contracted Gauss-Codazzi relation \cite{opac-b1121838}
\be
\label{GC-contracted sec}
\mathcal{R} - \big(\textrm{Tr}K\big)^2 + \textrm{Tr}K^2
=
h^{\mu\rho} h^{\nu\sigma} \hspace{-.1cm} \perp\hspace{-.1cm} R_{\mu\nu \rho \sigma} \,,
\ee
where $\mathcal{R}$ is the Ricci scalar, which provides the intrinsic curvature of $\gamma$, and $\perp\hspace{-.1cm} R_{\mu\nu \rho \sigma}  
= h_\mu^{\;\;\alpha}h_\nu^{\;\;\beta} h_\rho^{\;\;\gamma}h_\sigma^{\;\;\lambda} R_{\alpha\beta \gamma \lambda}$ is the Riemann tensor of $g_{\mu\nu} $ projected on $\gamma$.
Performing explicitly the contractions, the r.h.s. of (\ref{GC-contracted sec}) reads
\be
\label{hhR}
h^{\mu\rho} h^{\nu\sigma} 
\hspace{-.1cm} \perp\hspace{-.1cm} R_{\mu\nu \rho \sigma} 
=
R - 2 \,n^\mu n^\nu R_{\mu\nu}
=
\,-\, 2 \,n^\mu n^\nu G_{\mu\nu}\,,
\ee
where $G_{\mu\nu}$ is the Einstein tensor of $g_{\mu\nu}$.

At any given point of $\gamma$, two principal curvatures $\kappa_1$ and $\kappa_2$ can be introduced, which are the eigenvalues of the extrinsic curvature. 
Thus, the mean curvature is given by $(\kappa_1 + \kappa_2)/2=\textrm{Tr}K/2$.

Many gravitational backgrounds occurring in the AdS/CFT correspondence are conformally related to asymptotically flat spacetimes (e.g. the asymptotic AdS$_4$ black holes and the domain wall geometries that will be introduced in \S\ref{sec static backs}).
Motivated by this fact, let us assume that the metric $g_{\mu\nu}$ of the background space is conformal to $\tilde{g}_{\mu\nu} $, namely
\be
\label{conformal metric sec2}
g_{\mu\nu}  = e^{2\varphi} \, \tilde{g}_{\mu\nu} \,,
\ee
where $\tilde{g}_{\mu\nu} $ defines an Euclidean asymptotically flat  space $\widetilde{\mathcal{M}}_3$ and $\varphi$ is a function of the coordinates. 
The surface $\gamma$ can be also seen as embedded into $\widetilde{\mathcal{M}}_3$ and therefore we can define the induced metric $\tilde{h}_{\mu\nu}$ and the extrinsic curvature $\widetilde{K}_{\mu\nu}$ characterising this embedding through $ \tilde{g}_{\mu\nu} $ as above. 
Denoting by $\tilde{n}_\mu$ the unit vector normal to the surface $\gamma \subset \widetilde{\mathcal{M}}_3$, we have $n_{\mu} = e^{\varphi} \tilde{n}_{\mu}$ (and therefore $n^{\mu} = e^{-\varphi} \tilde{n}^{\mu}$), and this implies that $h_{\mu\nu} = e^{2\varphi} \tilde{h}_{\mu\nu}$.
Considering the determinants (restricted to the tangent vectors) $h$ and $\tilde{h}$ of the induced metrics, we find that $h=e^{4\varphi} \tilde{h}$.
This leads us to conclude that the area elements $d\mathcal{A}  = \sqrt{h} \, d\Sigma$ and $d\tilde{\mathcal{A}}  = \sqrt{\tilde{h}} \, d\Sigma$ with $d\Sigma = d\sigma_1 d\sigma_2$ (we denoted by $\sigma_i$ some local coordinates) are related as $d\mathcal{A} = e^{2\varphi} d\tilde{\mathcal{A}}$.
Being the metrics $g_{\mu\nu}$ and  $\tilde{g}_{\mu\nu} $ conformally related, the corresponding extrinsic curvatures $K_{\mu\nu} $ and $\widetilde{K}_{\mu\nu}$ obey the following relation
\be
\label{Kmunu law}
K_{\mu\nu} 
=
e^{\varphi}  \big(  \widetilde{K}_{\mu\nu} + \tilde{h}_{\mu\nu} \,\tilde{n}^\lambda \partial_\lambda \varphi  \big)\,.
\ee

From the transformation rules given above,  it is not difficult to realise that the following combination is Weyl invariant 
\be
\label{2-dim comb}
\textrm{Tr} \mathcal{K}^2  \,d\mathcal{A} 
\,=\,
\left(\textrm{Tr} K^2 -\frac{1}{2} \big(\textrm{Tr} K\big)^2  \right) d\mathcal{A} \,.
\ee

By employing the Gauss-Codazzi relation (\ref{GC-contracted sec}),  together with (\ref{hhR}) to eliminate $\textrm{Tr} K^2$ in (\ref{2-dim comb}), 
the Weyl invariance of the combination (\ref{2-dim comb}) can be recast as
\be
\label{weyl inv static}
\left(
  \, \frac{1}{2} \big(\textrm{Tr} K \big)^2 - \mathcal{R}  - 2\,n^\mu n^\nu G_{\mu\nu} 
\right) d\mathcal{A} 
=
\left(
\, \frac{1}{2} \big(\textrm{Tr} \widetilde{K}\big)^2 - \widetilde{\mathcal{R}} - 2\, \tilde{n}^\mu \tilde{n}^\nu \widetilde{G}_{\mu\nu} 
\right) d\tilde{\mathcal{A}} \,,
\ee
where the tilded quantities refer to the asymptotically flat metric $\tilde{g}_{\mu\nu} $.
In the left and right side of (\ref{weyl inv static}), the same surface $\gamma$ is embedded either in $\mathcal{M}_3$ or in $\widetilde{\mathcal{M}}_3$ respectively.
The formulas for the change of $\mathcal{R}$ and $G_{\mu\nu} $ under a Weyl transformation are given respectively by
\bea
\label{intrinsic R weyl sec}
& &  \mathcal{R} \,=\,
e^{- 2 \varphi}
\big(
\widetilde{\mathcal{R}} 
-  2  \,\widetilde{\mathcal{D}}^2 \varphi 
\big)\,,
\\
\label{G weyl sec}
& &  G_{\mu\nu} \,=\,
\widetilde{G}_{\mu\nu} 
- 
\widetilde{\nabla}_\mu  \widetilde{\nabla}_\nu \varphi
+ \widetilde{\nabla}_\mu \varphi\, \widetilde{\nabla}_\nu  \varphi
+ \tilde{g}_{\mu\nu} \widetilde{\nabla}^2\varphi\,,
\eea
where $\widetilde{\mathcal{D}}_\mu$ is the covariant derivative constructed through $\tilde{h}_{\mu\nu}$ and $\widetilde{\mathcal{D}}^2$ the corresponding Laplacian operator.
By first plugging (\ref{intrinsic R weyl sec}) and (\ref{G weyl sec}) into (\ref{weyl inv static}) (using also that $n^{\mu} = e^{-\varphi} \tilde{n}^{\mu}$ and $d\mathcal{A}= e^{2\varphi} d\tilde{\mathcal{A}} $) and then integrating the resulting equation over $\gamma$, we find
\be
\label{0= combination static}
0\,=\,
\int_{\gamma} 
\left(
\widetilde{\mathcal{D}}^2 \varphi
-\widetilde{\nabla}^2\varphi 
+ \tilde{n}^\mu \tilde{n}^\nu \, \widetilde{\nabla}_\mu \widetilde{\nabla}_\nu \varphi
- \big( \tilde{n}^\lambda \partial_\lambda \varphi \big)^2
- \frac{1}{4} \big(\textrm{Tr} \widetilde{K}\big)^2
\right)
 d\tilde{\mathcal{A}}
 +
 \frac{1}{4}  \int_{\gamma} 
  \big(\textrm{Tr} K\big)^2 \,d\mathcal{A}\,.
\ee
Adding the area to both sides of this identity, it becomes
\be
\label{area generic J2}
\mathcal{A}[\gamma]
\,=\,
\int_{\gamma} 
\left(
\widetilde{\mathcal{D}}^2 \varphi
-\widetilde{\nabla}^2\varphi 
+ e^{2\varphi}
+ \tilde{n}^\mu \tilde{n}^\nu \, \widetilde{\nabla}_\mu \widetilde{\nabla}_\nu \varphi
- \big( \tilde{n}^\lambda \partial_\lambda \varphi \big)^2
- \frac{1}{4} \big(\textrm{Tr} \widetilde{K}\big)^2
\right)
 d\tilde{\mathcal{A}}
 +
 \frac{1}{4}  \int_{\gamma} 
  \big(\textrm{Tr} K\big)^2 \,d\mathcal{A}\,.
  \hspace{.1cm}
\ee
We remark that (\ref{area generic J2}) holds for a generic two dimensional surface embedded into the three dimensional Euclidean space given by (\ref{conformal metric sec2}).
The first term is a total derivative and therefore it vanishes if $\gamma$ is a closed surface without boundaries.

When $\gamma$ has a boundary (which could be made by many disjoint components), the first term in (\ref{area generic J2}) is a boundary term
\be
\label{area generic J2 bdy}
\mathcal{A}[\gamma]
\,=\,
\oint_{\partial \gamma} 
\tilde{b}^\mu  \partial_\mu \varphi\, d\tilde{s} 
-\int_{\gamma} 
\left(
\, \frac{1}{4} \big(\textrm{Tr} \widetilde{K}\big)^2
+\widetilde{\nabla}^2\varphi 
- e^{2\varphi}
- \tilde{n}^\mu \tilde{n}^\nu \, \widetilde{\nabla}_\mu \widetilde{\nabla}_\nu \varphi
+ \big( \tilde{n}^\lambda \partial_\lambda \varphi \big)^2
\right)
 d\tilde{\mathcal{A}}
 +
 \frac{1}{4}  \int_{\gamma} 
  \big(\textrm{Tr} K\big)^2 \,d\mathcal{A}\,.
  \hspace{.1cm}
\ee

In our case the three dimensional metric $g_{\mu\nu}$ is asymptotically $\mathbb{H}_3$ and $\partial \gamma$ lies close to the boundary. 
The three dimensional Euclidean hyperbolic space $\mathbb{H}_3$ is characterised by the metric $ds^2 = z^{-2} (dz^2 + d\boldsymbol{x}^2)$, where $z >0$ and  $d\boldsymbol{x}^2$ is the space element of $\mathbb{R}^2$.
Thus, let us consider a system of coordinates $(z,\boldsymbol{x})$ in $\mathcal{M}_3$, where $z>0$, the boundary of $\mathcal{M}_3$ is given by $z=0$  and $\boldsymbol{x}$ is the position vector in the $z=0$ plane.
The boundaries of the surfaces $\gamma_\varepsilon$  belong to the plane $z=\varepsilon $.
Taking $\varphi = -\log(z) + O(z^a)$ with $a>1$ when $z \to 0$ in (\ref{conformal metric sec2}), we have  that $g_{\mu\nu}$ is asymptotically $\mathbb{H}_3$ while $\tilde{g}_{\mu\nu}$ is asymptotically flat. 
Considering the surfaces $\gamma_\varepsilon$ and this behaviour for $\varphi $ in (\ref{area generic J2 bdy}), we need to know $\tilde{b}^z$ at $z=\varepsilon$.
In \S\ref{app vector b_mu} we report the analysis of  \cite{AM, Graham:1999pm}, which shows that $\tilde{b}^z = -1+ o(\varepsilon)$ as $\varepsilon \to 0$.
We remark that the latter condition holds also for a surface intersecting orthogonally the plane $z=0$ which is  not necessarily minimal.  
Thus, from (\ref{area generic J2 bdy}) we have that the area of the surfaces $\gamma_\varepsilon$ reads
\be
\label{area generic J2 bis}
\mathcal{A}[\gamma_\varepsilon]
\,=\,
\frac{P_A}{\varepsilon} - \mathcal{F}_A + o(1)\,,
\ee
 where the area law term comes from the boundary integral in (\ref{area generic J2 bdy}) and the $O(1)$ term is given by
\be
\label{FA generic surface}
\mathcal{F}_A \,\equiv\,
\int_{\gamma_A} 
\left(\,
\frac{1}{4} \big(\textrm{Tr} \widetilde{K}\big)^2
+ \big( \tilde{n}^\lambda \partial_\lambda \varphi \big)^2
+ \widetilde{\nabla}^2\varphi 
- e^{2\varphi}
- \tilde{n}^\mu \tilde{n}^\nu \, \widetilde{\nabla}_\mu \widetilde{\nabla}_\nu \varphi
\right)
 d\tilde{\mathcal{A}}
 -
 \frac{1}{4}  \int_{\gamma_A} 
  \big(\textrm{Tr} K\big)^2 \,d\mathcal{A}\,.
\ee

Let us  specialize (\ref{FA generic surface}) to the minimal area surfaces $\hat{\gamma}_A$ entering in the computation of the holographic entanglement entropy \cite{Ryu:2006bv, Ryu:2006ef}. 
For minimal area surfaces we have
\be
\label{minimality condition}
\textrm{Tr} K  = 0
\hspace{.5cm} \Longleftrightarrow \hspace{.5cm}
\big( \textrm{Tr}\widetilde{K} \big)^2 = 4 (\tilde{n}^\lambda \partial_\lambda \varphi )^2\,,
\ee
where the right side of the equivalence comes from (\ref{Kmunu law}).
Thus (\ref{area generic J2 bis}) becomes 
\be
\label{area generic RT}
\mathcal{A}[\hat{\gamma}_\varepsilon]
\,=\,
\frac{P_A}{\varepsilon} - F_A + o(1)\,,
\ee
with
\be
\label{FA sec}
F_A
 \,=\, 
\int_{\hat{\gamma}_A}   
\left[\,
  \frac{1}{2} \big(\textrm{Tr} \widetilde{K}\big)^2  
  +
\widetilde{\nabla}^2\varphi 
- e^{2\varphi}
- \tilde{n}^{\mu} \tilde{n}^{\nu} \,\widetilde{\nabla}_\mu \widetilde{\nabla}_\nu \varphi
\,\right]
   d\tilde{\mathcal{A}}\,,
\ee
where the first term of the integrand can be also written in terms of $\varphi$ like the other ones through (\ref{minimality condition}).
The formula (\ref{FA sec}) is the main result of this section. It is worth remarking that it holds for any smooth entangling curve $\partial A$, including the ones made by many disjoint components. 

A two dimensional surface can be defined implicitly as a real constraint $\mathcal{C}=0$, being $\mathcal{C}$ a function of the three coordinates $x^\mu$. 
The unit vector $\tilde{n}_\mu$ normal to this surface is obtained from this constraint as follows
\be
\label{unit vector C}
\tilde{n}_\mu  
= \frac{\partial_\mu \mathcal{C}}{\sqrt{\tilde{g}^{\alpha\beta} \,\partial_\alpha \mathcal{C} \,\partial_\beta \mathcal{C}}}\,.
\ee
Since the global sign of $\mathcal{C}$ is unspecified, the orientation of the vector $\tilde{n}_\mu$ is a matter of choice as well. Notice that this sign does not change (\ref{FA sec}) because only quadratic terms in $\tilde{n}^\mu  $ occur.

In \S\ref{sec app higher dims} we briefly discuss the application of the method employed above to the higher dimensional case.


\subsection{Some static backgrounds}
\label{sec static backs}

In this manuscript we will consider three examples of static asymptotically AdS$_4$ metrics: AdS$_4$,  asymptotically AdS$_4$ black holes and some domain wall geometries. We will not consider geometries which are singular when $z\to \infty$.
For the backgrounds we are interested in, (\ref{conformal metric sec2}) holds with $\varphi = -\log(z)$. 
Hence, the first and the last term of the integrand in (\ref{FA sec}) become respectively
\be
\label{explicit exp}
\big(\textrm{Tr} \widetilde{K}\big)^2  = \frac{4(\tilde{n}^{z})^2}{z^2} \,,
\qquad
\tilde{n}^\mu \tilde{n}^\nu \, \widetilde{\nabla}_\mu \widetilde{\nabla}_\nu \varphi
=
\frac{(\tilde{n}^{z})^2}{z^2}+\frac{1}{z} \, \widetilde{\Gamma}^z_{\mu\nu} \, \tilde{n}^\mu \tilde{n}^\nu \,,
\ee
where the first expression is obtained through (\ref{minimality condition}) and $\widetilde{\Gamma}^z_{\mu\nu}$ in the second expression are some components of the Christoffel connection compatible with  $\tilde{g}_{\mu\nu}$.
In the following we specify (\ref{FA sec}) to these three backgrounds.

\subsubsection{AdS$_4$: the Willmore energy}
\label{sec:ads4static}

The simplest bulk geometry to study is AdS$_4$, which is given by 
\be
\label{ads4 metric}
ds^2 
=
\frac{1}{z^2} \big( -dt^2+ dz^2 + d\boldsymbol{x}^2  \big)\,,
\ee
where the AdS radius has been set to one and $d\boldsymbol{x}^2 $ is infinitesimal spacetime interval of $\mathbb{R}^2$ at $z=0$.
Comparing (\ref{ads4 metric}) with (\ref{ansatz t=cost}) and (\ref{conformal metric sec2}), we have that $g_{\mu\nu} $ is the metric of $\mathbb{H}_3$ and $\tilde{g}_{\mu\nu} $ is the flat metric of $\mathbb{R}^3$. 
The latter fact leads to important simplifications in the general formulas given in \S\ref{sec general case}.
Indeed,  $\widetilde{\nabla}^2\varphi - e^{2\varphi} = 0$ and all the components of $\widetilde{\Gamma}^z_{\mu\nu}$ vanish.
Thus, for a generic surfaces $\gamma_A$ the expression (\ref{FA generic surface}) reduces to   \cite{Babich:1992mc, AM}
\be
\label{FA generic surface AdS4}
\mathcal{F}_A \,=\,
\frac{1}{4} \bigg(
\int_{\gamma_A} 
\big(\textrm{Tr} \widetilde{K}\big)^2 \,d\tilde{\mathcal{A}}
 -
 \int_{\gamma_A} \big(\textrm{Tr} K\big)^2 \,d\mathcal{A} \bigg)\,.
\ee
For the minimal area surfaces $\hat{\gamma}_A$, which satisfy the condition (\ref{minimality condition}),  it simplifies further to
\be
\label{FA ads4 sec}
F_A
 \,=\, 
   \frac{1}{4} 
\int_{\hat{\gamma}_A}   
\big(\textrm{Tr} \widetilde{K}\big)^2  \,    d\tilde{\mathcal{A}}
\,=\,
\int_{\hat{\gamma}_A}   
\frac{(\tilde{n}^{z})^2}{z^2} \,    d\tilde{\mathcal{A}}\,,
\ee
which can be found also by specifying (\ref{FA sec}) to $\tilde{g}_{\mu\nu} = \delta_{\mu\nu}$.
Notice that (\ref{FA ads4 sec}) does not depend on the choice of the coordinate system in the $z=0$ plane but, for explicit computations, this coordinate system must be fixed in order to write $\tilde{n}^{z}$ and $d\tilde{\mathcal{A}}$ (see \S\ref{sec app normal vector}).

\begin{figure}[t] 
\vspace{-.5cm}
\hspace{-.25cm}
\includegraphics[width=1.03\textwidth]{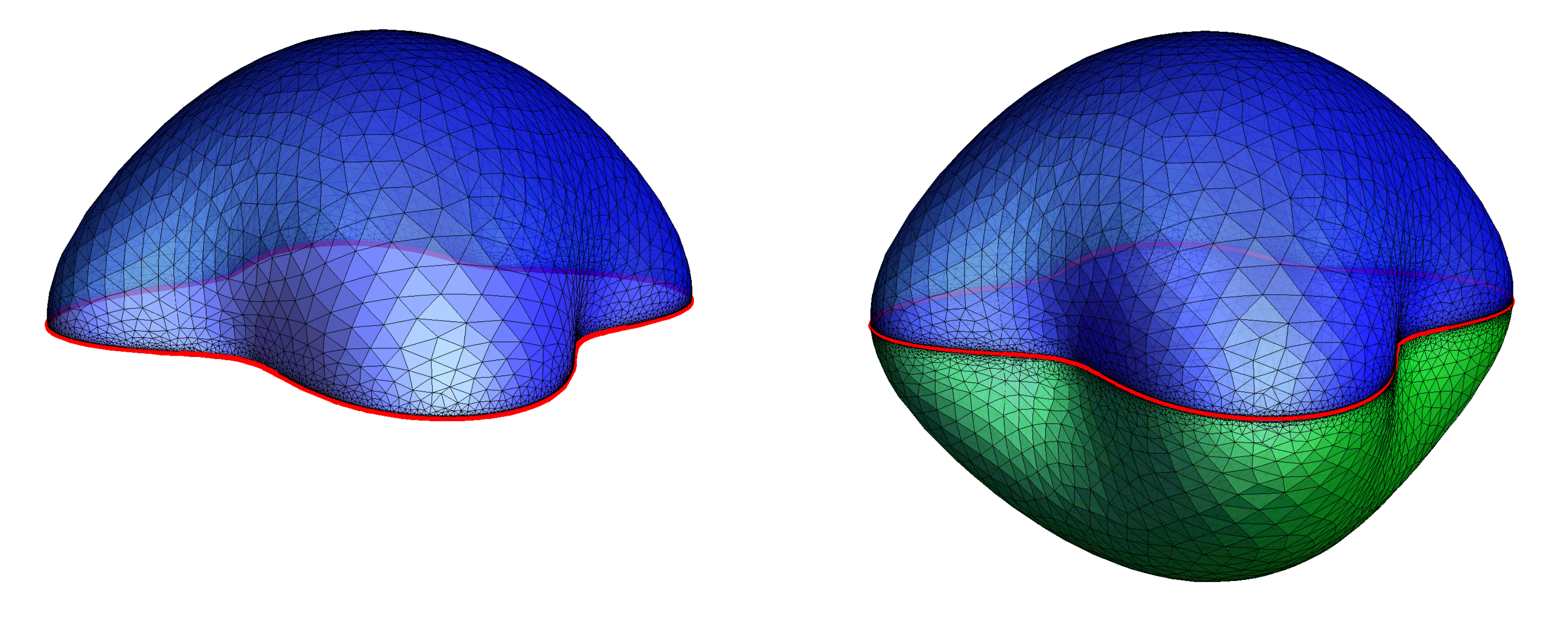}
\vspace{-.3cm}
\caption{\label{3Ddoubling}
Left: A minimal area surface $\hat{\gamma}_A$ for AdS$_4$ whose boundary at $z=0$ (entangling curve) is given by  the red curve.
Right: The closed surface $\hat{\gamma}^{\textrm{\tiny (d)}}_A$ embedded in  $\mathbb{R}^3$, obtained from $\hat{\gamma}_A$
by attaching $\hat{\gamma}_A$ (blue part) and its reflected copy $\hat{\gamma}^{\textrm{\tiny (r)}}_A$ (green part) along $\partial A$ (red curve), which is an umbilic line for $\hat{\gamma}^{\textrm{\tiny (d)}}_A$ \cite{Babich:1992mc}.
}
\end{figure}

Following \cite{Babich:1992mc}, we find it convenient to introduce a closed surface $\hat{\gamma}^{\textrm{\tiny (d)}}_A$ embedded in  $\mathbb{R}^3$  obtained by ``doubling'' $\hat{\gamma}_A$.
In particular, $\hat{\gamma}^{\textrm{\tiny (d)}}_A$ is the union $\hat{\gamma}^{\textrm{\tiny (d)}}_A = \hat{\gamma}_A \cup \hat{\gamma}^{\textrm{\tiny (r)}}_A $, where $ \hat{\gamma}^{\textrm{\tiny (r)}}_A $ is the surface with $z<0$ obtained by reflecting  the minimal surface $\hat{\gamma}_A$ with respect to the plane $z=0$.
The entangling curve $\partial A$ is a particular curve on the closed surface $\hat{\gamma}^{\textrm{\tiny (d)}}_A$ and in \cite{Babich:1992mc} it has been found that the two principal curvatures are equal on this curve (i.e. $\partial A$ is an {\it umbilic} line).
The set of  closed oriented compact surfaces given by $\hat{\gamma}^{\textrm{\tiny (d)}}_A$ as $A$ varies within the set of domains with smooth $\partial A$ is strictly included into the set of the Riemann surfaces embedded in $\mathbb{R}^3$. 
Indeed, they are symmetric with respect to the $z=0$ plane and their intersection with such plane is an umbilic closed curve. 
In Fig.\,\ref{3Ddoubling} we show a minimal surface $\hat{\gamma}_A$ and the corresponding closed surface $\hat{\gamma}^{\textrm{\tiny (d)}}_A$ (the red curve on $\hat{\gamma}^{\textrm{\tiny (d)}}_A$   along which $\hat{\gamma}_A$ and $\hat{\gamma}^{\textrm{\tiny (r)}}_A$ match is an umbilic line).
It is worth remarking that already among the connected domains $A$ one can find cases such that $\hat{\gamma}^{\textrm{\tiny (d)}}_A$ has genus two or higher\footnote{We are grateful to Veronika Hubeny for pointing this issue out to us.}.

The formula in (\ref{FA ads4 sec}) tells us that $F_A$ is related to the Willmore energy of $\hat{\gamma}_A \subset \mathbb{R}^3$.
Given an oriented, smooth and closed two dimensional surface $\Sigma_g$ with genus $g$ embedded in $\mathbb{R}^3$, the Willmore energy functional evaluated on $\Sigma_g$ is defined as \cite{Thomsen, blaschke, willmorebound,willmorebook}
\be
\label{willmore functional def}
\mathcal{W}[\Sigma_g] \equiv   \frac{1}{4} 
\int_{\Sigma_g}   
\big(\textrm{Tr} \widetilde{K}\big)^2  \,    d\tilde{\mathcal{A}}\,.
\ee
In terms of the principal curvatures $\kappa_1$ and $\kappa_2$ of the surface $\Sigma_g$, the Willmore energy (\ref{willmore functional def}) is the integral of $[(\kappa_1+\kappa_2)/2]^2$ (i.e. the square of the mean curvature) over $\Sigma_g$.
The Willmore energy of a round sphere with radius $R$ is $4\pi$, independently of the radius.
Surfaces extremizing the functional (\ref{willmore functional def}) are called Willmore surfaces.  
It is possible to prove that, for a generic surface $\Sigma_g$ (see Theorem 7.2.2 in \cite{willmorebook})
\be
\label{bound willmore any surface}
\mathcal{W}[\Sigma_g] \,\geqslant\, 4\pi\,,
\ee
where the bound is saturated only by round spheres, for which every point is umbilic. 
Considering domains $A$ with the same perimeter, from (\ref{FA generic surface AdS4}) one can realise that the surface $\hat{\gamma}^{\textrm{\tiny (d)}}_A$ is also a critical point of the functional (\ref{willmore functional def}) \cite{Babich:1992mc}. 
Among the large number of papers in the mathematical literature about the Willmore functional, let us mention \cite{White:1973abc, Bryant:1984abc, Simon:1993abc, Kuwert:2004abc, Riviere:2010abc, Mondino:2013abc, willmoreboundproof, MondinoMalchiodi}.

Given (\ref{FA ads4 sec}) and (\ref{willmore functional def}), one concludes that, when the bulk geometry is AdS$_4$, the term $F_A$ is the Willmore energy of the surface $\hat{\gamma}_A$ embedded in $\mathbb{R}^3$ \cite{Babich:1992mc}. The surface $\hat{\gamma}_A$ lies in the part $z\geqslant 0$ of $\mathbb{R}^3$ and its boundary is at $z=0$. 
Considering the closed surface $\hat{\gamma}^{\textrm{\tiny (d)}}_A \subset \mathbb{R}^3$ introduced above, it is straightforward to observe that
\be
\label{FA from willmore}
F_A = \frac{1}{2}\, \mathcal{W}\big[ \hat{\gamma}^{\textrm{\tiny (d)}}_A \big] \,.
\ee

From (\ref{FA from willmore}) and (\ref{bound willmore any surface}), it is straightforward to realise that for a simply connected domain $A$ we have
\be
F_A \, \geqslant\,  2\pi \,,
\ee
where the bound of $2\pi$ is saturated only when $A$ is a disk. 
Thus, the disk maximises the holographic entanglement entropy for AdS$_4$ among the domains having the same perimeter (the problem of finding the shape which maximises  $S_A$ in higher dimensions has been addressed in \cite{Astaneh:2014uba}).
We remark that the bound (\ref{bound willmore any surface}) applies also for $A$ made by disjoint domains (see \S\ref{sec HMI}).

It is interesting to observe that, considering a domain $A$ and another one $A'$ obtained by rescaling $A$ through a factor $\lambda$ keeping the same shape, i.e. the same ratios of the various geometric parameters, we have that $F_A = F_{A'}$.
Indeed, the minimal surface $\hat{\gamma}_{A'}$ can be found by rescaling $\hat{\gamma}_{A}$ through the same factor $\lambda$ and in the integrand of (\ref{FA ads4 sec})  we have that $z \to \lambda z$, $d\tilde{\mathcal{A}} \to \lambda^2 d\tilde{\mathcal{A}}$, while $\tilde{n}^{z}$ remains invariant.
Thus, $F_A$ is obtained from $F_{A'}$ through a straightforward change of variables. 
Since this result comes from the fact that $(z,x,y) \to \lambda (z,x,y) $ is an isometry of $\mathbb{H}_3$, it does not hold for the spacetimes occurring in \S\ref{sec bh gen} and \S\ref{sec dw}, which do not have this isometry.

A generalisation of the Willmore energy functional (\ref{willmore functional def}) is the Helfrich energy functional \cite{Helfrich:1973abc}, whose role is very important in the study of the cell membranes \cite{cell-review}. 
Considering the surfaces $\gamma_A$ intersecting the boundary $z=0$ orthogonally, in \S\ref{app helfrich} we have briefly discussed the surface $\hat{\gamma}_A^{\textrm{\tiny (H)}}$ whose part restricted to $z\geqslant \varepsilon$ has an area given by (\ref{area generic J2 bis}) where the $O(1)$ term of the expansion given in (\ref{FA generic surface AdS4}) is the  Helfrich energy of $\hat{\gamma}_A^{\textrm{\tiny (H)}}$ as surface embedded in $\mathbb{R}^3$.

\subsubsection{Black holes}
\label{sec bh gen}

The asymptotically AdS$_4$ charged black hole (Reissner-Nordstr\"om-AdS black hole) \cite{Romans:1991nq, London:1995ib, Chamblin:1999tk} is given by
\be
\label{bh 4dim}
ds^2 
=
\frac{1}{z^2} \left( -\,f(z)\,dt^2+ \frac{dz^2}{f(z)} +  d\boldsymbol{x}^2  \right)\,,
\qquad
f(z) = 1 - M z^{3} + Q^2 z^4\,,
\ee
where $M$ is the mass and $Q$ is the charge of the black hole. 
The Hawking temperature of this black hole vanishes in the extremal case, for which the two horizons coincide and the emblacking function becomes $f(z)=1-4(z/z_h)^3+3(z/z_h)^4$.
The Schwarzschild-AdS black hole corresponds to the uncharged case $Q=0$ and for this geometry the horizon is $z_h = 1/\sqrt[3]{M}$.

\begin{figure}[t] 
\vspace{-.5cm}
\hspace{-.25cm}
\includegraphics[width=1.02\textwidth]{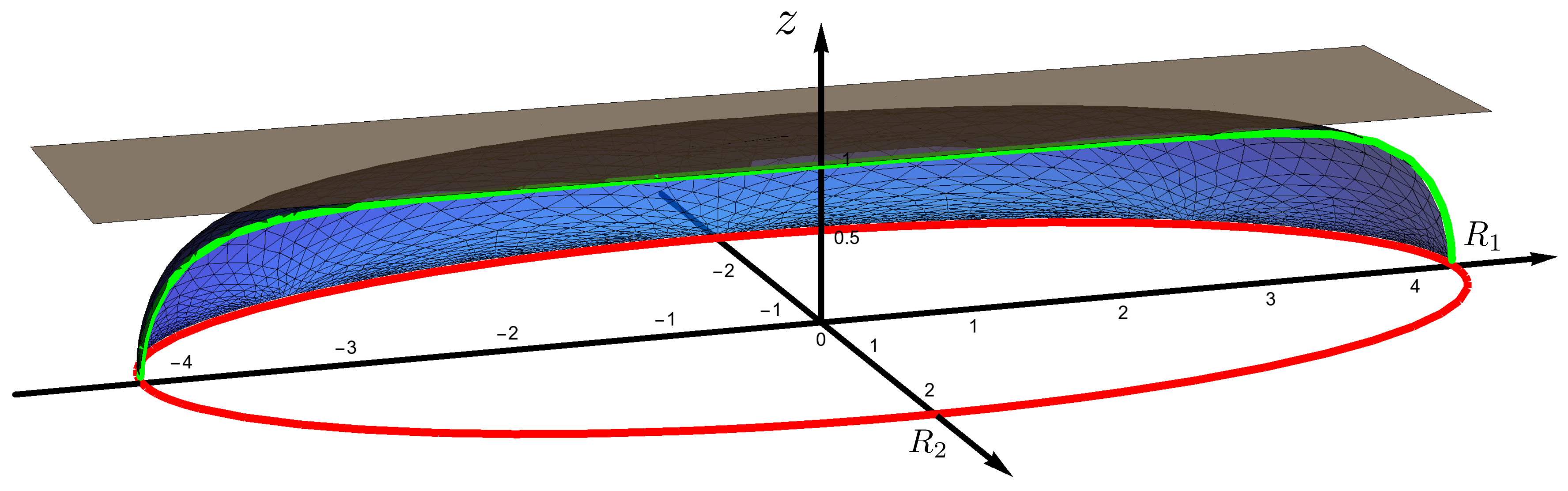}
\vspace{-.1cm}
\caption{\label{3DellipseBH}
Minimal area surface $\hat{\gamma}_A$ for a Schwarzschild-AdS black hole.
The entangling curve $\partial A$ is an ellipse with semi-major axis $R_1$ and semi-minor axis $R_2$ (the red curve is plotted at $z=\varepsilon$).
Here $\varepsilon=0.01$ and the grey plane corresponds to the horizon at $z_h=1$.
Only half of the surface is shown in order to highlight a section of the surface (green curve) which reaches the highest value $z_\ast < z_h$ of the coordinate $z$ for the whole surface.
In this case $z_\ast$ is the intersection between the green curve and the $z$ axis. 
}
\end{figure}

Comparing (\ref{conformal metric sec2}) and (\ref{bh 4dim}), we have that $\varphi = -\log(z)$ and $\tilde{g}_{\mu\nu}$ is provided by the metric within the parenthesis in (\ref{bh 4dim}).
In this case all the terms occurring in (\ref{FA sec}) are non trivial. In particular, we get
\be
\label{liouville bh}
\widetilde{\nabla}^2\varphi  - e^{2\varphi} 
\,=\, \frac{f(z) - z f'(z)/2-1}{z^2}\,,
\qquad
\tilde{n}^\mu \tilde{n}^\nu  \, \widetilde{\nabla}_\mu \widetilde{\nabla}_\nu \varphi
\,=\,
\frac{(\tilde{n}^z)^2}{z^2} \left(  1 - \frac{z  f'(z)}{2f(z)}\,  \right) ,
\ee
where we recall that $\tilde{n}^z= f(z)\, \tilde{n}_z$.
Combining these results with the expression for $(\textrm{Tr} \widetilde{K})^2 $ in (\ref{explicit exp}) we find that (\ref{FA sec}) becomes
\be
\label{FA bh gen}
F_A 
    \,=\,
  \int_{\hat{\gamma}_A} 
  \frac{1}{z^2} \left[ \left( 1+ \frac{z  f'(z)}{2f(z)} \right) (\tilde{n}^{z})^2
+ f(z) - \frac{z f'(z)}{2}-1\, 
 \right]
   d\tilde{\mathcal{A}}  \,.
\ee
The choice of the system of coordinates in the $z=0$ plane enters in the explicit expressions of $\tilde{n}^{z}$ and  $d\tilde{\mathcal{A}} $ (see \S\ref{sec app normal vector}).
In Fig.\,\ref{3DellipseBH} we show a minimal area surface $\hat{\gamma}_A$ for which the entangling curve $\partial A$ is an ellipse and the bulk geometry is the Schwarzschild-AdS black hole. 
Denoting by $z_\ast$ the highest value of the coordinate $z$ reached by the points of $\hat{\gamma}_A$, for a static asymptotically AdS black hole we have that $z_\ast < z_h$, i.e. the minimal surface does not penetrate the horizon \cite{Barbon:2008sr, Hubeny:2012ry}.

As first consistency check of (\ref{FA bh gen}), we observe that for  $f(z) =1$ identically  the expression (\ref{FA ads4 sec}) for AdS$_4$ is recovered, as expected.

When the domain $A$ is very large, we expect a minimal area surface $\hat{\gamma}_A$ close to a cylindrical surface $\hat{\gamma}_A^{\textrm{\tiny cyl}}$  whose horizontal cross section is $\partial A$ and having only one base at constant $z=z_\ast \lesssim z_h$. 
Hence, we expect that $F_A$ is also close to the integral in (\ref{FA bh gen}) evaluated on $\hat{\gamma}_A^{\textrm{\tiny cyl}}$, that will be denoted by $F_A^{\textrm{\tiny cyl}}$.
The latter quantity is the sum of two contributions: the integral over the base and the one over the vertical part of the cylinder, whose height is $z_\ast \lesssim z_h$. 
As for the former term, whose integration domain is horizontal, we have $\tilde{n}_{z}=1/\sqrt{f(z_\ast)}$ and therefore the integral turns out to be proportional to the area of $A$.
Instead, on the vertical part of $\hat{\gamma}_A^{\textrm{\tiny cyl}}$  we have $\tilde{n}_{z}=0$ and the corresponding integral is proportional to $P_A$.
The sum of these terms reads
\be
\label{FA_cyl}
F_A^{\textrm{\tiny cyl}} = 
  \frac{2f(z_\ast)-1}{z_\ast^2} \, \textrm{Area}(A)
 + P_A 
  \int_0^{z_\ast}
  \frac{1}{z^2} \left[  f(z) - \frac{z f'(z)}{2}-1\, \right]
   dz\,,
\ee
where  the term containing $\textrm{Area}(A)$ dominates when $A$ is large. 
Since $z_\ast$ is close to the horizon, one easily finds that the leading term in (\ref{FA_cyl}) is $F_A^{\textrm{\tiny cyl}} =  - \, \textrm{Area}(A)/z_h^2 + \dots $ (see e.g. \cite{Barbon:2008sr, Tonni:2010pv, Liu:2013una}).

\subsubsection{Domain wall geometries}
\label{sec dw}

\begin{figure}[t] 
\vspace{-.5cm}
\hspace{-.25cm}
\includegraphics[width=1.02\textwidth]{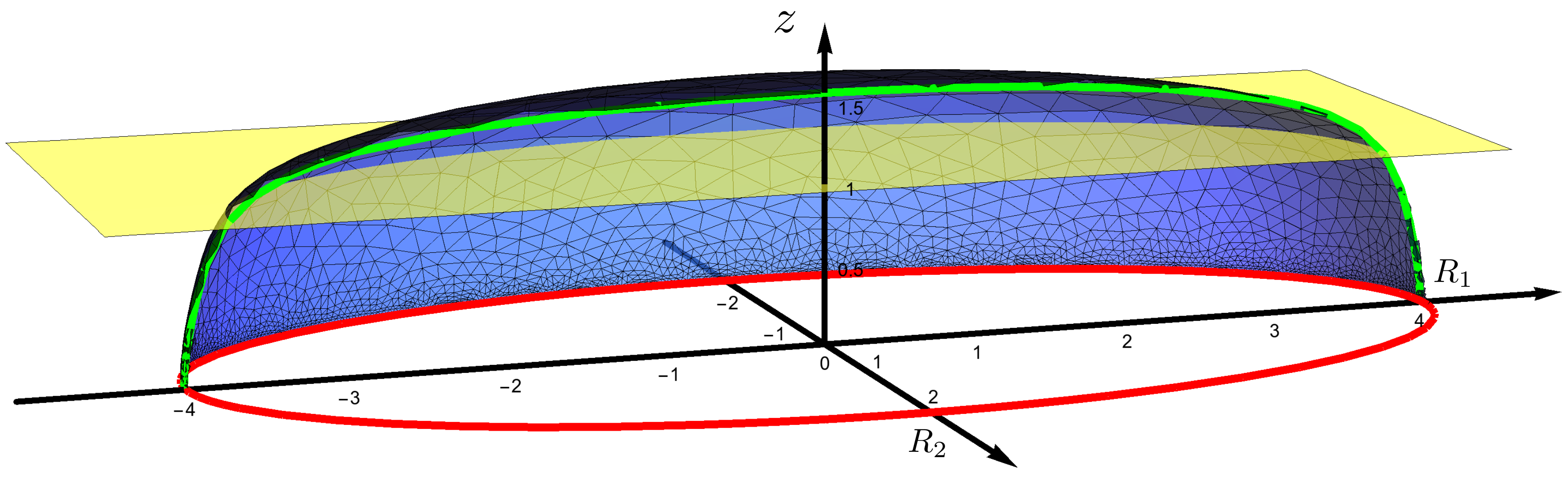}
\vspace{-.1cm}
\caption{\label{3DellipseRG}
Minimal area surface $\hat{\gamma}_A$ for the domain wall geometry (\ref{holog rg 4dim}) with $\alpha=2$ and $\gamma=1$.
The yellow plane corresponds to $z_{\textrm{\tiny RG}} =1$.
The entangling curve $\partial A$ is an ellipse whose semi axis $R_2<R_1$ (the red curve is plotted at $z=\varepsilon$ and here $\varepsilon=0.01$).
The green curve is a section whose intersection with the $z$ axis provides the highest value $z_\ast$ for the coordinate $z$ on the surface. When the domain $A$ is very large $z_\ast \gg z_{\textrm{\tiny RG}}$ and the deep IR region is probed, where the asymptotic geometry is AdS$_4$ with radius  $L_{\textrm{\tiny IR}} =1/(1+\gamma\alpha)$. 
}
\end{figure}

Asymptotically AdS$_4$ static backgrounds have been introduced also to provide a holographic dual description of a RG flow of the boundary theory \cite{Freedman:1999gp, Girardello:1998pd, Girardello:1999bd}. 
The holographic entanglement entropy for these geometries has been already studied in \cite{Albash:2010mv, Myers:2010tj, Myers:2012ed, Liu:2012eea,  Liu:2013una}, mainly for the infinite strip and for the disk. 

The example that we are going to consider is given by the following four dimensional bulk metric\footnote{We are grateful to Rob Myers for addressing our attention to this metric and for useful discussions about it.}
\be
\label{holog rg 4dim}
ds^2
\,=\, 
 \frac{1}{z^2} \left(\frac{- \,dt^2 +d\boldsymbol{x}^2}{p(z)} + dz^2 \, \right) ,
\qquad
p(z) = \big[1+(z/z_{\textrm{\tiny RG}})^\alpha\big]^{2\gamma} ,
\ee
where $z>0$ and  $\alpha >0$ to guarantee a well defined $z\to 0 $ behaviour.
The background (\ref{holog rg 4dim}) has a crossover scale $z_{\textrm{\tiny RG}}$ separating the ultraviolet (UV) region $z \gg z_{\textrm{\tiny RG}}$ from the infrared (IR) region $z \ll z_{\textrm{\tiny RG}}$, where the metric  (\ref{holog rg 4dim})  asymptotes  to AdS$_4$ with different radii.
Indeed, when $z/z_{\textrm{\tiny RG}} \ll 1$ we easily recover AdS$_4$ with unit radius $L_{\textrm{\tiny UV}} =1$, while for $z/z_{\textrm{\tiny RG}} \gg 1$, by introducing the variable $u / L_{\textrm{\tiny IR}} = z^{1+\gamma\alpha}/(z_{\textrm{\tiny RG}}^{\gamma\alpha} L_{\textrm{\tiny UV}} )$, we get  AdS$_4$ with radius $L_{\textrm{\tiny IR}} =1/(1+\gamma\alpha)$.
The null energy condition for the four dimensional metric $g_{MN}$ in (\ref{holog rg 4dim}) specified to null vector $\ell^M=(-\sqrt{p(z)} , 1,0,0)$ provides the condition $p [z\,p''+p'] - z(p')^2 \geqslant 0$, which tells us that $\gamma >0$, once the explicit expression for $p(z)$ in (\ref{holog rg 4dim}) is substituted. 
Thus, since $\gamma \alpha > 0$, we have that  $L_{\textrm{\tiny IR}} < L_{\textrm{\tiny UV}} $.
Plotting the Ricci scalar of (\ref{holog rg 4dim}) normalized by its value at large $z/z_{\textrm{\tiny RG}}$ in terms of $z/z_{\textrm{\tiny RG}}$, one observes that the smooth transition between the two asymptotic AdS$_4$ is faster as $\alpha$ increases for a given $\gamma$.

The metric (\ref{holog rg 4dim}) can be written also as $ds^2 = \zeta^{-2} [- \,dt^2 +d\boldsymbol{x}^2 + d\zeta^2/P(\zeta)]$ (see e.g. \cite{Liu:2012eea, Liu:2013una}), where $\zeta = z\sqrt{p(z)}$ and $P(\zeta) = (1+z\,p'(z)/[2p(z)])^2$. In terms of this holographic coordinate the above null energy condition becomes simply $P'(\zeta) \geqslant 0$.
Notice that we cannot write $z=z(\zeta)$ analytically for generic values of the parameters.

Denoting by $z_\ast$ the highest value of the coordinate $z$ for the minimal area surface $\hat{\gamma}_A$, we have that $\hat{\gamma}_A$ probes the UV regime when $z_\ast \ll z_{\textrm{\tiny RG}}$ and the IR regime when $z_\ast \gg z_{\textrm{\tiny RG}}$.

As for the term $F_A$ of the holographic entanglement entropy given in (\ref{FA sec}) for this gravitational background, by specifying (\ref{explicit exp}) for the metric (\ref{holog rg 4dim}) we find 
\be
\label{rg parts}
\widetilde{\nabla}^2\varphi  - e^{2\varphi} 
\,=\,  \frac{p'(z)}{z\, p(z)}\,,
\qquad
\tilde{n}^{\mu} \tilde{n}^{\nu} \,\widetilde{\nabla}_\mu \widetilde{\nabla}_\nu \varphi
\,=\,
\frac{(\tilde{n}^z)^2}{z^2} +\frac{p'(z)}{2z\,p(z)}\Big[ 1 - (\tilde{n}^z)^2 \Big]\,.
\ee
Notice that a coordinate system must be chosen to evaluate $\widetilde{\Gamma}^z_{\mu\nu} = -\tfrac{1}{2} \partial_z \tilde{g}_{\mu\nu}$ and to implement the normalisation condition for the vector $\tilde{n}^z$. Nevertheless, the expressions we give here hold for both cartesian and polar coordinate systems in the $z=0$ plane.
By employing (\ref{rg parts}), the formula (\ref{FA sec}) for (\ref{holog rg 4dim}) becomes 
\be
\label{FA holog rg}
F_A 
    \,=\,
  \int_{\hat{\gamma}_A} 
  \frac{1}{z^2} \left[\left( 1+ \frac{z \, p'(z)}{2 \,p(z)} \right) (\tilde{n}^{z})^2 
+ \frac{z\,p'(z)}{2\, p(z)} \, 
 \right]
   d\tilde{\mathcal{A}}  \,.
\ee

Let us restrict to $\alpha >1$ to guarantee the finiteness of (\ref{FA holog rg}). 
When $p(z) =1$ identically (\ref{FA holog rg}) reduces to (\ref{FA ads4 sec}) for AdS$_4$, as expected. 
In Fig.\,\ref{3DellipseRG} we show a minimal surface $\hat{\gamma}_A$ whose entangling curve $\partial A$ is an ellipse (the same one of Fig.\,\ref{3DellipseBH}) and for which the bulk spacetime is the domain wall geometry (\ref{holog rg 4dim}). The parameters of the ellipse and the scale $z_{\textrm{\tiny RG}}$ are such that $z_\ast > z_{\textrm{\tiny RG}}$.


\section{Time dependent backgrounds}
\label{sec time-dep}

The holographic entanglement entropy can be computed also for asymptotically AdS time dependent backgrounds by employing the prescription given in \cite{Hubeny:2007xt}.
In these cases, the area functional to extremize must be evaluated on a class of two dimensional surfaces $\gamma_A$ (i.e. such that $\partial \gamma_A = \partial A$) which is larger than the one occurring in the static case. 
Indeed, the covariance of the proposal removes the restriction to the constant time slice, that is natural in the static case. 
Thus, for the time dependent backgrounds the surfaces $\gamma_A$ to consider in the extremization process are embedded into the whole four dimensional Lorentzian spacetime. 
\\
In this section we extend the analysis performed in \S\ref{sec general case} to four dimensional  time dependent bulk spacetimes.

\subsection{General case}
\label{sec time-dep general}

Consider a two dimensional spacelike surface $\gamma$ embedded in a four dimensional Lorentzian spacetime $\mathcal{M}_4$ characterized by the metric $g_{MN}$. 
Given two unit vectors $n^{(i)}$ (with $i\in \{1,2\}$) normal to $\gamma$ and orthogonal between them, the induced metric on $\gamma$ reads
\be
\label{induced metric time-dep}
h_{MN}
\,=\,
g_{MN}- \sum_{i=1}^2 \epsilon_i\,n^{(i)}_{M} n^{(i)}_{N} \,,
\ee
where $\epsilon_i=g_{MN} \, n^{(i)M}\,n^{(i)N}$ is either $-1$ or $+1$. Notice that $h_{MN} n^{(i)N}=0$.
For each unit normal vector $n^{(i)}$, we can compute the corresponding  extrinsic curvature and the associated traceless combination, which are respectively
\be
\label{extrinsic curv time}
K^{(i)}_{MN}
\,=\,
h_{M}^{\;\;\;A} h_{N}^{\;\;\;B} \, \nabla_{A} n_B^{(i)} \,,
\qquad
\mathcal{K}^{(i)}_{MN} \,\equiv\,  K^{(i)}_{MN}-\frac{\mathrm{Tr} K^{(i)} }{2} \, h_{MN} \,.
\ee
We recall that $K^{(i)}_{MN} n^{(i)N}=0$ for $i\in \{1,2\}$.

In this case we need to consider the following Gauss-Codazzi equation \cite{opac-b1121838}
\be
\label{GCtime}
h_{M}^{\;\;\;A}h_{N}^{\;\;\;B}h_{R}^{\;\;\;C}h_{S}^{\;\;\;D}
R_{ABCD} 
\,=\,
\mathcal{R}_{MNRS}
- \sum_{i=1}^2 \epsilon_i \Big[K^{(i)}_{MR} K^{(i)}_{NS}  - K^{(i)}_{MS} K^{(i)}_{NR}\,\Big] \,,
\ee
and, following the analysis done in \S\ref{sec general case} for the static case, let us take the contraction given by
\be
\label{contracted GC}
\mathcal{R}
- \sum_{i=1}^2 \epsilon_i \Big[\big(\mathrm{Tr} K^{(i)}\big)^2 - \mathrm{Tr} (K^{(i)})^2\Big]
\,=\,
h^{MR} h^{NS} \hspace{-.1cm} \perp\hspace{-.1cm} R_{MNRS} \,.
\ee
By employing (\ref{induced metric time-dep}), the r.h.s. of (\ref{contracted GC}) can be expanded in terms of the orthogonal vectors $n^{(1)}$ e $n^{(2)}$, finding
\be
\label{Riema}
h^{MR} h^{NS}  \hspace{-.1cm} \perp\hspace{-.1cm} R_{MNRS}
\,=\,
2 \,\epsilon_1 \epsilon_2 \,R(n^{(1)}, n^{(2)} , n^{(1)} , n^{(2)})
-2 \sum_{i=1}^2 \epsilon_i  \,R(n^{(i)},n^{(i)}) +R \,,
\ee
where, in order to avoid a proliferation of indices, we have adopted the notation such that a scalar quantity with parenthesis stands for the contraction of the corresponding tensor with the vectors within the parenthesis in the specified order. 
Let us rewrite the r.h.s. of (\ref{Riema}) by replacing the contraction involving the Riemann tensor with the same contraction of the Weyl tensor according to the following formula\footnote{
We recall that, for a $q\geqslant 4$ dimensional spacetime (in our case $q=4$), the Weyl tensor is defined as \cite{Wald}
\be
 W_{ik\ell m}
 \,=\,
 R_{ik\ell m} - \frac{1}{q-2}\big(
 R_{i\ell}g_{km}  
- R_{im}g_{k\ell} 
- R_{k\ell}g_{im}
+ R_{km}g_{i\ell} \big)
+ \frac{1}{(q-1)(q-2)} \,R \big(
g_{i\ell}g_{km} - g_{im}g_{k\ell} \big) \,.
\nonumber
\ee
}
\be
\label{weyl and riemann contracted}
\epsilon_1 \epsilon_2 \,W(n^{(1)}, n^{(2)} , n^{(1)} , n^{(2)})
\,=\,
\epsilon_1 \epsilon_2 \,R(n^{(1)}, n^{(2)} , n^{(1)} , n^{(2)})
-\frac{1}{2} \sum_{i=1}^2 \epsilon_i \,R(n^{(i)},n^{(i)})
+ \frac{R}{6} \,.
\ee
The reason to prefer the Weyl tensor to the Riemann tensor in our analysis is that the former one changes in a nice way under conformal transformations \cite{Wald}. 
Thus, (\ref{Riema}) becomes
\be
\label{hhR weyl G version}
h^{MR} h^{NS}  \hspace{-.1cm} \perp\hspace{-.1cm} R_{MNRS}
\,=\,
2\,\epsilon_1 \epsilon_2 \,W(n^{(1)}, n^{(2)} , n^{(1)} , n^{(2)})
- \sum_{i=1}^2 \epsilon_i \, G(n^{(i)},n^{(i)})
- \frac{R}{3}\,,
\ee
where we have also employed the definition of the Einstein tensor $G_{MN}$ of the metric $g_{MN}$.

In order to follow the procedure discussed in \S\ref{sec general case} for the static case, we need to construct a Weyl invariant expression suggested by the contracted Gauss-Codazzi equation (\ref{contracted GC}).
From (\ref{Kmunu law}), we have that $\textrm{Tr}(\mathcal{K}^{(i)})^2\, d \mathcal{A}$ is Weyl invariant.
Hence, in this case we need to consider
\begin{subequations}
\bea
\sum_{i =1}^2 \epsilon_i \, \textrm{Tr}(\mathcal{K}^{(i)})^2\, d \mathcal{A}
&=&
\sum_{i =1}^2 \epsilon_i 
\bigg(\textrm{Tr}(K^{(i)})^2-\frac{1}{2}\big(\mathrm{Tr}K^{(i)}\big)^2\bigg)d \mathcal{A}
\\
\label{weyl inv explicit time-dep}
&=&
\bigg[\,
\frac{1}{2} \sum_{i =1}^2 \epsilon_i \big(\mathrm{Tr}K^{(i)}\big)^2 - \mathcal{R} 
+ h^{MR} h^{NS} \hspace{-.1cm} \perp\hspace{-.1cm}  R_{MNRS}
\bigg]d \mathcal{A}\,,
\eea
\end{subequations}
where in the last step we have eliminated the $\sum_i \epsilon_i  \,\mathrm{Tr}(\widetilde K^{(i)})^2$  by means of the contracted Gauss-Codazzi equation (\ref{contracted GC}).
By employing  (\ref{hhR weyl G version}), the Weyl invariant expression in (\ref{weyl inv explicit time-dep}) can be written as follows
\be
\label{weyl inv combination time}
\bigg(\,
\frac{1}{2} \sum_{i=1}^2 \epsilon_i  \big(\mathrm{Tr} K^{(i)}\big)^2
- \mathcal{R}
+ 2\,\epsilon_1 \epsilon_2 \,W(n^{(1)}, n^{(2)} , n^{(1)} , n^{(2)})
- \sum_{i=1}^2 \epsilon_i \,G(n^{(i)},n^{(i)})
- \frac{1}{3} R
\bigg) d\mathcal{A}\,.
\ee
Let us  first write explicitly the Weyl invariance of (\ref{weyl inv combination time}) and then integrate the resulting equation on a surface $\gamma$. 
Given the transformation property of the Weyl tensor, the two terms containing it cancel in the equation provided by the Weyl invariance of (\ref{weyl inv combination time}).
Then, we need the following transformation rules for the four dimensional Ricci scalar and Einstein tensor respectively
\bea
\label{Rtrans 4dim}
R & = & 
e^{- 2 \varphi}
\Big[
\widetilde{R} 
- \,6 \big( \widetilde{D}^2 \varphi  +  \widetilde{D}^S  \varphi\, \widetilde{D}_S  \varphi \big)
\Big]\,,
\\
\label{Gtrans 4dim}
G_{MN} 
&=&
\widetilde{G}_{MN} 
- 2
\big(
\widetilde{D}_M  \widetilde{D}_N  \varphi
- \widetilde{D}_M \varphi\, \widetilde{D}_N  \varphi
- \tilde{g}_{MN} \widetilde{D}^2\varphi
\big)
+\tilde{g}_{MN} 
 \widetilde{D}^S  \varphi\, \widetilde{D}_S  \varphi \,,
\eea
where $\widetilde{D}_M $ is covariant derivative compatible with $\tilde{g}_{MN}$.
By employing (\ref{intrinsic R weyl sec}), (\ref{Rtrans 4dim}) and (\ref{Gtrans 4dim}) into the equation for the Weyl invariance of  (\ref{weyl inv combination time}), one finds that
\bea
\label{0= combination time dep}
& & 
\hspace{-.8cm}
0 \,=\,
\int_{\gamma} 
\Bigg[\,
\widetilde{\mathcal{D}}^2 \varphi
+
\sum_{i =1}^2 \epsilon_i \, n^{(i)M} n^{(i)N}
\Big(
\widetilde{D}_M  \widetilde{D}_N  \varphi
- \widetilde{D}_M \varphi\, \widetilde{D}_N  \varphi
\Big)
-   \widetilde{D}^2  \varphi 
 - \frac{1}{4}  \sum_{i=1}^2    \epsilon_i \big(\mathrm{Tr} \widetilde{K}^{(i)}\big)^2
\, \Bigg] d\tilde{\mathcal{A}}
 \nonumber \\
 & &
 +\,
 \frac{1}{4}  \sum_{i=1}^2  \int_{\gamma} 
 \epsilon_i  \big(\mathrm{Tr} K^{(i)}\big)^2 \,d\mathcal{A} \,.
\eea
At this point, one adds the area $\mathcal{A}[\gamma] $ to both sides of (\ref{0= combination time dep}).
Then, by specialising the resulting expression to the class of surfaces given by $\gamma_\varepsilon$ and using the divergence theorem (see also  \cite{AM}) we find again the expansion $\mathcal{A}[\gamma_\varepsilon] = P_A/\varepsilon - \mathcal{F}_A + o(1)$ with
\bea
\label{FA generic surface time-dep}
\mathcal{F}_A &\equiv&
\int_{\gamma_A} 
\bigg[\, \frac{1}{4}  \sum_{i=1}^2    \epsilon_i  \big(\mathrm{Tr} \widetilde{K}^{(i)}\big)^2
 +
\widetilde{D}^2  \varphi  - e^{2\varphi} 
+
\sum_{i =1}^2  \epsilon_i \,n^{(i)M} n^{(i)N}
\Big(
\widetilde{D}_M \varphi\, \widetilde{D}_N  \varphi
- \widetilde{D}_M  \widetilde{D}_N  \varphi
\Big) 
\,\bigg]  d\tilde{\mathcal{A}}
 \nonumber \\
 & &
 - \,
 \frac{1}{4}  \sum_{i=1}^2  \int_{\gamma_A} 
 \epsilon_i  \big(\mathrm{Tr} K^{(i)}\big)^2 \,d\mathcal{A}\,.
\eea

We find it useful to give $\mathcal{F}_A$ also in terms of the energy-momentum tensor $T_{MN}$ of the bulk metric $g_{MN}$.
By employing the traceless tensors $\mathcal{K}^{(i)}_{MN}$ in (\ref{extrinsic curv time}) and the expression (\ref{hhR weyl G version}), the contracted Gauss-Codazzi equation (\ref{contracted GC}) can be written as 
\be
\label{contracted GC traceless K}
\epsilon_1 \epsilon_2 \,
W(n^{(1)}, n^{(2)} , n^{(1)} , n^{(2)})
+ \frac{1}{2} \sum_{i=1}^2 \epsilon_i \, \mathrm{Tr} (\mathcal{K}^{(i)})^2
\,=\,
\frac{1}{4} \sum_{i=1}^2 \epsilon_i \big(\mathrm{Tr} K^{(i)}\big)^2
- \frac{1}{2}\, \mathcal{R}
- \frac{1}{2} \sum_{i=1}^2 \epsilon_i \, G(n^{(i)},n^{(i)})
- \frac{1}{6} \,R \,,
\phantom{x}
\ee
where the l.h.s. is Weyl invariant, once multiplied by the area element $d\mathcal{A}$.
The Einstein equations with negative cosmological constant for the bulk metric $g_{MN}$ relate its Einstein tensor and the corresponding energy-momentum tensor as follows
\be
\label{einstein eqs}
G_{MN} \,=\,3\,g_{MN} +T_{MN}\,,
\ee
where we have absorbed the factor $8\pi G_N$ into the definition of the bulk energy-momentum tensor.
Taking the proper contractions of the Einstein equations (\ref{einstein eqs}), one finds that the combination involving the Einstein tensor and the Ricci scalar occurring in the r.h.s. of (\ref{contracted GC traceless K}) can be written as 
\be
\label{from G to T}
\frac{1}{2} \sum_{i=1}^2  \epsilon_i \,G(n^{(i)},n^{(i)}) + \frac{1}{6} \,R
\,=\,
 1 + \frac{1}{2} \sum_{i=1}^2  \epsilon_i \,T(n^{(i)},n^{(i)}) - \frac{1}{6} \,T \,,
\ee
where $T$ is the trace of the energy-momentum tensor.
By plugging (\ref{from G to T}) into (\ref{contracted GC traceless K}), integrating the resulting expression on a surface $\gamma$ and then exploiting the Weyl invariance of the terms coming from the l.h.s. of (\ref{contracted GC traceless K}), we find 
\bea
\label{area time dep general}
& &\hspace{-.7cm}
\mathcal{A}[\gamma] \,=\,
\frac{1}{2} \int_\gamma \widetilde{\mathcal{R}}\,d\tilde{\mathcal{A}} - \frac{1}{2} \int_\gamma \mathcal{R}\,d \mathcal{A}
+  \frac{1}{4}  \sum_{i=1}^2  \int_{\gamma}  \epsilon_i \big(\mathrm{Tr} K^{(i)}\big)^2 \,d\mathcal{A}
- \int_{\gamma}  \bigg(  \frac{1}{2} \sum_{i=1}^2  \epsilon_i \,T(n^{(i)},n^{(i)})- \frac{1}{6} \,T \bigg) d\mathcal{A} 
\nonumber
\\
& & \hspace{.6cm}
- \int_{\gamma} 
\bigg[\,
\frac{1}{4}  \sum_{i=1}^2   \epsilon_i \big(\mathrm{Tr} \widetilde{K}^{(i)}\big)^2
 -\frac{1}{2} \sum_{i=1}^2  \epsilon_i \,\widetilde{G}(\tilde{n}^{(i)},\tilde{n}^{(i)}) - \frac{1}{6} \,\widetilde{R}
\,\bigg]  d\tilde{\mathcal{A}} \,,
\eea
where $\mathcal{A}[\gamma] $ originates from the first term in the r.h.s. of (\ref{from G to T}).

When $\gamma$ has a boundary, the Gauss-Bonnet theorem allows us to simplify the first two terms in the r.h.s. of (\ref{area time dep general}) as follows
\be
\frac{1}{2} \int_{\gamma} \widetilde{\mathcal{R}}\,d\tilde{\mathcal{A}} 
- \frac{1}{2} \int_{\gamma} \mathcal{R}\,d \mathcal{A}
\,=\,
 \oint_{\partial \gamma}  \kappa \, ds
 - \oint_{\partial \gamma}  \tilde{\kappa}\, d\tilde{s}
 \,=\,
-  \oint_{\partial \gamma}  \tilde{b}^\mu  \partial_\mu \varphi\, d\tilde{s}  \,,
\ee
where in the last step we have employed the transformation law for the geodesic curvature under Weyl transformations, which reads
\be
\kappa = e^{-\varphi} 
\big(
\tilde{\kappa} - \tilde{b}^\mu \partial_\mu \varphi 
\big) \,.
\ee
Restricting our analysis to the class of surfaces given by $\gamma_\varepsilon$, we can easily adapt to the time dependent case the steps followed in the static backgrounds to obtain (\ref{area generic J2 bis}) from (\ref{area generic J2}), as done also above to write (\ref{FA generic surface time-dep}).
The final result it (\ref{area generic J2 bis}) with the $O(1)$ term given by
\bea
\label{area time dep general T}
& &\hspace{-.7cm}
\mathcal{F}_A \,=\,
\int_{\gamma_A} 
\bigg[\,
\frac{1}{4}  \sum_{i=1}^2   \epsilon_i \big(\mathrm{Tr} \widetilde{K}^{(i)}\big)^2
 - \frac{1}{2} \sum_{i=1}^2  \epsilon_i \,\widetilde{G}(\tilde{n}^{(i)},\tilde{n}^{(i)})
 - \frac{1}{6} \,\widetilde{R}
\,\bigg]
 d\tilde{\mathcal{A}}
\\
& & \hspace{.6cm}
+ \int_{\gamma_A}  \bigg( \,  \frac{1}{2} \sum_{i=1}^2  \epsilon_i \,T(n^{(i)},n^{(i)}) 
 - \frac{1}{6} \,T \bigg) d\mathcal{A}
-  \frac{1}{4}  \sum_{i=1}^2  \int_{\gamma_A}  \epsilon_i \big(\mathrm{Tr} K^{(i)}\big)^2\,d\mathcal{A}\,.
\nonumber
\eea
By using (\ref{from G to T}), (\ref{Rtrans 4dim}) and (\ref{Gtrans 4dim}), it is not difficult to check that (\ref{FA generic surface time-dep})  is recovered from  (\ref{area time dep general T}). 

It is worth recalling that (\ref{FA generic surface time-dep}) and (\ref{area time dep general T}) hold for a generic surface $\gamma_A$ ending orthogonally on the boundary at $z=0$.
For a given domain $A$, the extremal area surface $\hat{\gamma}_A$ is the solution of the following equations
\be
\label{minimality condition time-dep}
\textrm{Tr} K^{(i)}  = 0
\hspace{.5cm} \Longleftrightarrow \hspace{.5cm}
\big( \textrm{Tr}\widetilde{K}^{(i)} \big)^2 = 4 \big(\tilde{n}^{(i) M} \partial_M \varphi \big)^2 \,,
\ee
where the second expression comes from (\ref{Kmunu law}) properly adapted to the case we are considering. 
Specifying (\ref{FA generic surface time-dep}) and (\ref{area time dep general T}) to extremal area surfaces we find respectively
\bea
\label{FA final time-dep}
&& \hspace{-.8cm}
F_A \;=\;
\int_{\hat{\gamma}_A} 
\bigg(\, \frac{1}{2}  \sum_{i=1}^2    \epsilon_i  \big(\mathrm{Tr} \widetilde{K}^{(i)}\big)^2
 +
\widetilde{D}^2  \varphi  - e^{2\varphi} 
-
\sum_{i =1}^2  \epsilon_i \,\tilde{n}^{(i)M} \tilde{n}^{(i)N} \widetilde{D}_M  \widetilde{D}_N  \varphi
\,\bigg)
 d\tilde{\mathcal{A}}
 \\
&& \hspace{-.8cm}
 \phantom{F_A} \;=\;
 \int_{\hat{\gamma}_A} 
\bigg(\,
\frac{1}{4}  \sum_{i=1}^2   \epsilon_i \big(\mathrm{Tr} \widetilde{K}^{(i)}\big)^2
 - \frac{1}{2} \sum_{i=1}^2  \epsilon_i \,\widetilde{G}(\tilde{n}^{(i)},\tilde{n}^{(i)})
 - \frac{1}{6} \,\widetilde{R}
\,\bigg) d\tilde{\mathcal{A}}\,
 + \int_{\hat{\gamma}_A}  \bigg( \,  \frac{1}{2} \sum_{i=1}^2  \epsilon_i \,T(n^{(i)},n^{(i)}) 
 - \frac{1}{6} \,T \bigg) d\mathcal{A}\,.
 \nonumber
\eea

In explicit computations, the vectors $n^{(i)}$ must be chosen.
Taking $n^{(1)}$ timelike and $n^{(2)}$ spacelike, i.e. $\epsilon_1=-1$ and $\epsilon_2=1$, the sums in (\ref{FA generic surface time-dep}) and (\ref{area time dep general T}) become differences of two terms.
Further simplifications occur if the following lightlike vectors are introduced
\be
\label{ell vectors def}
\ell^{(\pm)}= \frac{n^{(1)} \pm n^{(2)}}{\sqrt{2}} \,.
\ee
Indeed,  $\big[T(n^{(1)},n^{(1)})- T(n^{(2)},n^{(2)})\big]/2 = T(\ell^{(-)},\ell^{(+)})$ and a similar expression holds for the terms involving the Einstein tensor.
By employing that $ \widetilde{K}_{MN}^{(\pm)} = \big[\widetilde{K}_{MN}^{(1)}  \pm \widetilde{K}_{MN}^{(2)}\big]/\sqrt{2} $ are the extrinsic curvatures defined through the null vectors in (\ref{ell vectors def}), one finds that (\ref{FA final time-dep}) becomes
\be
\label{holog MI transition}
 F_A \,=\,
 -\int_{\hat{\gamma}_A} 
\bigg(\,\frac{1}{2}\,
 \mathrm{Tr} \widetilde{K}^{(-)}\, \mathrm{Tr} \widetilde{K}^{(+)}
 - \widetilde{G}\big(\tilde{\ell}^{(-)},\tilde{\ell}^{(+)}\big)
 + \frac{1}{6} \,\widetilde{R}
\,\bigg) d\tilde{\mathcal{A}}\,
 - \int_{\hat{\gamma}_A}  \bigg(   T\big(\ell^{(-)},\ell^{(+)}\big) + \frac{1}{6} \,T \bigg) d\mathcal{A}\,.
\ee

In order to check the consistency of (\ref{FA generic surface time-dep}), let us recover the formula (\ref{FA generic surface}) for static backgrounds.\\
A generic static asymptotically AdS$_4$ spacetime is given by
\be
\label{static metric ADM}
ds^2\,=\, -N^2 dt^2+g_{\mu\nu} dx^\mu dx^\nu \,,
\ee
where $N$ and $g_{\mu\nu}$ are functions of the space coordinates $x^\mu= (z,\boldsymbol{x})$, being $\boldsymbol{x}$ the position vector in the $z=0$ plane.
The three dimensional Euclidean metric $g_{\mu\nu}$ is conformally related to $\tilde{g}_{\mu\nu}$ as in (\ref{conformal metric sec2}).
In this case, the timelike and spacelike unit vectors mentioned above are $n^{(1)}_M = (N,0,0,0)$ and  $n^{(2)}_M = (0,n_\mu)$ respectively, where $n_\mu$ is the three dimensional spacelike unit vector introduced in \S\ref{sec general case}.

A direct computation tells us that $K^{(1)}_{MN} = 0$ identically, which implies that the minimality equation for $n^{(1)}_M$ is trivially satisfied.
Since $\varphi$ is independent of time, we have $\widetilde{K}^{(1)}_{MN} = 0$ and $ \tilde{n}^{(1)M} \tilde{n}^{(1)N} \widetilde{D}_M \varphi\, \widetilde{D}_N  \varphi =0 $.
As for the Laplacian term, notice that $\widetilde{D}^2  \varphi $ specified to the static metric (\ref{static metric ADM}) provides $\widetilde{\nabla}^2\varphi $ plus an extra term which is canceled by the remaining term containing $n^{(1)M} $, namely
\be
\widetilde{D}^2  \varphi 
+ n^{(1)M} n^{(1)N}  \widetilde{D}_M  \widetilde{D}_N  \varphi
\,=\, \widetilde{\nabla}^2\varphi \,.
\ee
The spacelike vector $\tilde{n}^{(2)}$ provides all the other terms in (\ref{FA generic surface}). 
Indeed, the terms $ n^{(2)M} n^{(2)N}  \widetilde{D}_M \varphi\, \widetilde{D}_N  \varphi $ and
$ n^{(2)M} n^{(2)N} \widetilde{D}_M  \widetilde{D}_N  \varphi $ in (\ref{FA generic surface time-dep}) for the static metric (\ref{static metric ADM}) become $ ( \tilde{n}^\lambda \partial_\lambda \varphi )^2$ and $\tilde{n}^\mu \tilde{n}^\nu \, \widetilde{\nabla}_\mu \widetilde{\nabla}_\nu \varphi$ respectively.
Finally, it is immediate to see that in $K^{(2)}_{MN}$ only the spatial part $K^{(2)}_{\mu\nu}$ is non vanishing and therefore $(\mathrm{Tr} K^{(2)} )^2$ reduces to $(\mathrm{Tr} K)^2$ (the same observation holds for $\widetilde{K}^{(2)}_{MN}$).


\subsection{Vaidya-AdS backgrounds}
\label{sec vaidya bg}

In order to test the result of the section \S\ref{sec time-dep general}, let us consider the dynamical background given by the Vaidya-AdS metric \cite{Vaidya:1951zz, Bonnor:1970zz}. In Poincar\'e coordinates, it reads 
\be
\label{vaidya metric}
ds^{2} \,=\, \frac{1}{z^2}
\Big(-f(v,z)\,dv^{2}-2\, dv\, dz+d\boldsymbol{x}^{2} \Big)\,,
\qquad
f(v,z)=1-M(v) z^3\,,
\ee
where $v$ is the outgoing Eddington-Filkenstein coordinate which becomes the time coordinate $t$ of the boundary theory at $z=0$.
The metric (\ref{vaidya metric}) is a solution of the Einstein equations (\ref{einstein eqs}) with an energy-momentum tensor $T_{MN}$ having only one non vanishing component
\be
T_{vv} = z^2 M'(v)  \,.
\ee
The null energy condition (i.e. $T_{RS} N^R N^S \geqslant 0$ for any null vector $N^R$ \cite{Wald, HawkingEllis}) imposes that $M'(v)  \geqslant 0$.
Choosing $M(v)$ such that $M(v) \to 0$ when $v\to -\infty $ and $M(v) \to M$ when $v\to +\infty $,  the metric (\ref{vaidya metric}) describes the formation of a black hole of mass $M$ through the gravitational collapse of a null shell of matter. 
For $M(v) = M$ constant in time, a coordinate transformation brings (\ref{vaidya metric}) into the usual metric of the Schwarzschild-AdS black hole. 
These backgrounds provide the simplest examples of holographic thermalization.

The holographic entanglement entropy for the Vaidya-AdS backgrounds (\ref{vaidya metric}) must be computed through the covariant prescription of \cite{Hubeny:2007xt}. 
The result depends also on the boundary time coordinate $t$.
Keeping the entangling curve $\partial A$ fixed, the expansion (\ref{hee ads4 intro}) holds, where $F_A = F_A(t)$.

In order to specify the result of \S\ref{sec time-dep} for $F_A$ to this case, we find it more convenient to consider the expression (\ref{holog MI transition}).
Since $g^{vv} =0$, the trace of the energy-momentum tensor vanishes. 
Moreover, $\widetilde{R}=6z M(v)$, while for the Einstein tensor we need to choose a coordinate system in the $z=0$ plane. For example, the only non vanishing components of $\widetilde{G}_{MN}$ are $\widetilde{G}_{xx}=\widetilde{G}_{yy}=-3zM(v)$ in cartesian coordinates and $\widetilde{G}_{\rho\rho}=\widetilde{G}_{\theta\theta}/\rho^2=-3zM(v)$ in polar coordinates.
In order to simplify the term of (\ref{holog MI transition}) containing the extrinsic curvatures, it is useful to employ first the extremal surface conditions (\ref{minimality condition time-dep}) and then the vectors (\ref{ell vectors def}). The final result reads
\be
\label{FA vaidya}
F_A 
\,=\,
 -\int_{\hat{\gamma}_A} 
\bigg(\,\frac{2\,\tilde{\ell}^{(-)z} \tilde{\ell}^{(+)z}}{z^2}
 - \widetilde{G}\big(\tilde{\ell}^{(-)},\tilde{\ell}^{(+)}\big)
 +z \,M(v) + M'(v) \, \tilde{\ell}^{(-)v}\tilde{\ell}^{(+)v}
\,\bigg) d\tilde{\mathcal{A}}\,.
\ee
When $M(v)$ is constant, this formula simplifies, providing (\ref{FA ads4 sec}) and (\ref{FA bh gen}) for $M=0$ or $M>0$ respectively.

A simple mass profile $M(v)$ satisfying the null energy condition reads
\be 
\label{eq:kinkeq}
M(v)=\frac{M}{2}\big(1+\tanh(v/v_0) \big) \,,
\ee 
where the parameter $v_0$ determines the steepness of the transition between the two asymptotic regimes of AdS$_4$ (when $v\to -\infty$) and Schwarzschild-AdS$_4$ black hole with mass $M$ (when $v\to +\infty$). Indeed, it parameterises the thickness of the shell falling along $v=0$.
The holographic entanglement entropy for the Vaidya-AdS background (\ref{vaidya metric}) with the mass profile (\ref{eq:kinkeq}) has been largely studied during the last years (see e.g. \cite{AbajoArrastia:2010yt, Albash:2010mv, Balasubramanian:2010ce, Balasubramanian:2011at, Allais:2011ys, Callan:2012ip, Caceres:2012em, Hubeny:2013hz, Liu:2013iza, Keranen:2011xs, Alishahiha:2014cwa, Fonda:2014ula}).
In \S\ref{sec vaidya disk}, considering circular domains $A$, we check numerically that (\ref{FA vaidya}) reproduces the same results already found by subtracting the most divergent term from the area of the extremal surface (see Fig.\,\ref{fig:FAvaidya}). 
It would be interesting to find some analytic result from (\ref{FA vaidya}) in the thin shell limit $v_0 \to 0$, along the lines of \cite{Balasubramanian:2010ce, Liu:2013iza}.

\section{Some particular domains}
\label{sec examples}

After a brief explanation of the numerical methods employed in this manuscript, in this section we test the formulas for $F_A$ given above by first considering some simple cases of simply connected domains $A$ which have been largely studied in the literature: the infinite  strip and the disk.
Then, we extend the numerical analysis to the case of the elliptical entangling curves. 
These domains belong to a large class of spatial regions $A$ such that the corresponding minimal surface $\hat{\gamma}_A$ can be parameterised by $z=z(\boldsymbol{x})$, where $\boldsymbol{x} \in A$.
In this section we will also study $F_A$ for domains which do not belong to this class, since on the corresponding $\hat{\gamma}_A$ one can find pairs of distinct points having the same projection on the $z=0$ plane.

\subsection{Numerical methods}
\label{sec num methods}

The crucial numerical tool employed in this manuscript to study minimal surfaces $\hat{\gamma}_A$ for finite domains $A$ different from disks is {\it Surface Evolver} \cite{evolverpaper,evolverlink}, a multipurpose shape optimization program created by Ken Brakke \cite{evolverpaper} to address generic problems on energy minimizing surfaces.
In the context of AdS/CFT, it has been first employed in \cite{Fonda:2014cca} to get some numerical results about the shape dependence of the holographic mutual information in AdS$_4$.
Here we extend its application to other backgrounds. 

In Surface Evolver, a surface is implemented as a union of triangles (see e.g. Figs.\,\ref{3Ddoubling}, \ref{3DellipseBH}, \ref{3DellipseRG} and \ref{3DbananaRG}). 
Given the background metric $g_{\mu\nu}$, the boundary curve $\partial A$ in the plane $z=\varepsilon$ and an initial trial triangulated surface, the program evolves the surface towards a local minimum of the area functional by employing a gradient descent method (see the appendix B of \cite{Fonda:2014cca} for a very brief discussion).
The final stage of the evolution is a triangulated surface close to $\hat{\gamma}_A$.
The approximation improves as the number of triangles increases.
For any triangulated surface, one can read off both the area of the whole surface and all the unit normal vectors.
Let us denote by $\hat{\gamma}_A^{\textrm{\tiny SE}}$ the best approximation  of the minimal surface $\hat{\gamma}_A$ found with Surface Evolver.
Given the corresponding area $\mathcal{A}^{\textrm{\tiny SE}}$ and unit vectors $n_\mu^{\textrm{\tiny SE}}$, we can numerically compute the following two quantities
\be
\label{SE F_A}
 \widehat{F}_A^{\textrm{\tiny SE}} \,=\, -\big(\mathcal{A}^{\textrm{\tiny SE}}-P_A/\varepsilon\big) \,,
\hspace{.6cm} \qquad \hspace{.6cm}
\bar{F}_A^{\textrm{\tiny SE}} \,=\, F_A\big|_{\hat{\gamma}_A^{\textrm{\tiny SE}}} \,,
\ee
where $\bar{F}_A^{\textrm{\tiny SE}} $ is obtained from the analytic expression (\ref{FA sec}) evaluated on the triangulated surface $\hat{\gamma}_A^{\textrm{\tiny SE}}$ through its unit normal vector $\tilde{n}_\mu^{\textrm{\tiny SE}}$.
Both these expressions are finite in the limit $\varepsilon \to 0$.
Verifying that both the quantities in (\ref{SE F_A}) give the same values provides a strong check of the analytic formula (\ref{FA sec}).
Indeed, from the series expansion of the holographic entanglement entropy, we expect that $|\widehat{F}_A^{\textrm{\tiny SE}}  - \bar{F}_A^{\textrm{\tiny SE}} | =o(1)$ as $\varepsilon \to 0$.
Examples will be provided involving both the black hole and the domain wall geometry introduced in \S\ref{sec static backs} (see \cite{Fonda:2014cca} for AdS$_4$).

We perform the numerical analysis through Surface Evolver whenever the  partial differential equation defining $\hat{\gamma}_A$ cannot be simplified (e.g. for the elliptic domains in Figs\,\ref{fig:FABH}, \ref{fig:FABHext}, \ref{fig:FARG}, \ref{fig:FARG2} and for the non convex domains of Fig.\,\ref{fig:FAbanana}).
For highly symmetric regions $A$, the corresponding minimal area equation simplifies to an ordinary differential equation in one variable. 
This happens for the infinite strip (\S\ref{sec: strip}), the disk (\S\ref{sec:disk}) and the annulus (\S\ref{sec app annulus}).
For these domains, more standard softwares (e.g. {\it Mathematica}) can be employed to study numerically the corresponding ordinary differential equations.

\subsection{Strip}
\label{sec: strip}

When $A$ is an elongated strip with sides having lengths $\ell$ and $L$ with $\ell \ll L$, it is convenient to adopt cartesian coordinates $\{z,x,y\}$ which can be always chosen such that $A=\{(x,y)\,; |x| \leqslant \ell/2 \,, |y| \leqslant L/2\}$.
Since $L\gg \ell$, we can assume that $z=z(x)$ and therefore $z_y = 0$ and $\tilde{n}_y = 0$. 
Moreover, the symmetry of the domain with respect to the the $y$ allows us to consider $0\leqslant x\leqslant \ell/2$ only.
In \cite{Fonda:2014cca} a numerical analysis through Surface Evolver has been done where the elongated strip is approximated through various smooth domains.

\subsubsection{Black holes}
\label{sec bh strip}

Let us first address the case of the black holes characterised by the metric (\ref{bh 4dim}), which includes AdS$_4$ as special case when $f(z) =1$ identically.

The area functional evaluated for the class of surfaces given by $\gamma_\varepsilon$ reads \cite{Ryu:2006bv, Ryu:2006ef}
\be
\label{area functional strip}
\mathcal{A}[\gamma_\varepsilon] = 2L \int_{0}^{\ell/2-\omega}  \frac{1}{z^2} \, \sqrt{1+\frac{z_x^2}{f(z)}}\; dx \,,
\ee
where $f(z)$ is the emblacking factor in (\ref{bh 4dim}) and the parameter $\omega$ is defined by $z(\ell/2-\omega)=\varepsilon$.
Since the integrand of (\ref{area functional strip}) does not depend on $x$ explicitly, we can simplify the problem of finding the extremum of (\ref{area functional strip}) by writing the following first integral
\be
\label{const motion strip bh}
z^2 \, \sqrt{1+\frac{z_x^2}{f(z)}} = z_\ast^2 \,,
\qquad
z(0) =  z_\ast \,,
\ee
being $z_\ast$ the highest value reached from the minimal surface along the holographic direction. 
The expression (\ref{const motion strip bh}) is a first order ordinary differential equation and therefore much easier to solve with respect to the equation of motion coming from (\ref{area functional strip}).
By isolating $z_x$ in (\ref{const motion strip bh}) (we recall that $z_x<0$), the first order differential equation becomes
\be
\label{z_x strip}
z_x \,=\, - \,\frac{\sqrt{(z_\ast^4-z^4) f(z)}}{z^2} \,,
\ee
which can be solved through separation of the variables, getting the relation between $\ell$ and $z_\ast$, namely
\be
\label{ell-zast strip}
\frac{\ell}{2} 
\,=\, \int_0^{z_\ast} \frac{z^2}{\sqrt{ (z_\ast^4-z^4) f(z)}}\, dz \,.
\ee

As for the finite term $F_A$ of the holographic entanglement entropy for the strip in this black hole background, it is obtained simply by specifying (\ref{FA bh gen}) to this case. By using (\ref{normal vectors up}) and (\ref{area tilde element}) for the vector $\tilde{n}^z$ and the area element $d\tilde{\mathcal{A}} $ respectively, one finds 
\begin{subequations}
\bea
\label{FA bh strip}
F_A 
    &=&
  2L \int_0^{\ell/2}
  \frac{1}{z^2}\left[\left( f(z) + \frac{z f'(z)}{2}\right)\frac{1}{1+z_x^2/f(z)} 
+ f(z) - \frac{z f'(z)}{2} -1 \, 
 \right]
   \sqrt{1+\frac{z_x^2}{f(z)}}\; dx
   \\
   \rule{0pt}{.65cm}
   \label{FA bh strip 2}
  &=&
     \frac{2L}{z_\ast^2} \int_0^{\ell/2}
\left[\,  f(z) + \frac{z f'(z)}{2}
+ \frac{z_\ast^4}{z^4} \left(f(z) - \frac{z f'(z)}{2} -1\right)
 \right] dx
 \\
    \rule{0pt}{.65cm}
   \label{FA bh strip 3}
  &=&
     \frac{2L}{z_\ast^2} \int_0^{z_\ast}
\left[\,  f(z) + \frac{z f'(z)}{2}
+ \frac{z_\ast^4}{z^4} \left(f(z) - \frac{z f'(z)}{2} -1\right) \right]
\frac{z^2}{\sqrt{ (z_\ast^4-z^4) f(z)}}\, dz \,,
\eea
\end{subequations}
where (\ref{FA bh strip 2}) and (\ref{FA bh strip 3}) has been obtained  by employing (\ref{const motion strip bh}) and (\ref{z_x strip}) respectively.  
Thus $F_A/L$ is a complicated function of $\ell$ that could be found by first performing the integral (\ref{FA bh strip 3}) explicitly and then by finding $z_\ast(\ell)$ from (\ref{ell-zast strip}).

A major simplification occurs  for AdS$_4$.
Indeed, when $f(z)=1$ identically the integrand in (\ref{FA bh strip 2})  becomes equal to 1. 
Moreover, also the integral (\ref{ell-zast strip}) can be performed explicitly in this case. Thus, for AdS$_4$ we have that
\be
\label{ads4 FA and size strip}
\textrm{AdS$_4$\;:}
\qquad
\frac{\ell}{2} 
\,=\, \frac{\sqrt{\pi}\;\Gamma(\tfrac{3}{4})}{\Gamma(\tfrac{1}{4})}\, z_\ast  \,,
\qquad
F_A 
\,=\, \frac{L \,\ell}{z_\ast^2}
\,=\,  \frac{2\sqrt{\pi}\;\Gamma(\tfrac{3}{4}) \, L}{\Gamma(\tfrac{1}{4})\, z_\ast}  \,,
\ee
which is the result of \cite{Maldacena:1998im, Rey:1998ik, Ryu:2006ef}.

\subsubsection{Domain wall geometries}
\label{sec rg strip}

When the bulk geometry is (\ref{holog rg 4dim}) and the domain $A$ in the boundary is the elongated strip described above, the area functional for the class of surfaces $\gamma_\varepsilon$ reads
\be
\label{area functional strip rg}
\mathcal{A}[\gamma_\varepsilon] =  2L \int_{0}^{\ell/2-\omega}  \frac{\sqrt{1+z_x^2 \,p(z)}}{z^2 \, p(z)} \, dx  \,,
\ee
where $\omega$ has been already introduced below (\ref{area functional strip}).
Since the metric (\ref{holog rg 4dim}) on a  constant time slice can be written like a black hole metric at $t=\textrm{const}$ with a proper $f(z)$, one could employ the results of \S\ref{sec bh strip}.
Nevertheless, we find instructive to provide explicitly the analysis also in this coordinates.

Since the integrand in (\ref{area functional strip rg}) does not depend explicitly on $x$, we can write the following conserved quantity
\be
\label{const motion strip rg}
z^2 \, p(z)\sqrt{1+z_x^2\, p(z)} = z_\ast^2\, p(z_\ast)  \,,
\ee
which allows us to find $z_x$ (we recall that $z_x<0$) 
\be
\label{z_x strip rg}
z_x \,=\, - \,\frac{\sqrt{z_\ast^4 \, p(z_\ast)^2 - z^4\, p(z)^2 }}{z^2\, p(z)}  \,.
\ee
By separating the variables in this first order differential equation, one finds the relation between $\ell$ and $z_\ast$
\be
\label{ell-zast strip rg}
\frac{\ell}{2} 
\,=\, \int_0^{z_\ast} \frac{z^2\, p(z)}{\sqrt{ z_\ast^4\, p(z_\ast)^2 - z^4\, p(z)^2}}\, dz  \,.
\ee

The finite term $F_A$ in the holographic entanglement entropy for these domain wall geometries with $A$ given by the elongated strip is obtained by 
specializing (\ref{FA holog rg}) to this domain. 
By employing (\ref{normal vectors up rg}) and (\ref{area element rg})  for the vector $\tilde{n}^z$ and the area element $d\tilde{\mathcal{A}} $ respectively, one gets
\begin{subequations}
\bea
\label{FA strip rg}
F_A 
    &=&
  2L \int_0^{\ell/2}
  \left[ \left( 1+ \frac{z \, p'(z)}{2 \,p(z)} \right) \frac{1}{1+z_x^2\, p(z)}
+ \frac{z\,p'(z)}{2\, p(z)} \, 
 \right]
\frac{\sqrt{1+z_x^2\, p(z)}}{z^2\, p(z)}\,dx
   \\
   \rule{0pt}{.65cm}
   \label{FA strip rg 2}
  &=&
     \frac{2L}{z_\ast^2\, p(z_\ast)} \int_0^{\ell/2}
\left[\,  1+  \frac{z \, p'(z)}{2 \,p(z)}  \left( 1+ \frac{z_\ast^4\, p(z_\ast)^2}{z^4\,p(z)^2} \right)
 \right] dx
 \\
    \rule{0pt}{.65cm}
   \label{FA strip rg 3}
  &=&
     \frac{2L}{z_\ast^2\, p(z_\ast)} \int_0^{z_\ast}
\left[\,  1+  \frac{z \, p'(z)}{2 \,p(z)}  \left( 1+ \frac{z_\ast^4\, p(z_\ast)^2}{z^4\,p(z)^2} \right)
 \right]
 \frac{z^2\, p(z)}{\sqrt{ z_\ast^4\, p(z_\ast)^2 - z^4\, p(z)^2}}\, dz  \,,
\eea
\end{subequations}
where (\ref{FA strip rg 2}) and (\ref{FA strip rg 3}) have been found through (\ref{const motion strip rg}) and (\ref{z_x strip rg}) respectively.  
From (\ref{FA strip rg 2}) it is straightforward to check that the AdS$_4$ result for $F_A$ in (\ref{ads4 FA and size strip}) is recovered when $p(z)=1$ identically.

\subsubsection{Vaidya-AdS backgrounds}
\label{sec vaidya strip}

Let us consider the elongated strip and the gravitational background in the bulk given by the Vaidya-AdS metric (\ref{vaidya metric}).
Choosing the cartesian coordinate system in the boundary as explained in the beginning of \S\ref{sec: strip}, the profile of the surfaces $\gamma_A$  can be described by the two functions $z(x)$ and $v(x)$.
In this case, the area functional to extremize reads
\be
\mathcal{A}[\gamma_\varepsilon] \,=\, 
2L \int_0^{\ell/2-\omega}
\frac{\sqrt{1-2v' z' -f(v,z) (v')^2}}{z^2} \; dx  \,.
\ee
In order to apply the formula (\ref{FA vaidya}), we need the vectors and the area element discussed in \S\ref{app:unit vectors vaidya}.
Considering the explicit expression $f(v,z)=1-M(v)z^3$, the formula (\ref{FA vaidya}) becomes 
\be
\label{FA vaidya strip exp}
\frac{F_A}{2L} =
 \int_0^{\ell/2}
\frac{
2 \,v' z' \left(z^3 M -2\right)-2 \left((v')^2-1\right)-2 (z')^2
-z^3 \big[\big(z M'+ z^3 M^2 -3M\big)(v')^2  +4 M\big]}{
z^2\,\sqrt{1-2 v' z' -(1-Mz^3) (v')^2 }}  \;dx  \,,
\ee
where $M=M(v)$.
As consistency check of (\ref{FA vaidya strip exp}), we notice that (\ref{FA bh strip}) can be recovered in the special case of $M(v)$ constant.

\subsection{Disk}
\label{sec:disk}

When $A$ is a disk of radius $R$, it is convenient to adopt the cylindrical coordinates $\{z,\rho, \theta \}$, with the origin of the polar coordinates $\{\rho, \theta \}$ in the $z=0$ plane given by the center of the disk.
The symmetry of the domain tells us that $z=z(\rho)$. This means that $z_\theta = 0$ and $\tilde{n}_\theta = 0$. 
The disk is more complicated than the strip considered in \S\ref{sec: strip} because the coordinate $\rho$ is not cyclic and therefore the ordinary differential equation to study  is a second order one.

\subsubsection{Black holes}
\label{sec disk bh}

\begin{figure}[t] 
\vspace{-.2cm}
\begin{center}
\hspace{-.5cm}
\includegraphics[width=1.\textwidth]{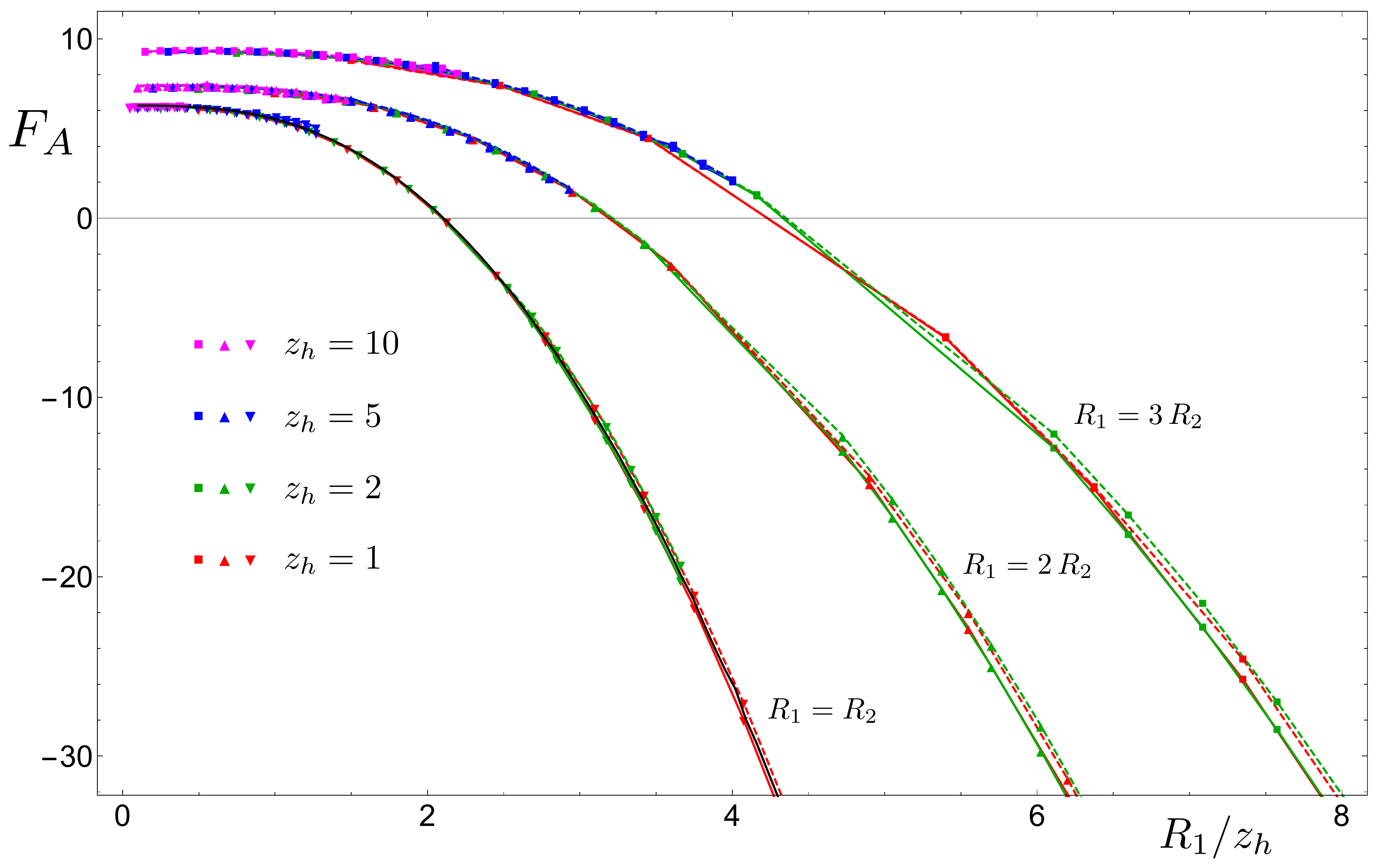}
\end{center}
\vspace{-.4cm}
\caption{\label{fig:FABH}
The quantity $F_A$ for a Schwarzschild-AdS black hole when the entangling curve $\partial A$ is an ellipse with semi-major axis $R_1$ and semi-minor axis $R_2$. The computations have been done with Surface Evolver (here $\varepsilon=0.01$) through the two ways given in (\ref{SE F_A}) (solid and dashed colored lines respectively). For the disks (bottom curve) the expression (\ref{FA bh disk}) holds and it can be analyzed with Mathematica (solid black line). 
}
\end{figure}

\begin{figure}[t] 
\vspace{-.2cm}
\begin{center}
\hspace{-.5cm}
\includegraphics[width=1.\textwidth]{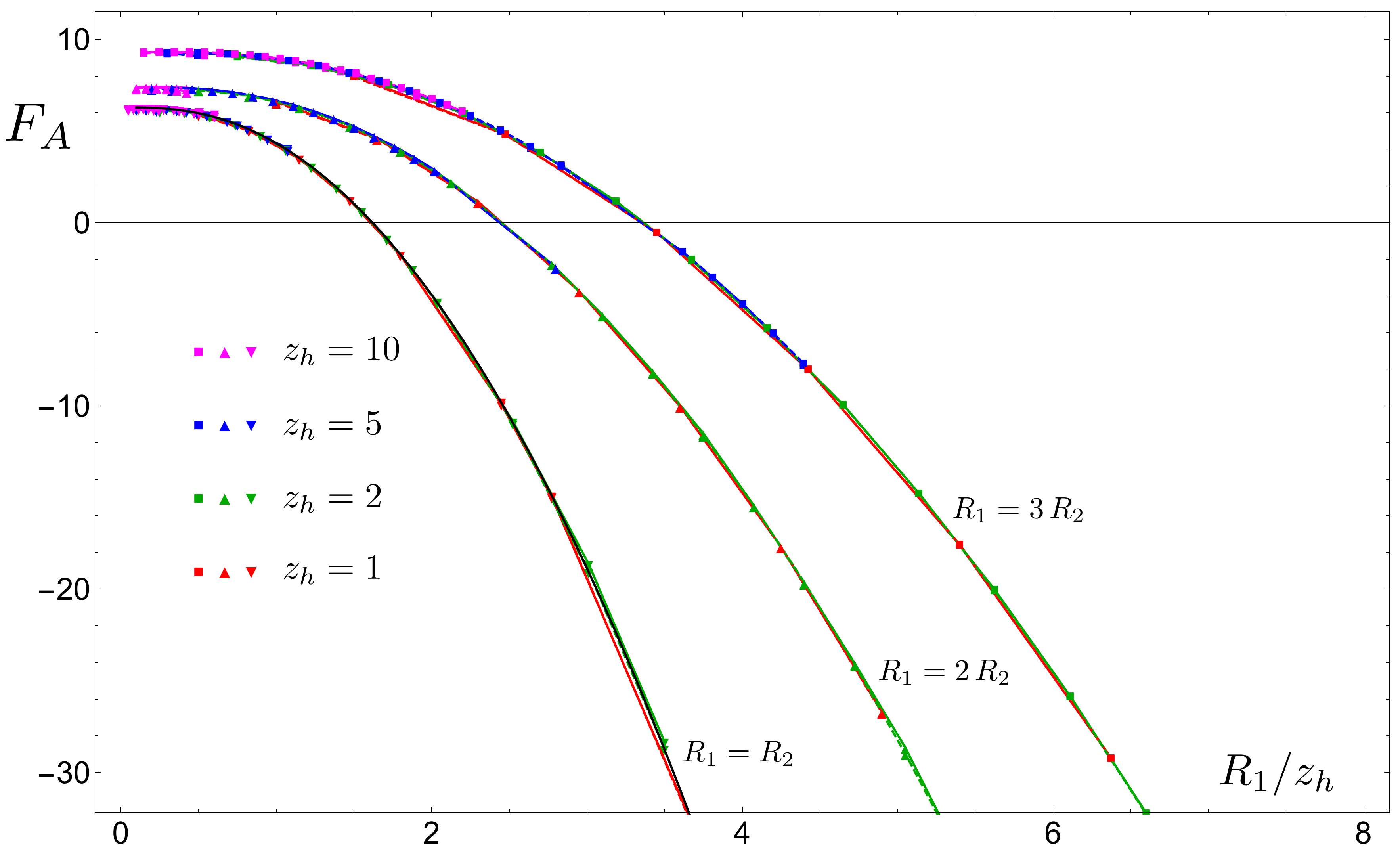}
\end{center}
\vspace{-.4cm}
\caption{\label{fig:FABHext}
The quantity $F_A$ for the extremal black hole with the entangling curves $\partial A$ given by ellipses with semi-axis $R_1\geqslant R_2$.
The computations have been done with Surface Evolver (here $\varepsilon=0.01$) in the two ways given in (\ref{SE F_A}) (solid and dashed colored lines respectively). 
For the disks also the expression (\ref{FA bh disk hat}), which can be studied with Mathematica, is shown (solid black line). 
}
\end{figure}

Let us consider the black hole metric  (\ref{bh 4dim}) at constant time slice with polar coordinates $\{\rho, \theta\}$ in the $z=0$ plane.
Given the ansatz $z=z(\rho)$, the area functional for the surfaces $\gamma_\varepsilon$ reads
\be
\label{area functional bh disk}
\mathcal{A}[\gamma_\varepsilon] = 
2\pi \int_0^{R-\omega} \frac{\rho}{z^2} \, \sqrt{1+\frac{z_\rho^2}{f(z)}}\, d\rho  \,,
\ee
where $\omega $ is defined by the condition $z(R-\omega) = \varepsilon$.
As already remarked, in this case the integrand explicitly depends on $\rho$ and therefore we cannot write a first integral as done for the strip. 
The equation of motion is a second order ordinary differential equation and its analytic solution is not known for a non trivial $f(z)$.

As for the finite term $F_A$ in the holographic entanglement entropy,  by employing the proper expressions in (\ref{normal vectors up}) and (\ref{area tilde element}) for the vector $\tilde{n}^z$ and the area element $d\tilde{\mathcal{A}} $ respectively, (\ref{FA bh gen}) becomes 
\be
\label{FA bh disk}
F_A 
    \,=\,
  2\pi \int_0^R
\frac{1}{z^2}\left[ \left( f(z)+ \frac{z  f'(z)}{2} \right) \frac{f(z)}{f(z) + z_\rho^2}
+ f(z) - \frac{z f'(z)}{2}-1\, 
 \right]
   \frac{\sqrt{f(z) + z_\rho^2}}{\sqrt{f(z)}}\, \rho\, d\rho \,.
\ee
This expression holds for both the Schwarzschild-AdS black hole and the charged black hole.
It can be employed only once the solution $z(\rho)$ of the extremal area equation is known. 
In (\ref{FA bh disk}) the profile $z(\rho)$ satisfies the boundary condition $z(R) = 0$.
Since the second order ordinary differential equation providing $z(\rho)$ is quite complicated for non trivial $f(z)$, we have to rely on numerical methods.

An important special case of (\ref{FA bh disk}) is AdS$_4$, for which $f(z) =1$ identically.
In this case the profile $z(\rho)$ is known analytically  and it is given by the hemisphere.
By simplifying (\ref{FA bh disk}) first and then employing the explicit solution for the profile, the result of \cite{Ryu:2006bv, Ryu:2006ef} is recovered, namely
\be
\label{FA ads4 disk}
F_A 
\,=\,2\pi \int_0^R \frac{\rho\,d\rho}{z^2\, \sqrt{1+z_\rho^2}}
\,=\,  2\pi \,,
\qquad
\hspace{.5cm}
\qquad
z(\rho) = \sqrt{R^2-\rho^2} \,.
\ee

Let us restrict to the Schwarzschild-AdS black hole, i.e. $f(z) = 1-(z/z_h)^3$, where $z_h$ the position of the event horizon, 
and perform the following rescaling
\be
\label{rho z hat def bh}
\hat{\rho} \equiv \frac{\rho}{z_h} \,,
\qquad
\hat{z} \equiv \frac{z}{z_h} \,.
\ee
In terms of $\hat{\rho}$ and $\hat{z} $, we have $f(z) = 1 - \hat{z}^3 \equiv \hat{f}(\hat{z})$  and $z \, f'(z) = \hat{z} \,\hat{f}'(\hat{z})$, where $\hat{f}'(\hat{z}) \equiv \partial_{\hat{z}}\hat{f}(\hat{z})$.
Moreover, $z_\rho = \hat{z}_{\hat{\rho}}$ and, denoting by $\mathcal{L}$ the integrand of (\ref{area functional bh disk}), we have that $\mathcal{L} = \hat{\mathcal{L}}/z_h$, where
\be
\hat{\mathcal{L}} = 
 \frac{\hat\rho}{\hat{z}^2} \, \sqrt{1+ \frac{\hat{z}_{\hat{\rho}}^2}{\hat{f}(\hat{z})}} \,.
\ee
It is straightforward to observe that the equation of motion $\tfrac{d}{d\rho}\big(\tfrac{\partial \mathcal{L}}{\partial z_\rho}\big) = \tfrac{\partial \mathcal{L}}{\partial z}$ can be written as the equation of motion for $\hat{\mathcal{L}} $, i.e. $\tfrac{d}{d\hat{\rho}}\big(\tfrac{\partial \hat{\mathcal{L}}}{\partial \hat{z}_{\hat{\rho}}}\big) = \tfrac{\partial \hat{\mathcal{L}}}{\partial \hat{z}}$.
Thus, the profile of the minimal area surface is given by $\hat{z}=\hat{z}(\hat{\rho})$. 
As for the boundary conditions for this differential equation, from $z(R)=0$ and (\ref{rho z hat def bh}) one finds that $\hat{z}(R/z_h) =0$.
By employing these observations and performing the rescaling (\ref{rho z hat def bh}), we can conclude that (\ref{FA bh disk}) for the Schwarzschild-AdS black hole can be written as
\be
\label{FA bh disk hat}
F_A  \,= \, 2\pi
\int_0^{R/z_h}
\left[ \left( \hat{f}(\hat{z})+ \frac{\hat{z}  \hat{f}'(\hat{z})}{2} \right) 
\frac{\hat{f}(\hat{z}) }{\hat{f}(\hat{z}) + \hat{z}_{\hat{\rho}}^2}
+ \hat{f}(\hat{z}) - \frac{\hat{z}  \hat{f}'(\hat{z})}{2} -1\, 
 \right]
   \frac{\sqrt{\hat{f}(\hat{z}) + \hat{z}_{\hat{\rho}}^2}}{\hat{z}^2\, \sqrt{\hat{f}(\hat{z}) }}\, \hat{\rho}\, d\hat{\rho} \,.
\ee
From this expression we read that $F_A= F_A(R/z_h)$, which is given by  the bottom curve in Fig.\,\ref{fig:FABH}.

For the extremal black hole, where $f(z)=1-4(z/z_h)^3+3(z/z_h)^4$ and the inner and outer horizons coincide, one can repeat the same reasoning finding again that $F_A= F_A(R/z_h)$ given by (\ref{FA bh disk hat}) with $\hat{f}(\hat{z})=1-4\hat{z}^3+3\hat{z}^4 $ (see the bottom curve in Fig.\,\ref{fig:FABHext}).
In the non extremal case the analysis can be done in the same way but the outcome is slightly different because of the occurrence of two independent parameters. 
Indeed, by performing the rescaling $\hat{\rho} = \sqrt[3]{M}\, \rho$ and $\hat{z} = \sqrt[3]{M}\, z$, and repeating the steps explained above, one finds that 
$F_A=F_A(R\sqrt[3]{M}, Q^3/M^2)$, whose explicit expression is given by (\ref{FA bh disk hat}) properly adapted to the rescaling entering in this case.

\subsubsection{Domain wall geometries}
\label{sec disk rg}

\begin{figure}[t] 
\vspace{-.2cm}
\begin{center}
\hspace{-.5cm}
\includegraphics[width=1.\textwidth]{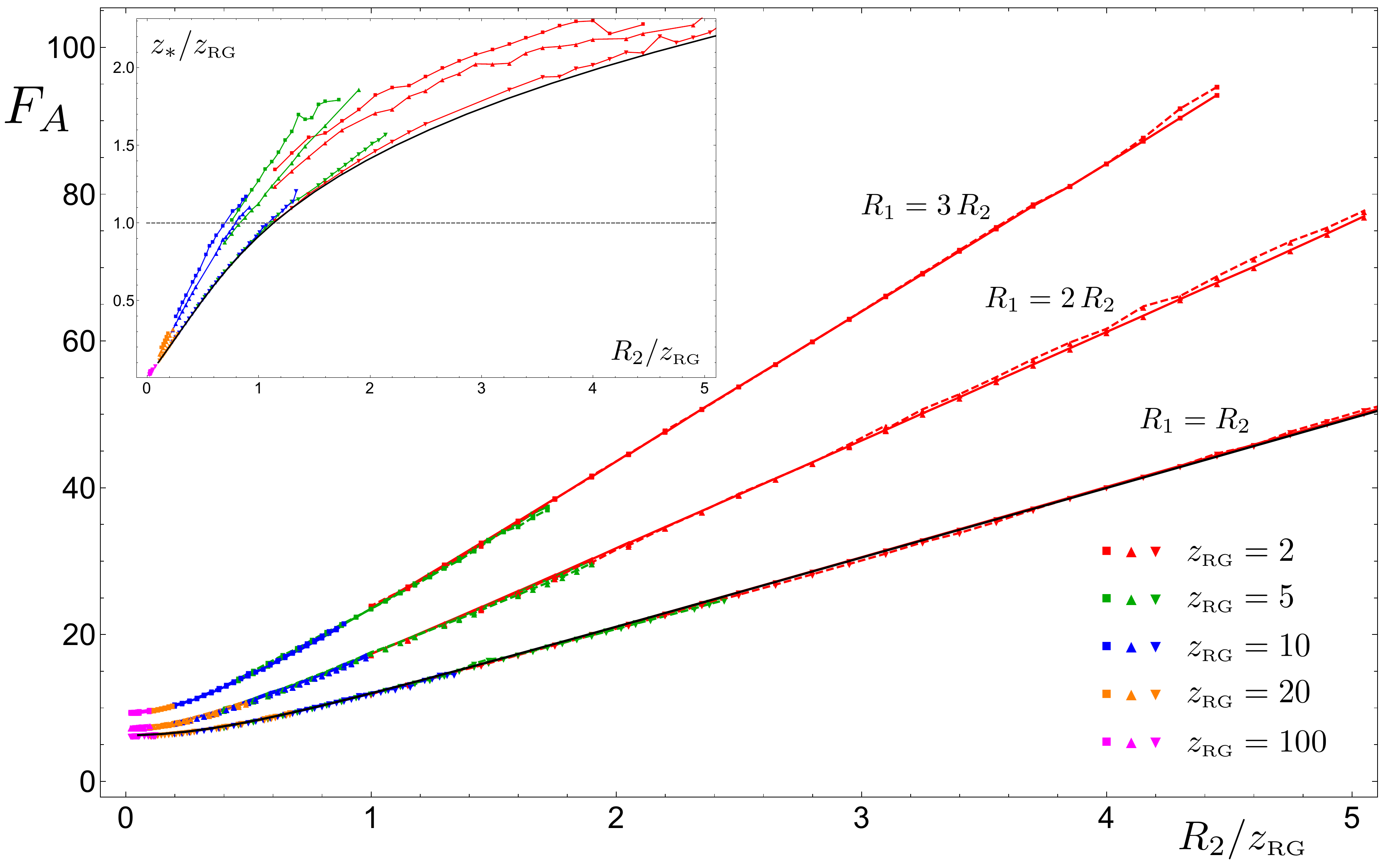}
\end{center}
\vspace{-.4cm}
\caption{\label{fig:FARG}
The quantity $F_A$ for the domain wall geometry (\ref{holog rg 4dim}) with $\alpha=2$ and $\gamma=1$. 
The entangling curves $\partial A$ are ellipses with semi-axis $R_1\geqslant R_2$.
The computations have been done with Surface Evolver (here $\varepsilon=0.01$) in the two ways given in (\ref{SE F_A}) (solid and dashed colored lines respectively). 
For the disks (bottom curve), the expression (\ref{FA disk rg}) holds, which can be studied with Mathematica (solid black line).
In the inset we show the highest point $z_\ast$ of the surfaces corresponding to all the points in the main plot, with the same colour code. 
}
\end{figure}

\begin{figure}[t] 
\vspace{-.2cm}
\begin{center}
\hspace{-.5cm}
\includegraphics[width=1.\textwidth]{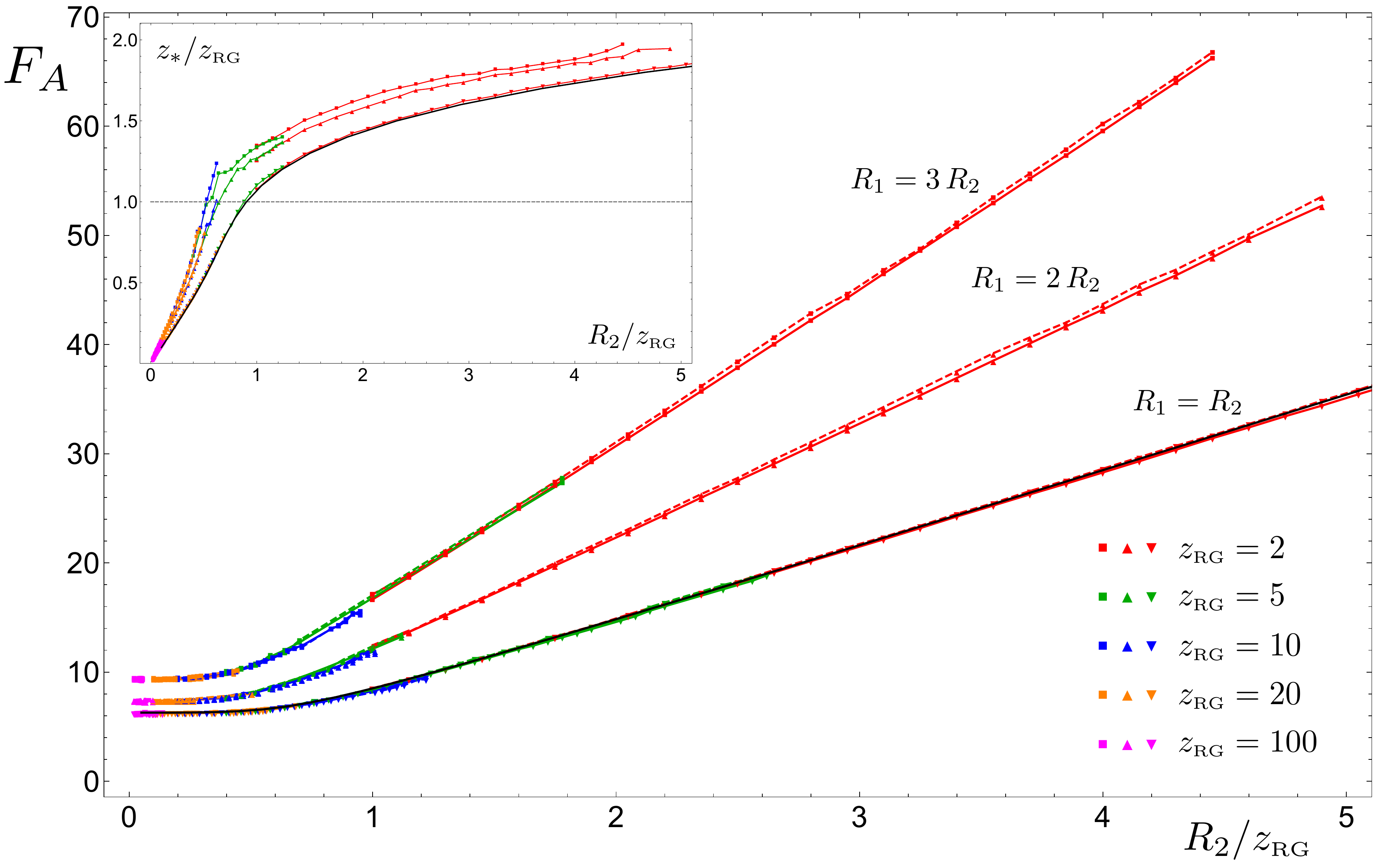}
\end{center}
\vspace{-.4cm}
\caption{\label{fig:FARG2}
The quantity $F_A$ for the domain wall geometry (\ref{holog rg 4dim}) with $\alpha=4$ and $\gamma=1$. 
The entangling curves $\partial A$ are ellipses with semi-axis $R_1\geqslant R_2$.
The computations have been done with Surface Evolver (here $\varepsilon=0.01$) in the two ways given in (\ref{SE F_A}) (solid and dashed colored lines respectively). 
For the disks (bottom curve), also the expression (\ref{FA disk rg}) is shown (solid black line), which can be studied with Mathematica.
In the inset we provide the highest point $z_\ast$ of the surfaces corresponding to all the points in the main plot, by adopting the same colour code. 
}
\end{figure}

Given a disk $A$ in the $z=0$ plane with radius $R$, in this subsection we consider the background (\ref{holog rg 4dim}).
Since $z=z(\rho)$, the area functional evaluated on the class of surfaces $\gamma_\varepsilon$ associated to the disk is given by
\be
\label{area functional disk rg}
\mathcal{A}[\gamma_\varepsilon] = 
2\pi \int_{0}^{R-\omega}  \frac{\rho}{z^2 \, p(z)} \,
\sqrt{1+z_\rho^2 \,p(z)}
\, d\rho \,.
\ee
As already remarked above, also in this case we can observe that, since the integrand depends explicitly on $\rho$, we cannot write a first integral. 
The equation of motion to solve remains an ordinary differential equation of the second order and its analytic solution is not known for a non trivial $p(z)$.

\begin{figure}[h!] 
\vspace{-.7cm}
\begin{center}
\hspace{-.27cm}
\includegraphics[width=.921\textwidth]{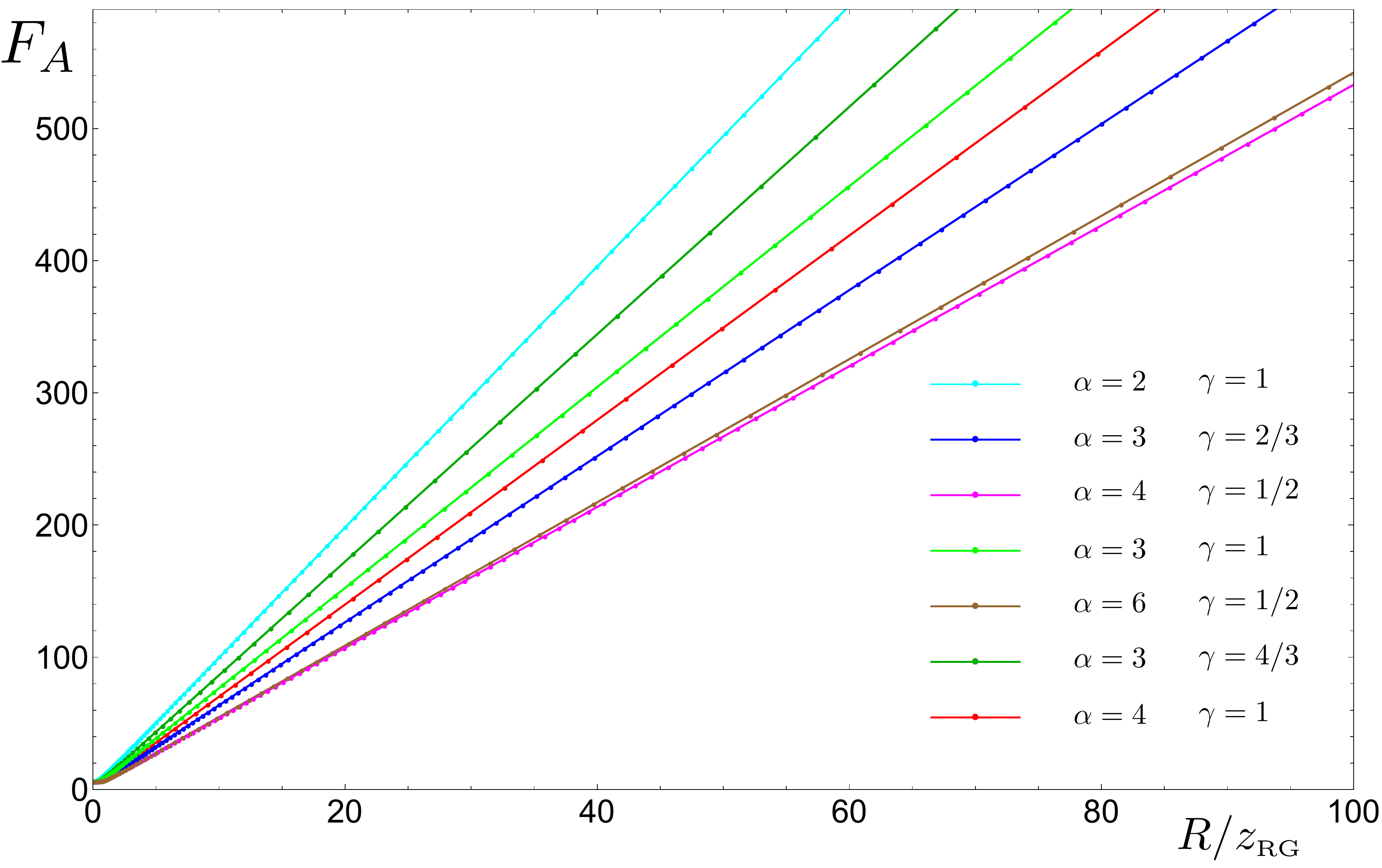}
\\
\vspace{.1cm}
\hspace{-.55cm}
\includegraphics[width=.91\textwidth]{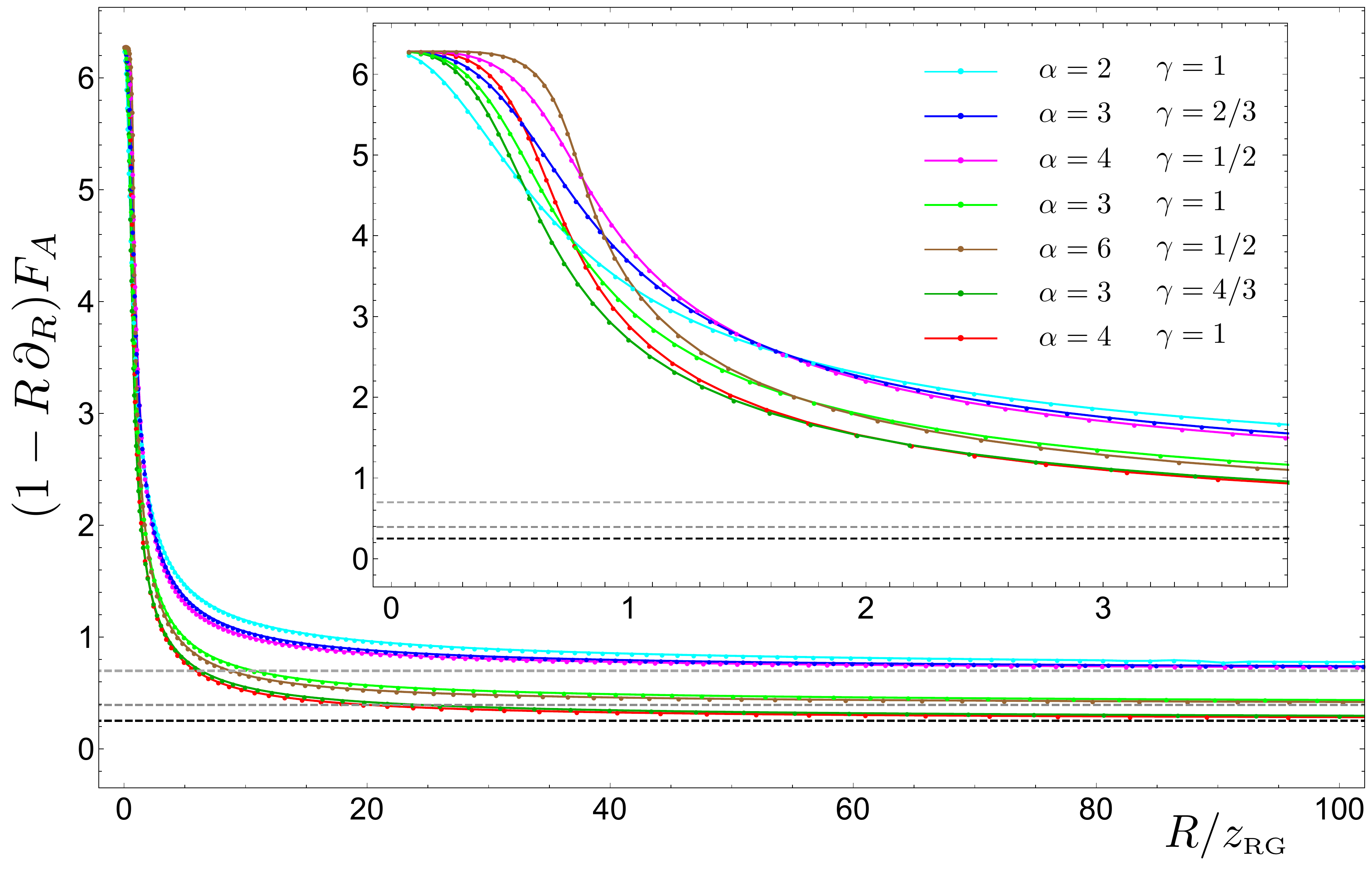}
\end{center}
\vspace{-.3cm}
\caption{\label{fig:FAdiskCfunction}
The quantity $F_A$ of a disk for the domain wall geometry (\ref{holog rg 4dim}) with various $\alpha$ and $\gamma$.
Top: Plot with larger values of $R$ with respect to Figs.\,\ref{fig:FARG} and \ref{fig:FARG2}. 
Bottom: The function $4G_N C$ from (\ref{C function def disk}) in terms of $R/z_{\textrm{\tiny RG}}$. In the UV regime $4G_N C_{\textrm{\tiny UV}} = 2\pi $ and in the IR regime $4G_N C_{\textrm{\tiny IR}} = 2\pi/(1+\alpha\gamma)^2 $ (dashed lines).
In the inset we show a zoom of the main plot for small values of $R/z_{\textrm{\tiny RG}}$.
These data correspond to $z_{\textrm{\tiny RG}}=1$ but they have been checked also through other values of $z_{\textrm{\tiny RG}}$.
}
\end{figure}

The finite term $F_A$ for the holographic entanglement entropy of a disk can be obtained from (\ref{FA holog rg}).
Indeed, by employing the proper expressions for the vector $\tilde{n}^z$ and the area element $d\tilde{\mathcal{A}} $ given in (\ref{normal vectors up rg}) and (\ref{area element rg})  respectively, one finds that (\ref{FA holog rg}) becomes
\be
\label{FA disk rg}
F_A 
    \,=\,
  2\pi \int_0^R
 \left[
 \left( 1+ \frac{z \, p'(z)}{2 \,p(z)} \right) \frac{1}{1+z_\rho^2\, p(z)}
+ \frac{z\,p'(z)}{2\, p(z)} \, 
 \right]
   \frac{\sqrt{1+z_\rho^2\, p(z)}}{z^2\, p(z)}\,\rho\,d\rho  \,.
\ee
This expression needs the explicit form of $z(\rho)$, which can be found by solving numerically the second order ordinary differential equation coming from the variation of (\ref{area functional disk rg}).
In order to check the consistency of this expression, notice that for $R/z_{\textrm{\tiny RG}} \ll 1$ we have that $p(z) \to 1$ (i.e. $p'(z) \to 0$) and in this limit 
(\ref{FA disk rg}) becomes (\ref{FA ads4 disk}) for AdS$_4$, as expected.

An analysis similar to the one made for the black hole in \S\ref{sec disk bh} leads us to observe that $F_A  = F_A (R/z_{\textrm{\tiny RG}})$. 
In particular, one first introduces the following rescaling
\be
\label{rho z hat def}
\hat{\rho} \equiv \frac{\rho}{z_{\textrm{\tiny RG}}}  \,,
\qquad
\hat{z} \equiv \frac{z}{z_{\textrm{\tiny RG}}}  \,.
\ee
in terms of which $p(z) = (1+ \hat{z}^\alpha )^{2\gamma} \equiv \hat{p}(\hat{z})$.
Then, we also have  $z \, p'(z) = \hat{z} \,\hat{p}'(\hat{z})$, where $\hat{p}'(\hat{z}) = \partial_{\hat{z}}\hat{p}(\hat{z})$, and $z_\rho = \hat{z}_{\hat{\rho}}$.
The differential equation obtained by extremizing (\ref{area functional disk rg}) gives $\hat{z}=\hat{z}(\hat{\rho})$. 
Indeed, denoting by $\mathcal{L}$ the integrand of (\ref{area functional disk rg}), we have that $\mathcal{L} =  \hat{\mathcal{L}} / z_{\textrm{\tiny RG}}$, where
\be
\hat{\mathcal{L}} = 
\frac{\hat{\rho}}{\hat{z}^2\, \hat{p}(\hat{z})}\, \sqrt{1+ \hat{z}_{\hat{\rho}}^2\, \hat{p}(\hat{z})} \,.
\ee
The equation of motion for $\mathcal{L}$ can be written as the equation of motion for $\hat{\mathcal{L}}$ and the boundary condition is $\hat{z}(R/z_{\textrm{\tiny RG}}) =0$, as one can see from $z(R)=0$ and (\ref{rho z hat def}).
These observations allow us to write (\ref{FA disk rg}) in terms of (\ref{rho z hat def}), finding
\be
\label{FA disk rg hat}
F_A 
    \,=\,
2\pi \int_0^{R/z_{\textrm{\tiny RG}}}
 \left[
 \left( 1+ \frac{\hat{z}\,\hat{p}'(\hat{z})}{2\, \hat{p}(\hat{z})}  \right) \frac{1}{1+\hat{z}_{\hat{\rho}}^2\, \hat{p}(\hat{z})}
+ \frac{\hat{z}\,\hat{p}'(\hat{z})}{2\, \hat{p}(\hat{z})} \, 
 \right]
   \frac{\sqrt{1+ \hat{z}_{\hat{\rho}}^2\, \hat{p}(\hat{z})}}{\hat{z}^2\, \hat{p}(\hat{z})}\,\hat{\rho}\,d\hat{\rho} \,,
\ee   
which tells us that $F_A  = F_A (R/z_{\textrm{\tiny RG}})$.

The bottom curves in Figs.\,\ref{fig:FARG} and \ref{fig:FARG2} provide a check of the expressions (\ref{FA disk rg}) and (\ref{FA disk rg hat}) against numerical results obtained through Surface Evolver (coloured lines) and Mathematica (black line).
Further observations can be made from these curves. 
In particular, an interesting quantity to compute is $C= -(1-R\,\partial_R ) S_A $ when $A$ is a disk of radius $R$ because for $2+1$ dimensional field theories  it plays a role similar to the one of the central charge in $1+1$ dimensions \cite{Liu:2012eea, Casini:2012ei}. 
It is straightforward to observe that the leading term proportional to $R$ giving the area law in (\ref{hee ads4 intro}) does not contribute to $C$ and therefore we have 
\be
\label{C function def disk}
C= \frac{1}{4G_N}\, \big(1-R\,\partial_R \big) F_A = \frac{1}{4G_N}\,  \big(1-R_{\textrm{\tiny RG}} \,\partial_{R_{\textrm{\tiny RG}}} \big) F_A  \,,
\qquad
R_{\textrm{\tiny RG}} \equiv \frac{R}{z_{\textrm{\tiny RG}}}\,.
\ee

When $R_{\textrm{\tiny RG}}  \ll 1$ the minimal surface probes AdS$_4$ with radius equal to one and therefore $4G_N \,C_{\textrm{\tiny UV}} =2\pi$.
In order to probe the IR regime  very large values of $R_{\textrm{\tiny RG}}$ must be considered. 
In Fig.\,\ref{fig:FAdiskCfunction} we have performed a numerical analysis of $F_A$ and of the $C$ function (\ref{C function def disk}) in terms of $R_{\textrm{\tiny RG}}$ (reported in the top panel and in the bottom panel respectively) by taking values of $R_{\textrm{\tiny RG}}$ much larger than the ones explored in Figs.\,\ref{fig:FARG} and \ref{fig:FARG2}, finding that the latter ones do not allow us to capture the correct IR behaviour.
Indeed, in the IR regime a linear behaviour occurs $F_A = a R_{\textrm{\tiny RG}} + 2\pi/(1+\alpha\gamma)^2 +\dots $, where the dots correspond to subleading terms \cite{Liu:2013una}.
Thus $4G_N \,C_{\textrm{\tiny IR}} = 2\pi/(1+\alpha\gamma)^2 = 2\pi L^2_{\textrm{\tiny IR}} $ and therefore $C_{\textrm{\tiny IR}} < C_{\textrm{\tiny UV}} $.
Let us stress that, despite the fact that already for the values of $R_{\textrm{\tiny RG}}$ in Figs.\,\ref{fig:FARG} and \ref{fig:FARG2} a linear behaviour seems to arise, it is not enough to get the expected value for $C_{\textrm{\tiny IR}}$, as one can appreciate by means of a comparison with the plot shown in the bottom panel of Fig.\,\ref{fig:FAdiskCfunction}.
While $C_{\textrm{\tiny IR}}$ depends only on the product $\alpha\gamma$ (the asymptotic values are highlighted by the horizontal dashed lines in the bottom panel of Fig.\,\ref{fig:FAdiskCfunction}), the slope $a$ of the linear behaviour in the IR regime depends on these parameters separately, as one can observe from the top panel of  Fig.\,\ref{fig:FAdiskCfunction}.

\subsubsection{Vaidya-AdS backgrounds}
\label{sec vaidya disk}

When the bulk background is the Vaidya-AdS metric (\ref{vaidya metric}) and $A$ is a disk of radius $R$, the rotational symmetry allows to describe the profile of $\gamma_A$ in terms of two functions, $z(\rho)$ and $v(\rho)$, once the polar coordinates $(t,\rho,\theta)$ have been chosen for the Minkowski space at $z=0$.

The area functional for $\gamma_\varepsilon$ in this case reads 
\be
\mathcal{A}[\gamma_\varepsilon] \,=\, 
2\pi \int_0^{R-\omega}
\frac{\sqrt{1-2v' z' -f(v,z) (v')^2}}{z^2} \; \rho\,d\rho \,.
\ee
Considering the explicit expression $f(v,z)=1-M(v)z^3$ and by employing the results discussed in \S\ref{app:unit vectors vaidya} for the unit vectors and the area element, the formula (\ref{FA vaidya}) for the finite term becomes
\be 
\label{FA vaidya disk}
\frac{F_A}{2\pi} = \int_0^R
\frac{
2 v' z' \big(z^3 M-2\big)-2 \big((v')^2-1\big)-2 (z')^2
-z^3 \big[(v')^2 \left(z M'+M  \left(z^3 M -3\right)\right)+4 M \big]}{
z^2\,\sqrt{1-2 v' z' -(1-M z^3) (v')^2 }} 
\;\rho \,d\rho \,,
\ee
where $M=M(v)$.
Notice that (\ref{FA vaidya disk}) reproduces (\ref{FA bh disk}) when $M(v)$ is constant.  

\begin{figure}[h!] 
\vspace{-.2cm}
\begin{center}
\hspace{-.5cm}
\includegraphics[width=.9\textwidth]{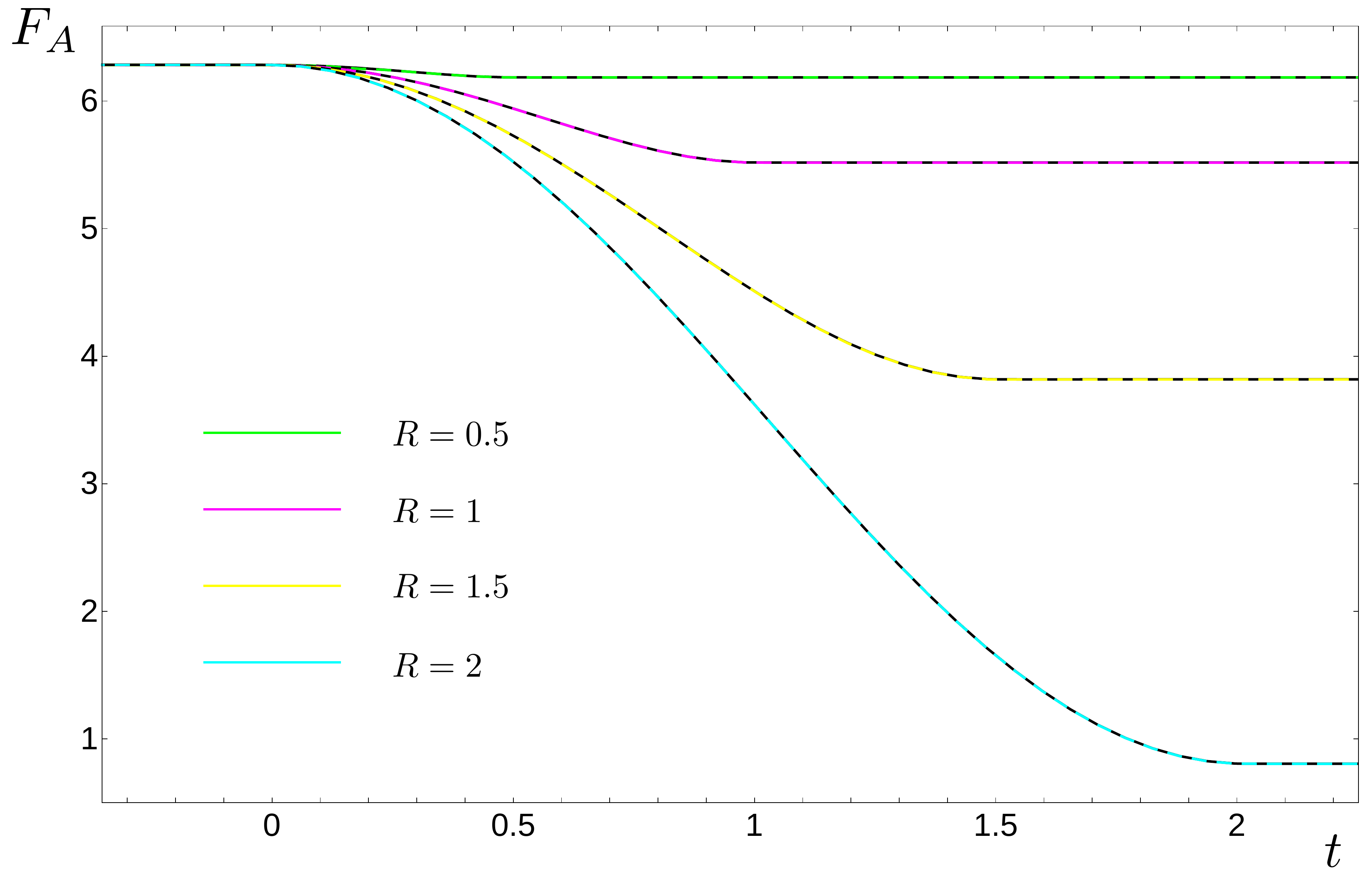}
\\
\hspace{-.5cm}
\includegraphics[width=.9\textwidth]{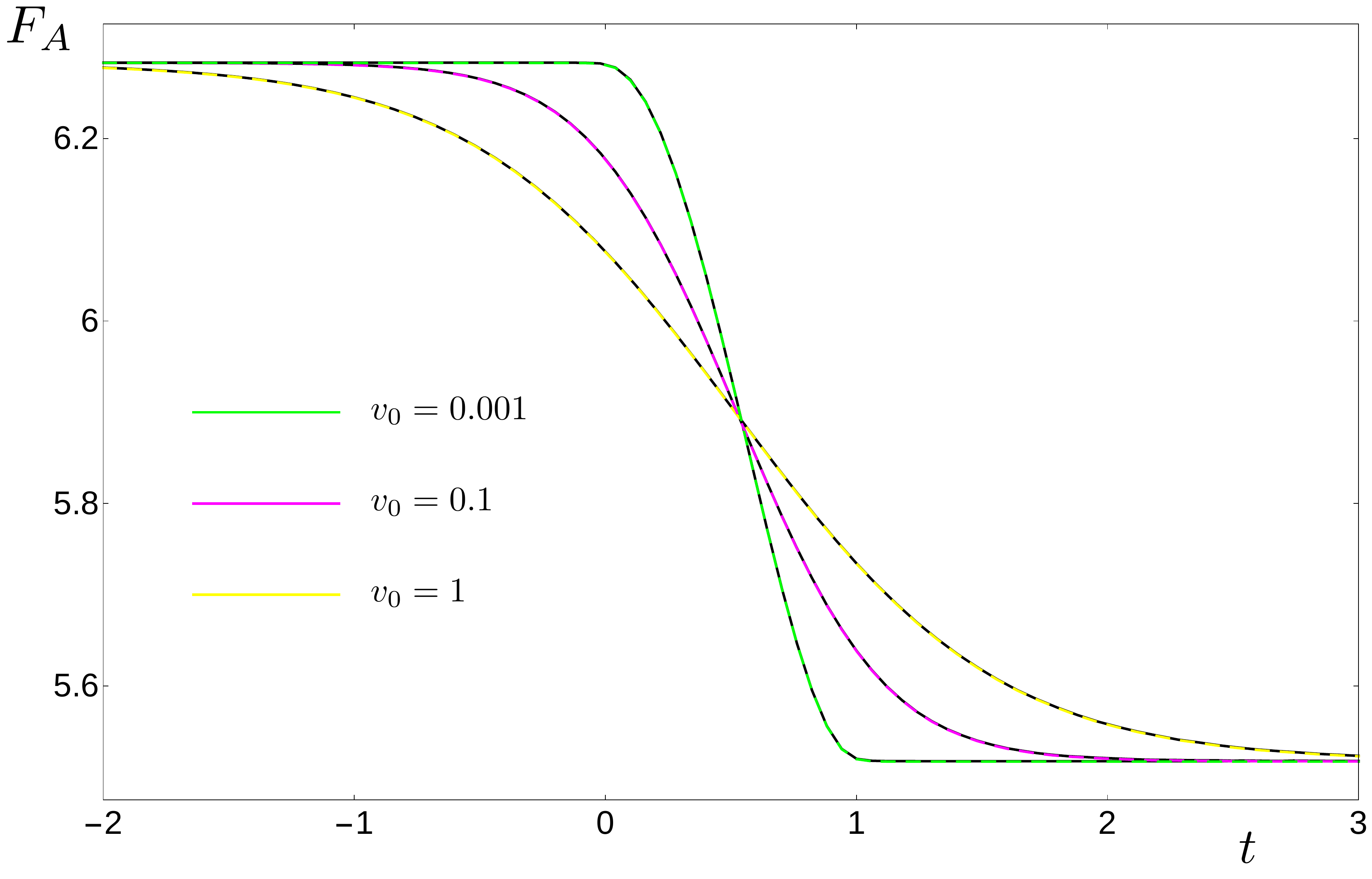}
\end{center}
\vspace{-.4cm}
\caption{\label{fig:FAvaidya}
The quantity $F_A$ for Vaidya-AdS backgrounds (\ref{vaidya metric}) with mass profile (\ref{eq:kinkeq}) as function of the boundary time $t$ when $A$ are disks of radius $R$.
Here $\varepsilon = 10^{-6}$ and these computations have been done with Mathematica.
The solid coloured lines correspond to the formula (\ref{FA vaidya disk}) while the dashed ones have been obtained 
by subtracting the term $2\pi R/\varepsilon$ from the area.
Top: $v_0=10^{-3}$ fixed.
Bottom: $R=1$ fixed.
}
\end{figure}

Choosing the mass profile (\ref{eq:kinkeq}), in Fig.\,\ref{fig:FAvaidya} we plot $F_A$ found in two ways: through our formula (\ref{FA vaidya disk}) (solid coloured lines) or through the usual method of subtracting the divergence from the area of the extremal surface. The good agreement of these results provides an important check for (\ref{FA vaidya disk}).

\subsection{Other domains}

\begin{figure}[t] 
\vspace{-.7cm}
\hspace{-.25cm}
\includegraphics[width=1.02\textwidth]{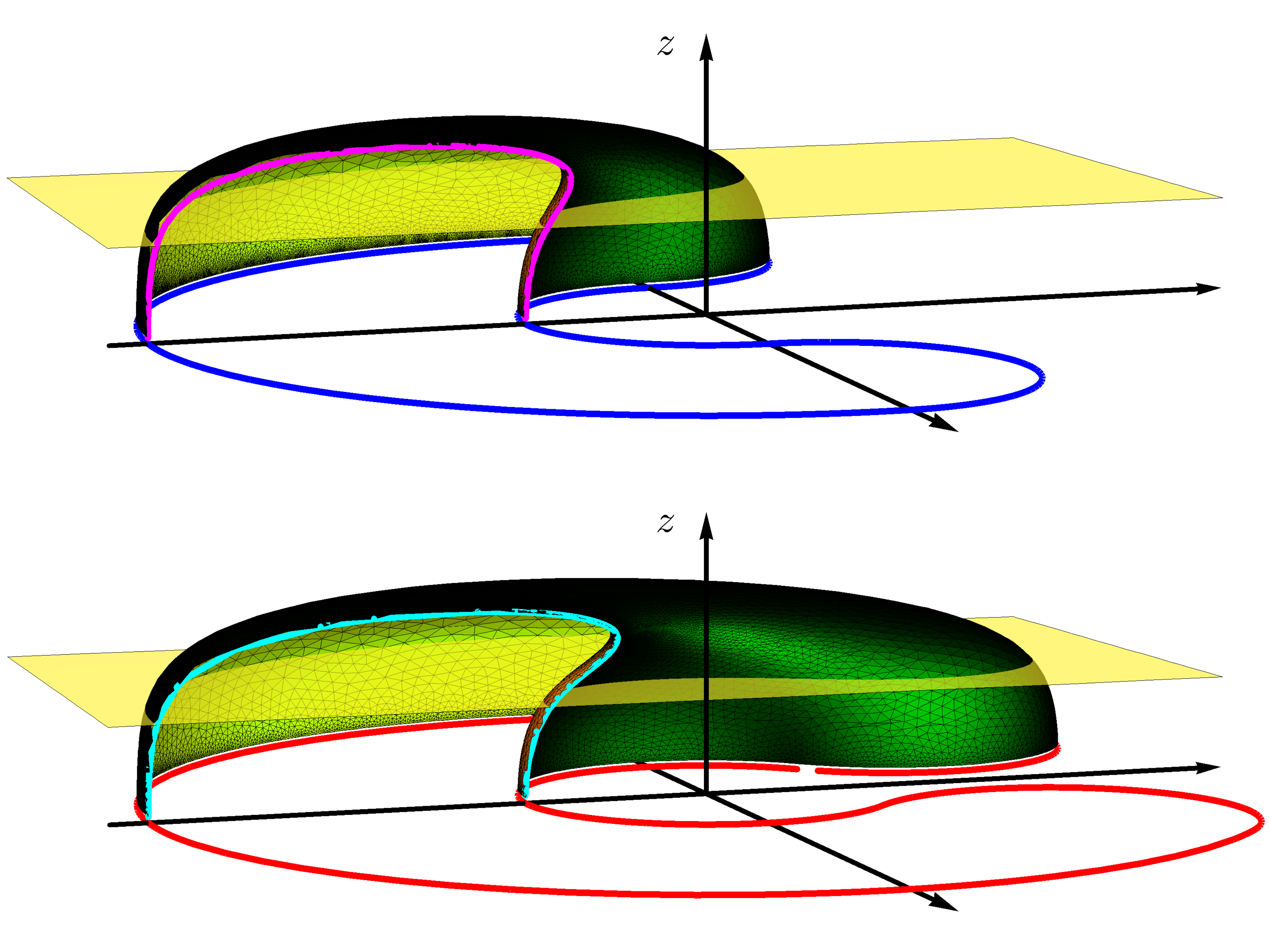}
\vspace{.2cm}
\caption{\label{3DbananaRG}
Minimal area surfaces $\hat{\gamma}_A$ for the domain wall geometry (\ref{holog rg 4dim}) with $\alpha=2$, $\gamma=1$.
The yellow plane corresponds to $z=z_{\textrm{\tiny RG}}$.
The entangling curves $\partial A$ (blu curve in the top panel and red curve in the bottom panel), which belong to the $z=0$ plane, are constructed by joining arcs of circumferences and they delimit two non convex domains.
The centers of the circumferences (the outer one has radius $R=3$ and the inner one $R/3$) form an opening angle of $\pi$ (top) and $1.54\pi$ (bottom). 
Here $z_{\textrm{\tiny RG}}=0.5$.
Only half of the surfaces $\hat{\gamma}_A$ are shown in order to highlight the section (magenta curve in the top panel and cyan curve in the bottom panel) reaching the highest value $z_\ast$ along the holographic direction. 
}
\end{figure}

\begin{figure}[t] 
\vspace{-.2cm}
\begin{center}
\hspace{-.5cm}
\includegraphics[width=1.\textwidth]{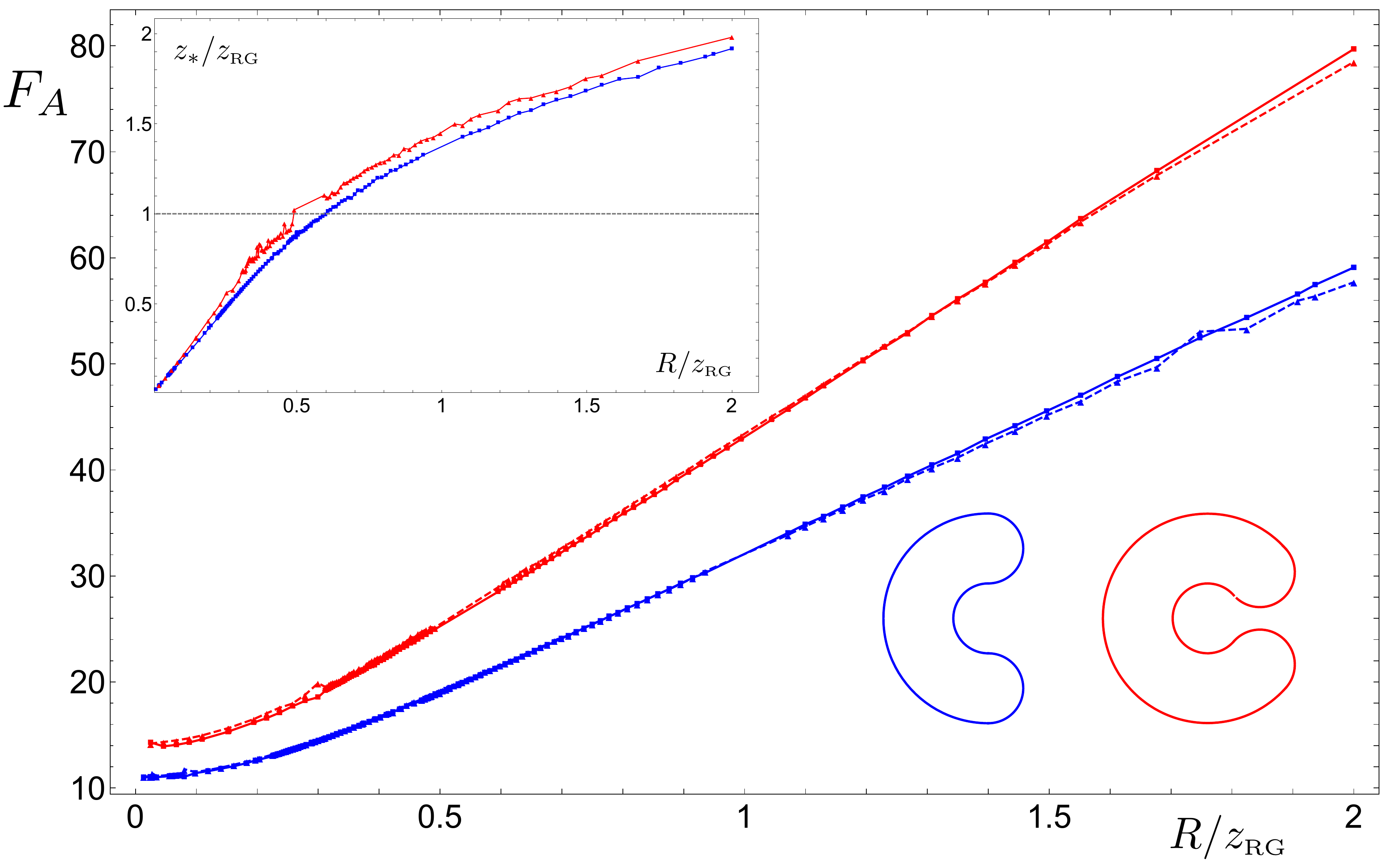}
\end{center}
\vspace{-.4cm}
\caption{\label{fig:FAbanana}
The quantity $F_A$ for the domain wall geometry (\ref{holog rg 4dim}) with $\alpha=2$ and $\gamma=1$.
The entangling curves are the blue  and the red ones in the bottom right part of the plot, which are obtained by joining arcs of circumferences  whose centers provide an opening angle given by $\pi$ and $1.54 \pi$ respectively. 
The radius of the external circumference is $R$ and the radius of the internal one is $R/3$
(see Fig.\,\ref{3DbananaRG} for two examples of minimal surfaces $\hat{\gamma}_A$ anchored to these entangling curves).
The numerical analysis has been done with Surface Evolver by taking $\varepsilon =0.03$, $R=3$ and moving $z_{\textrm{\tiny RG}}$ in the interval $(0.5, 70)$. 
Solid and dashed lines  correspond respectively to the two ways to find $F_A$ given in (\ref{SE F_A}).
In the inset we show $z_\ast/z_{\textrm{\tiny RG}}$ in terms of $R/z_{\textrm{\tiny RG}}$ corresponding to all the points in the main plot. 
}
\end{figure}

In the previous discussions we have considered domains $A$ which are highly symmetric because their symmetry usually allows to treat the problem of the minimal area surface analytically up to some point.

In order to study analytically the minimal area surface $\hat{\gamma}_A$ associated to a generic domain $A$, the first problem to address is the parameterisation of the class of surfaces $\gamma_A$.
Then, one has to solve the differential equation coming from the extremal area condition to get $\hat{\gamma}_A$ and finally compute the  area of $\hat{\gamma}_\varepsilon$. 
For simply connected domains $A$ with smooth boundary which do not have any particular symmetry, already the first step could be very difficult (see e.g. Fig.\,\ref{3DbananaRG}). 
Assuming that a convenient parameterisation for the surface has been found, the differential equation coming from the extremal area condition is usually a second order partial differential equation very difficult to solve. The main simplification introduced by highly symmetric domains (e.g. strips, disks and annuli) is that this differential equation reduces to an ordinary differential equation. 
The latter one could be difficult to solve anyway (e.g. for the black holes or for the domain wall geometries), but ordinary differential equations are much easier to study than partial differential equations, even from the numerical point of view.

The formulas for $F_A$ discussed in \S\ref{sec general case} and \S\ref{sec time-dep general} hold for a generic domain $A$ with smooth boundary, including the ones made by disjoint components.
In the latter case two or more local minima occur and the holographic prescription (\ref{RT formula intro}) requires to choose the global minimum, as we will discuss in \S\ref{sec HMI} for the case of two regions.
Nevertheless,  the formulas for $F_A$ discussed in \S\ref{sec general case} and \S\ref{sec time-dep general} involve the unit normal vector $\tilde{\boldsymbol{n}}$ and therefore one should know the analytic solution for $\hat{\gamma}_A$ in order to find it.
For instance, when $\hat{\gamma}_A$ can be parameterised as $z=z(x,y)$, the expression for $\textrm{Tr} \widetilde{K}$ contains all possible first and second order partial derivatives in a complicated way that we do not find interesting to report here.

The big advantage of the numerical analysis with Surface Evolver \cite{evolverpaper, evolverlink} is that the minimal area surface is obtained without going through this procedure of finding the convenient parameterisation first and then solving the differential equation (see \S\ref{sec num methods}). 
Moreover, as already remarked in \S\ref{sec num methods}, besides the area of the surface, also its unit normal vector $\tilde{\boldsymbol{n}}$ can be found and this allows us to check the formulas found in \S\ref{sec static backs} for non trivial domains.

Besides the cases of disks and strips discussed in \S\ref{sec: strip} and \S\ref{sec:disk}, we have considered $F_A$ also for more complicated simply connected domains, both convex and non convex. 
In particular we have studied regions $A$ delimited by ellipses for all the static backgrounds of \S\ref{sec static backs}.
For the domain wall geometries, we have considered also the non convex domains delimited by the blue and the red curves in Fig.\,\ref{3DbananaRG}.
Once the shape and all the relative ratios between the various geometrical parameters have been fixed, we have computed $F_A$ changing the total size of the region $A$. 
The numerical analysis has been done as explained in \S\ref{sec num methods}.
The area $\mathcal{A}_A$ for domains $A$ delimited by ellipses as small perturbations of circumferences has been already considered through the standard approach e.g. in \cite{Hubeny:2012ry, Allais:2014ata}  and by employing  the interesting method of \cite{Babich:1992mc, Ishizeki:2011bf, Kruczenski:2013bsa, Kruczenski:2014bla} (which is based on the solution of the cosh-Gordon equation in terms of algebraic curves) in \cite{Dekel:2015bla}.

When the bulk geometry is AdS$_4$, this rescaling of $A$ does not change $F_A$ because the Willmore energy is invariant, as already discussed in \S\ref{sec:ads4static}. 
On the other hand, for asymptotically AdS$_4$ black holes and domain wall geometries this invariance is broken and a non trivial behaviour is found under rescaling of $A$.

In Fig.\,\ref{fig:FABH} and Fig.\,\ref{fig:FABHext} we study this rescaling for the Schwarzschild-AdS black hole and the extremal Reissner-Nordstr\"om-AdS black hole respectively by employing both the formula  (\ref{FA bh gen}) and the usual way to get $F_A$ by subtracting the area law divergence, as explained in \S\ref{sec num methods}.
We show $F_A$ for $A$ given by disks or domains delimited by ellipses with semi-axis $R_1 \geqslant R_2$ having two different eccentricity.
Let us remind that the perimeter $P_A$ of an ellipse with semi-axis $R_1 \geqslant R_2$ is $P_A = 4R_1\, \mathbb{E} (1-R_2^2/R_1^2)$, where $\mathbb{E}$ is the complete elliptic integral of the second kind\footnote{We adopt the convention of Mathematica for the arguments of the elliptic integrals.}, and its area is $\textrm{Area}(A) =\pi R_1 R_2$.
For the disks we have employed also the simpler formula (\ref{FA bh disk}), which can be evaluated numerically by using Mathematica.
The plots in Fig.\,\ref{fig:FABH} and Fig.\,\ref{fig:FABHext} show that  $F_A$ is a function of $R_1/z_h$ for a given eccentricity.
It would be helpful to have data for large ellipses in order to check the behaviour $F_A =  - \, \textrm{Area}(A)/z_h^2 + \dots $ expected from (\ref{FA_cyl}).

As for the domain wall geometry (\ref{holog rg 4dim}),  in Figs.\,\ref{fig:FARG} and Fig.\,\ref{fig:FARG2}  we have considered the domains $A$ just mentioned having elliptical entangling curves for two different sets of parameters for the background ($(\alpha,\gamma)=(2,1)$ and $(\alpha,\gamma)=(4,1)$  respectively).
The expression of $F_A$ for the domain wall geometry is (\ref{FA holog rg}) and the numerical analysis has been done as mentioned above and explained in \S\ref{sec num methods}.
As the size of $A$ changes, the qualitative behaviour of $F_A$ for the domains delimited by ellipses is the same one found for the disks.
In particular, $F_A$ has a finite limit when $A$ is very small ($z_\ast \ll z_{\textrm{\tiny RG}}$). 
Nevertheless, we remark that the values of $R_{\textrm{\tiny RG}}$ explored in Figs.\,\ref{fig:FARG} and Fig.\,\ref{fig:FARG2} are too small to capture the correct IR behaviour, as we have seen in Fig.\,\ref{fig:FAdiskCfunction} for the disks.

The similarity between $F_A$ for the disk and the ones corresponding to domains delimited by ellipses observed in Figs.\,\ref{fig:FARG} and Fig.\,\ref{fig:FARG2}  motivated us to explore also the case of non convex domains. 
In particular, we have considered the non convex domains delimited by the red and blue curves in Fig.\,\ref{3DbananaRG}.
For both these cases the domains $A$ have the same shape and only one geometrical parameter (the opening angle) distinguishes them.
It is  worth remarking that, for these domains, finding a parameterisation for $\gamma_A$ is not easy, as one can immediately understand from Fig.\,\ref{3DbananaRG}, where the minimal area surfaces $\hat{\gamma}_A$ obtained with Surface Evolver are shown.

In Fig.\,\ref{fig:FAbanana} we show the results for $F_A$ corresponding to these non convex domains.
Interestingly, the qualitative behaviour of $F_A$ is again the same one observed for the disk, which led to the definition (\ref{C function def disk}) of the $C$ function.
This suggests that it could be worth generalising the definition (\ref{C function def disk}) introduced for the disks by interpreting $R$ as a global parameter of $A$ and exploring better whether proper $C$ functions can be defined from domains which are not disks \cite{Liu:2013una, Liu:2012eea}.  
We remark that in our numerical analysis for domains which are not disks we have not considered domains large enough to capture the IR behaviour of $F_A$.
Indeed, from the case of the disk, whose relevant plots are shown in Fig.\,\ref{fig:FAdiskCfunction}, we have learned that the values of $R_{\textrm{\tiny RG}}$ explored in Figs.\,\ref{fig:FARG}, \ref{fig:FARG2} and \ref{fig:FAbanana} are too small to probe the deep IR regime. 
We should push our numerical analysis to much higher values of $R_{\textrm{\tiny RG}}$ but, unfortunately, our present code is numerically unstable.
We hope to overcome this technical obstacle in the near future.

As for the time dependent backgrounds, it would be very interesting to check (\ref{FA vaidya}) by considering finite regions with smooth non circular boundaries.
This would be helpful to have a better understanding of the tsunami picture introduced in \cite{Liu:2013iza}.

\subsection{Infinite wedge}
\label{sec wedge}

An important class of domains $A$ to study is given by the ones whose boundary $\partial A$ contains some corners. 
In these cases, the entanglement entropy has also a logarithmic divergence besides the area law term.
The simplest example to address is the infinite wedge for the gravitational background given by AdS$_4$, whose corresponding minimal area surface has been first studied in \cite{Drukker:1999zq} and its area has been computed. 
The aim of this section is to show how the analytic result of  \cite{Drukker:1999zq} can be recovered through the formula (\ref{FA ads4 sec}).
Here we give only the main expressions to understand the result, but all the technical details of this computation have been reported  in \S\ref{sec app wedge}). 
It is worth recalling that (\ref{FA ads4 sec})  has been obtained by assuming smooth entangling curves for $\partial A$ and, under this hypothesis, it provides a finite result as $\varepsilon \to 0$.
Nevertheless, we find it interesting and non trivial that the Willmore energy (\ref{FA ads4 sec}) provides the expected logarithmic divergence for non smooth entangling curves.

Choosing polar coordinates $\{\rho, \theta\}$ in the $z=0$ plane such that the origin coincides with the tip of the wedge, the domain that we are going to consider is  $A= \{(\rho, \theta), |\theta| \leqslant \Omega/2 , \rho \leqslant L\}$, where $\Omega$ is the opening angle of the wedge and $L \gg 1$ its length along the edges. 
Since $L \gg 1$, we can employ the following ansatz for the minimal surface $\hat{\gamma}_A$ \cite{Drukker:1999zq} 
\be
\label{ansatz wedge}
z = \frac{\rho}{q(\theta)} \,,
\ee
where the function $q(\theta)$ can be found by imposing the extremal area condition.

Considering the boundary of $\hat{\gamma}_\varepsilon$, part of this curve  lies at $z=\varepsilon$, which will be denoted by $\partial\hat{\gamma}_\varepsilon^\parallel$, and the remaining part $\partial\hat{\gamma}_\varepsilon^\perp$ belongs to a vertical  cylinder.
From the projection of $\partial\hat{\gamma}_\varepsilon^\parallel$ on the $z=0$ plane, one finds that the range of the radial coordinate of $\hat{\gamma}_\varepsilon$ is $\rho_{\textrm{\tiny min}}\leqslant \rho \leqslant \rho_{\textrm{\tiny max}}$, where
\be
\label{rho minimax}
\varepsilon = \frac{\rho_{\textrm{\tiny min}}}{q_0} \,,
\qquad
\varepsilon = \frac{\rho_{\textrm{\tiny max}}}{q(\Omega/2-\omega)} \,,
\qquad
L = \rho_{\textrm{\tiny max}} \cos \omega \,,
\ee
being $\omega \sim 0^+$ the angle between the edge of $A$ and the straight line connecting the tip of the wedge to the intersection point between the circumference $\rho= \rho_{\textrm{\tiny max}} $ and the projection of $\partial\hat{\gamma}_\varepsilon^\parallel$ on the $z=0$ plane.

In \S\ref{sec app wedge} we show that for the wedge we are considering the formula (\ref{FA ads4 sec}) gives
\be
\label{F_A wedge willmore}
F_A \,=\,
2\, b(\Omega) \log(L/\varepsilon)  + O(1) \,,
\ee
where $b(\Omega)$ is the function found in \cite{Drukker:1999zq} 
\be
\label{cusp exact result}
b(\Omega)
\,=
\int_0^\infty d\zeta  \left(1-\sqrt{\frac{\zeta ^2+{q_0}^2+1}{\zeta ^2+2 {q_0}^2+1}}\,\right)
\,=\,
\frac{\mathbb{E}(\hat{q}_0^2)-\big(1-\hat{q}_0^2\big)\mathbb{K}(\hat{q}_0^2)}{\sqrt{1-2\hat{q}_0^2}} \,,
\ee
being $\hat{q}_0^2 \equiv q_0^2/(1+2q_0^2) \in [0,1/2]$ and the functions $\mathbb{K}$ and $\mathbb{E}$ are the complete elliptic integrals of the first and second kind respectively.

As for the contribution of $\partial \hat{\gamma}_\varepsilon$ to the holographic entanglement entropy, it is given by the contour integral in the r.h.s. of (\ref{area generic J2 bdy}), whose integrand is $\tilde{b}^z/z $ in our case.  
Such contour  integral is the sum of two contributions: the line integral over  $\partial\hat{\gamma}_\varepsilon^\parallel$ and the line integral over  $\partial\hat{\gamma}_\varepsilon^\perp$. 
Along  $\partial\hat{\gamma}_\varepsilon^\parallel$ we have $\tilde b^z \sim -1$ for the parts of the curve close to the edges of $A$, while it significantly deviates  from $-1$ for the part of $\partial \hat{\gamma}_\varepsilon$ close to the tip of the wedge (i.e. at $\rho\sim\rho_{\textrm{\tiny min}}$) without becoming infinitesimal.
Instead, along  $\partial\hat{\gamma}_\varepsilon^\perp$ we have $\tilde b^z \sim 0$ close to the boundary and becomes finite around $\theta=0$, which is also the point of $\hat{\gamma}_A$ with the highest value of $z$.
Considering these two contributions together and using that $L \gg 1$, one finds that
\be
\label{wedge bdy int total}
\int_{\partial \hat{\gamma}_\varepsilon} \frac{\tilde b^z}{z} \, d\tilde{s}
\,=\,
-\,\frac{2L}{\varepsilon}+O(1) \,.
\ee
Thus, while the area term for the infinite wedge comes from the boundary integral (\ref{wedge bdy int total}), the subleading logarithmic divergence (\ref{F_A wedge willmore}) is encoded into the Willmore energy (\ref{FA ads4 sec}) with the expected coefficient.


\section{Holographic mutual information}
\label{sec HMI}

In this section we briefly discuss the holographic mutual information and, in the case of AdS$_4$, some straightforward consequences of the formula (\ref{FA ads4 sec}).

Given two disjoint spatial domains $A_1$ and $A_2$ in the boundary, one can consider the entanglement entropy $S_{A_1 \cup A_2}$, which measures the entanglement between $A_1 \cup A_2$ and its complement.
A very useful quantity to introduce is the mutual information
\be
\label{MI def}
I_{A_1, A_2} \equiv S_{A_1} + S_{A_2} - S_{A_1 \cup A_2}  \,.
\ee
Since the divergent terms of the entanglement entropy $S_A$ depend on the entangling surface $\partial A$, they cancel in this combination and the mutual information (\ref{MI def}) is UV finite.

For the two disjoint domains $A_1$ and $A_2$,  the subadditivity property of the entanglement entropy reads
\be
\label{sub add}
S_{A_1} + S_{A_2} \geqslant S_{A_1 \cup A_2} \,,
\ee
which means that the mutual information (\ref{MI def}) is non negative $I_{A_1, A_2}  \geqslant 0$.

When $A_1$ and $A_2$ have a non vanishing intersection,  the strong subadditivity property of the entanglement entropy holds \cite{Lieb-Ruskai}
\be
\label{strong sub add}
S_{A_1} + S_{A_2} \geqslant S_{A_1 \cup A_2} + S_{A_1 \cap A_2} \,,
\ee
which tells us the mutual information increases as one of the two disjoint domains is enlarged
\be
\label{ssa MI}
I_{A_1 \cup A_0 , A_2} \geqslant I_{A_1 , A_2} \,.
\ee

The holographic entanglement entropy formula (\ref{RT formula intro}) of \cite{Ryu:2006bv, Ryu:2006ef} can be applied also for  disjoint domains and the strong subadditivity property for the holographic prescription has been proven in \cite{Headrick:2007km}.
For two disjoint regions $A_1$ and $A_2$,  let us introduce $\mathcal{I}_{A_1, A_2}$ as follows
\be
\label{holog MI def}
I_{A_1, A_2} \equiv \frac{\mathcal{I}_{A_1, A_2}}{4G_N} \,.
\ee
As already remarked above, the mutual information is UV finite and, from (\ref{hee ads4 intro}), we have that 
\be
\label{I2 holo def}
\mathcal{I}_{A_1, A_2} \,=\,
F_{A_1 \cup A_2} - F_{A_1}  - F_{A_2}  + o(1) \,,
\ee
where $F_{A_1 \cup A_2}$ is found by taking the global minimum $\hat{\gamma}_{A_1 \cup A_2}$ of the area functional among all the surfaces $\gamma_{A_1 \cup A_2}$ such that $\partial \gamma_{A_1 \cup A_2} = \partial( A_1 \cup A_2) =  \partial A_1 \cup \partial A_2$.
In this computation, it is well known that typically two local minima occur: a connected surface $\hat{\gamma}_{A_1, A_2}^{\textrm{\tiny  \,con}}$ joining $\partial A_1$ and $\partial A_2$ through the bulk and a disconnected configuration $\hat{\gamma}_{A_1} \cup \hat{\gamma}_{A_2}$ made by the two disjoint surfaces  found for the holographic entanglement entropy of $A_1$ and $A_2$ separately \cite{Gross:1998gk, Drukker:2005cu, Headrick:2010zt, Tonni:2010pv}.\\
Since $\hat{\gamma}_{A_1, A_2}^{\textrm{\tiny  \,con}}$ and $\hat{\gamma}_{A_1} \cup \hat{\gamma}_{A_2}$ have the same boundary, from (\ref{RT formula intro}) and (\ref{hee ads4 intro}) one gets
\be
\label{Will2 def}
F_{A_1 \cup A_2} =
\textrm{max}\big( F_{A_1, A_2}\, ,  F_{A_1}  + F_{A_2}  \big) \,,
\ee
where $F_{A_1, A_2} $ is defined as the $O(1)$ term in (\ref{hee ads4 intro}) when the global minimum is provided by the connected surface $\hat{\gamma}_{A_1, A_2}^{\textrm{\tiny  \,con}}$.
In (\ref{Will2 def}) the max occurs  because $F_A$ enters with a minus sign in the expansion of $S_A$.

When the two disjoint domains  $A_1$ and $A_2$ are very close, the connected surface $\hat{\gamma}_{A_1, A_2}^{\textrm{\tiny  \,con}}$ is the global minimum and $\mathcal{I}_{A_1, A_2}>0$, while, when the distance between the domains is large enough, the configuration made by the union of the two disconnected surfaces 
$\hat{\gamma}_{A_1} $ and $\hat{\gamma}_{A_2} $ becomes the global minimum and $\mathcal{I}_{A_1, A_2}=0$.
The transition between these two regimes occurs when 
\be
\label{HMI transition}
F_{A_1, A_2} = F_{A_1}  + F_{A_2}  \,.
\ee
Keeping the shapes of $\partial A_1$ and $\partial A_2$ and their relative orientation fixed, one can change the relative distance and find
where (\ref{holog MI transition}) holds. 
This is difficult for shapes which are not highly symmetric like disks or infinite strips (see \cite{Fonda:2014cca} for a numerical analysis). 
One could employ the expressions discussed in the previous sections to find some further results for the solution of (\ref{HMI transition}) in terms of the shapes of the domains. 

It is worth remarking that, while the disconnected configuration $\hat{\gamma}_{A_1} \cup \hat{\gamma}_{A_2}$ can be found for every distance between $\partial A_1$ and $\partial A_2$, the connected one does not exist for distances larger than a critical one, which is obviously bigger than the distance defined by (\ref{HMI transition}) \cite{Drukker:2005cu, Fonda:2014cca, Hirata:2006jx, Krtous:2014pva}.

Given the disjoint domains $A_1$ and $A_2$, let us enlarge $A_1$ getting $A_1 \cup A_0$ and consider the corresponding extremal area surfaces occurring in the computation of the holographic mutual information. 
Plugging the holographic formula (\ref{holog MI def}) into the strong subadditivity property (\ref{ssa MI}), one finds  that $\mathcal{I}_{A_1 \cup A_0 , A_2}  \geqslant \mathcal{I}_{A_1 , A_2}$ and, from (\ref{I2 holo def}), this tells us that 
\be
\label{holog ssa F_A}
F_{A_1 \cup A_0 \cup A_2} - F_{A_1 \cup A_2}
\,\geqslant\,
F_{A_1 \cup A_0} - F_{A_1}   \,.
\ee
Notice that, while in the l.h.s. of this inequality disjoint domains occur and therefore the maximisation in (\ref{Will2 def}) must be performed for both  terms, in the r.h.s. only connected domains are involved. 
When $A_1 \cup A_0 $ and $A_1$ are sufficiently far from $ A_2$, i.e.  their distances are such that  $\mathcal{I}_{A_1 \cup A_0 , A_2} = 0$ and $\mathcal{I}_{A_1 , A_2} = 0$, in (\ref{holog ssa F_A}) we have $F_{A_1 \cup A_0 \cup A_2} = F_{A_1 \cup A_0 }+ F_{A_2}$ and $F_{A_1 \cup A_2} = F_{A_1} + F_{A_2}$ and the inequality is trivially saturated.

Let us restrict to the part of the space of configurations where $\mathcal{I}_{A_1 \cup A_0 , A_2} > 0$ and $\mathcal{I}_{A_1 , A_2}>0$, where $\hat{\gamma}_{A_1 \cup A_0, A_2}^{\textrm{\tiny  \,con}}$ and $\hat{\gamma}_{A_1, A_2}^{\textrm{\tiny  \,con}}$ are the global minima
to consider for the holographic entanglement entropy $S_{A_1 \cup A_0 \cup A_2}$ and $S_{A_1 \cup A_2}$ respectively.
In this case (\ref{holog ssa F_A}) becomes
\be
\label{holog ssa F_A conn}
F_{A_1 \cup A_0 , A_2} - F_{A_1 , A_2}
\,\geqslant\,
F_{A_1 \cup A_0} - F_{A_1}   \,.
\ee
Considering a domain $A_0$ very small with respect to $A_1$, we can interpret the enlarging of $A_1$ by $A_0$ as a small perturbation of $A_1$. For this case, (\ref{holog ssa F_A conn}) tells us that  the variation of $F$ under such perturbation is bigger when $A_2$ occurs.
The inequality (\ref{holog ssa F_A conn}) is a non trivial property of the formulas for $F_A$ discussed in the previous sections.

\subsection{AdS$_4$}

When the gravitational background is AdS$_4$, we have that $2F_A$ is the Willmore energy of the closed surface  $\hat{\gamma}^{\textrm{\tiny (d)}}_A$ embedded in $\mathbb{R}^3$, as stated in (\ref{FA from willmore}).
We can employ some known results on the Willmore functional to find some properties of $F_A$, as done in \S\ref{sec:ads4static} for connected domains $A$.
For $A=A_1 \cup A_2$ made by two disjoint domains, the surface $\hat{\gamma}^{\textrm{\tiny (d)}}_A$ introduced in \S\ref{sec:ads4static} is connected when $\mathcal{I}_{A_1 , A_2}>0$ and disconnected when $\mathcal{I}_{A_1 , A_2}=0$.
In the former case we will denote the corresponding closed surface by $\hat{\gamma}_{A_1, A_2}^{\textrm{\tiny  \,con,(d)}}$, while in the latter case the two surfaces $\hat{\gamma}^{\textrm{\tiny (d)}}_{A_1} \cup \hat{\gamma}^{\textrm{\tiny (d)}}_{A_2}$ occur.
When $\mathcal{I}_{A_1 , A_2}>0$, the genus of  $\hat{\gamma}_{A_1, A_2}^{\textrm{\tiny  \,con,(d)}}$ is $g\geqslant 1$, depending on the shape of the entangling curve.

For domains $A_1$ and $A_2$ such that $\hat{\gamma}^{\textrm{\tiny (d)}}_A$ has genus one, we can apply the fact that for any $g=1$ closed surface embedded in $\mathbb{R}^3$, we have (see Theorem 7.2.4 in \cite{willmorebook})
\be
\mathcal{W}[\Sigma_1] \geqslant 2\pi^2 \,,
\ee
where the bound is saturated by a regular torus whose ratio between its radii is $\sqrt{2}$, which is known as the Clifford torus.
This claim has been conjectured by Willmore \cite{willmorebound} and proved only recently \cite{willmoreboundproof}.

Considering two disjoint disks for $A_1$ and $A_2$, if $\hat{\gamma}_{A_1, A_2}^{\textrm{\tiny  \,con,(d)}}$ were the Clifford torus, then the holographic mutual information would be $F_{A_1 \cup A_2} - F_{A_1}  - F_{A_2} =\pi^2 - 4\pi < 0$.
Thus, the Clifford torus does not occur among the genus one closed surfaces $\hat{\gamma}_{A_1, A_2}^{\textrm{\tiny  \,con,(d)}}$ providing the holographic mutual information of some configuration of two disks, which is always non negative.
Nevertheless, it is reasonable to ask whether the Clifford torus occurs anyway as local minimum of the area functional which is not a global one. 
For two disjoint disks it has been found that $F_{A_1 , A_2} - F_{A_1}  - F_{A_2} \geqslant  -0.7886 > \pi^2 - 4\pi $ \cite{Drukker:2005cu, Fonda:2014cca, Hirata:2006jx, Krtous:2014pva}.
Thus, half of the Clifford torus never occurs among the surfaces $\gamma_A \subset \mathbb{H}_3$ which are extremal points of the area functional.
This happens because not all the genus one surfaces can be spanned by considering $\gamma^{\textrm{\tiny (d)}}_A$ with varying $A$, but only those ones which are symmetric with respect to the plane $z=0$ and such that the curve $\partial A$ is umbilic. 
For regular tori, i.e. the ones obtained from two circumferences at fixed radii (and the Clifford torus is among them), the latter condition is not satisfied.

An interesting observation about AdS$_4$ that we find it worth remarking here concerns the strong subadditivity condition (\ref{holog ssa F_A}) for the holographic prescription. 
Choosing $A_0$ such that $A_1 \cup A_0$ has the same shape of $A_1$, namely $A_1 \cup A_0$ is a rescaling of $A_1$ by a factor greater than one, by employing the observation made in the last paragraph of \S\ref{sec:ads4static}, we have that the r.h.s. of (\ref{holog ssa F_A})  vanishes. This does not happen for the black holes and the domain wall geometries, where the invariance under scale transformations is broken by the occurrence of a scale.

Finding the minimal area surface $\hat{\gamma}_A$ such that $A$ is made by two equal disjoint disks is equivalent to obtain $\hat{\gamma}_A$ when $A$ is an annulus \cite{Krtous:2014pva, Fonda:2014cca}. In \S\ref{sec app annulus} we consider the latter domain, showing that the formula (\ref{FA ads4 sec}) specified to this case provides the analytic expression already found in through a direct computation of the area \cite{Drukker:2005cu, Hirata:2006jx, Dekel:2013kwa}.

\section{Conclusions}
\label{sec conclusions}

In this paper we have studied the holographic entanglement entropy (\ref{RT formula intro}) in the context of AdS$_4$/CFT$_3$ for domains $A$ having generic shapes.
When the entangling curve is smooth, the first non trivial term in the expansion $\varepsilon \to 0$ of the holographic entanglement entropy is the constant term $F_A$ (see (\ref{hee ads4 intro})).
This term is interesting because it depends on the whole minimal surface and, therefore, it allows to probe the IR part of the geometry when the corresponding domain $A$ is sufficiently large.

Our main results are (\ref{FA sec}) and (\ref{FA final time-dep}), where $F_A$ is given in terms of the unit vectors normal to the extremal area surface $\hat{\gamma}_A$ respectively for static and time dependent backgrounds  which are conformally related to asymptotically flat spacetimes.
These formulas has been applied for explicit backgrounds: among the static ones we have considered AdS$_4$, asymptotically AdS$_4$ black holes and domain wall geometries.
The latter ones provide an example of holographic RG flow.
In the simplest case of AdS$_4$ one finds that $F_A$ is given by the Willmore energy of $\hat{\gamma}_A$ viewed as surface embedded in $\mathbb{R}^3$ \cite{Babich:1992mc, AM}.
This allows us to easily prove that the disk maximises $S_A$ among the domains with the same perimeter. 
Among the time dependent spacetimes, we have considered the Vaidya-AdS metrics.

We have checked that our results reproduce the well known ones for highly symmetric domains like strips, disks and annuli.
As for less symmetric domains $A$, which are more difficult to treat (e.g. the ones delimited by ellipses or some non convex domains), our formulas have been tested numerically by employing Surface Evolver \cite{evolverpaper, evolverlink}.
An interesting outcome is obtained from the domain wall geometries. 
Indeed, from the holographic analysis of $F_A$ for the domains different from the disk, we have observed the same qualitative behaviour of $F_A$ for the disk, which provides the holographic $C$ function. 
Unfortunately, our numerics does not allow to probe the deep IR regime and therefore we cannot give conclusive statements. 
We hope that our analysis will be improved in the near future. 

Among other open issues that would be interesting to address in the future, let us mention the higher dimensional case, where the expansion of the entanglement entropy as $\varepsilon \to 0$ has more divergent terms whose coefficients depend on the geometry of $\partial A$ (see \cite{Astaneh:2014uba, Perlmutter:2015vma} for recent papers where the properties of the Willmore energy of $\partial A$ in $d=4$ have been employed to get some insights on entanglement entropy).

As for the time evolution of the holographic entanglement entropy through the Vaidya-AdS backgrounds, the result found here could lead to some deeper understanding of the entanglement tsunami picture \cite{Liu:2013iza}.
It would be also interesting to perform a numerical study of this time evolution for finite domains which are not disks, like the ones considered in this manuscript for static backgrounds.

\subsection*{Acknowledgments}

It is our pleasure to thank Luca Giomi for the collaboration in the initial part of this project and for useful discussions during its development. 
We are grateful in particular to Ken Brakke, Amit Dekel and Rob Myers for important discussions and correspondence. 
We acknowledge Pasquale Calabrese, Cristiano De Nobili, Matthew Headrick, Veronika Hubeny, Hong Liu, Esperanza Lopez, Andrea Malchiodi, Mukund Rangamani and Tadashi Takayanagi for interesting discussions and insightful comments.
We thank Veronika Hubeny, Rob Myers and Mukund Rangamani also for helpful comments on the draft.
DS and ET thank Galileo Galilei Institute for warm hospitality, financial support and the stimulating environment enjoyed during the program {\it Holographic Methods for Strongly Coupled Systems}, where part of this work has been done.
For the same reasons, ET is grateful also to Kavli Institute for Theoretical Physics (program {\it Entanglement in Strongly-Correlated Quantum Matter}).
PF thanks IPM (Teheran) for kind hospitality and support during part of this work.
ET has been supported by the ERC under Starting Grant  279391 EDEQS.

\appendix

\section{Unit normal vectors and area elements}
\label{sec app normal vector}

In this appendix we discuss some issues about the unit vectors and the area elements occurring in the main text. 
In particular, in \S\ref{app vector b_mu} we consider the construction of the vector $\tilde{b}^\mu$ introduced in \S\ref{sec general case} and its behaviour as $\varepsilon \to 0$.
In \S\ref{app:unit vectors backgrounds} we provide the explicit expressions of the unit vectors normal to the surfaces and of the area elements for the explicit backgrounds studied in the main text (see \S\ref{sec static backs} and \S\ref{sec vaidya bg}).

\subsection{Vector on $\partial \gamma_{\varepsilon}$ for smooth entangling curves}
\label{app vector b_mu}

In this subsection, following  \cite{AM, Graham:1999pm}, we discuss the construction and the properties of the vector $\tilde{b}^\mu$ for the surfaces $\gamma_A$ occurring in
\S\ref{sec general case}.
We are interested in the behaviour of $\gamma_A$ near the boundary $z=0$.

By adopting the Cartesian coordinate system in the $z=0$ plane, the entangling curve $\partial A$ can be written as $(z,x,y)=(0,x(\sigma),y(\sigma)) \in \partial A $ in parametric form. 
Let us introduce the following vectors
\be 
\label{tilde r and q def}
\tilde{r}^\mu (\sigma,z)
\equiv \big(\,z\, ,\,  x(\sigma)\, ,\, y(\sigma) \,\big) \,,
\qquad
\tilde{q}^\mu(\sigma)
=
\frac{1}{\sqrt{x'(\sigma)^2+y'(\sigma)^2}}\,
\big(  \,0\,,\, y'(\sigma)\, ,\,  -x'(\sigma) \, \big) \,,
\ee
where $\tilde{r}^\mu$ defines the vertical cylinder above $\partial A$ for $z>0$, while $\tilde{q}^\mu$ is the unit vector normal to $\partial A$.
Notice that $\tilde{q}^\mu$ cannot be defined at the vertices of a non smooth entangling curve.

We can employ (\ref{tilde r and q def}) to parameterize the surfaces $\gamma_A$ near the boundary at $z=0$. 
Indeed, the coordinates of a point belonging to the region of $\gamma_A$ close to $z=0$ can be written as
\be
\label{xts}
\tilde{p}^\mu(\sigma,z)
\,=\,
\tilde{r}^\mu(\sigma,z)+u(\sigma,z) \, \tilde{q}^\mu(\sigma)
\,=\,
\bigg(
z \, , \,
x(\sigma)+ \frac{y'(\sigma)\, u(\sigma,z)}{\sqrt{x'(\sigma)^2+y'(\sigma)^2}}  \, , \,
y(\sigma)- \frac{x'(\sigma)\, u(\sigma,z)}{\sqrt{x'(\sigma)^2+y'(\sigma)^2}}
\bigg) \,,
\ee
where $u(\sigma,z) \in \mathbb{R}$ is an arbitrary function.
Thus, $\gamma_A$ close to the boundary is described by its displacement from the vertical cylinder over $\partial A$.
The requirement $\partial \gamma_A = \partial A$ becomes $u(\sigma,0)=0$. 
Different functions $u$ provide different surfaces $\gamma_A$.
From \eqref{xts} we can easily find the following vectors tangent to the surface
\bea
\label{eq tangent m1}
& &
\tilde{m}^{(1)\mu} 
\,=\,
\partial_\sigma \tilde{p}^\mu(\sigma,z)
\,=\,
\Bigg(0\,,\,
(1- \kappa\,u)x'+\frac{y' u_\sigma}{\sqrt{x'^2+y'^2}} \,,\,
(1- \kappa\,u )y'-\frac{x' u_\sigma}{\sqrt{x'^2+y'^2}}
\Bigg) \,,
\\
\label{eq tangent m2}
\rule{0pt}{.7cm}
& &
\tilde{m}^{(2)\mu} 
\,=\,
\partial_z \tilde{p}^\mu(\sigma,z)
\,=\,
\Bigg(
1 \,,\,
\frac{y' u_z}{\sqrt{x'^2+y'^2}}
\,,\,
-\frac{x' u_z}{\sqrt{x'^2+y'^2}} 
\Bigg) \,,
\eea
where $\kappa(\sigma)$ is the geodesic curvature of the entangling curve $\partial A$, namely
\be
\label{eq geod curv}
\kappa(\sigma) = -\frac{x' y''-x'' y'}{(x'^2+y'^2)^{3/2}} \;.
\ee
Given the tangent vectors (\ref{eq tangent m1}) and (\ref{eq tangent m1}), the determinant of induced metric $h_{\mu\nu}$ reads
\be 
\label{eq det tilde h}
\det {h} 
= 
\frac{1}{z^4} \det {\tilde{h}} \,,
\qquad
\det {\tilde{h}} 
\,=\,
\det ( \tilde{m}^{(i)}_\mu \tilde{m}^{(j)\mu})
=
\big(
1-\kappa \,u
\big)^2
\left(
x'^2+y'^2
\right)
\left(
1+u_z^2
\right)
+
u_\sigma^2 \,,
\qquad
i,j \in \{1,2\} \,.
\ee
Another consistency condition to impose is the requirement that $\det \tilde{h} $ at $z=0$ provides the square of the line element of the entangling curve, i.e. $(\det \tilde{h} )|_{z=0} =  x'^2+y'^2$. By employing (\ref{eq det tilde h}) and $u(\sigma,0)=u_\sigma(\sigma,0)=0$, this condition implies that
\be
\label{vertical end}
u_z(\sigma,0)=0  \,,
\ee
which tells us that $\gamma_A$ intersects orthogonally the $z=0$ plane. 
Notice that $\tilde{m}^{(1)} \cdot \tilde{m}^{(2)} = u_\sigma u_z \neq 0$ for $z\neq 0$.

Considering the surface $\gamma_\varepsilon$ obtained by restricting $\gamma_A$ to $z\geqslant \varepsilon >0$, since $\tilde{m}^{(1)z} =0 $ the vector $\tilde{m}^{(1)\mu} $ belongs to the plane $z=\varepsilon$.
Thus, the vector $\tilde{b}^\mu$ introduced in \S\ref{sec general case} can be constructed as the linear combination of $\tilde{m}^{(1)\mu} $ and $\tilde{m}^{(2)\mu} $ which is orthogonal to $\tilde{m}^{(1)\mu} $, namely
\be 
\label{eq m mu cond}
\tilde{b}^{\mu} 
\,=\,
\tilde{m}^{(2)\mu} - 
\frac{ \tilde{m}^{(1)} \cdot \tilde{m}^{(2)}}
{ \tilde{m}^{(1)}\cdot \tilde{m}^{(1)}} \, 
\tilde{m}^{(1)\mu} \,,
\ee 
where we have neglected the normalization and the global sign.
Combining (\ref{eq tangent m1}), (\ref{eq tangent m2}) and (\ref{vertical end}) into (\ref{eq m mu cond}), at $z=\varepsilon$ one finds the following vector 
\be
\label{b_mu app}
\tilde{b}^{\mu} 
=
-
\left(
1- 2 u_2 \,\varepsilon^2 \,,\,
\frac{\varepsilon \,u_2 \,y'}{\sqrt{x'^2+y'^2}}
\, , \,
-  \frac{\varepsilon\,u_2 \, x'}{\sqrt{x'^2+y'^2}}
\right) + O(\varepsilon^3)\,,
\ee
where $u = u_2(\sigma) \,z^2/2 +O(z^3)$ is the first term of the expansion of $u$ as $z\to 0$.
The vector (\ref{b_mu app}) has unit norm up to $O(\varepsilon^2)$ terms. 
In particular, $\tilde{b}^{\mu}  \to (-1,0,0)$ when $\varepsilon  \to 0$.
Taking the vector product of $\tilde{b}^{\mu} $ and $\tilde{m}^{(1)\mu} $, we can easily find the unit vector normal to $\gamma_\varepsilon$ at $z=\varepsilon$, namely
\be
\tilde{n}^{\mu} 
\,=\,
\left(
\varepsilon \,u_2 \,,\,
- \frac{ y'}{\sqrt{x'^2+y'^2}}
\,,\,
 \frac{ x'}{\sqrt{x'^2+y'^2}}
\right) + O(\varepsilon^2)\,,
\ee
which tells us that $\tilde{n}^z = O(\varepsilon)$ when $u_2$ is non vanishing.

From (\ref{eq det tilde h}) it is straightforward to write the differential equation providing the extremal area condition, which turns out to be quite complicated. 
Nevertheless, by plugging the expansion $u = u_2(\sigma) \,z^2/2 +O(z^3)$ into it and expanding the result as $z \to 0$, the first non trivial order leads to 
\be
\label{equ2}
u_2(\sigma) = \kappa(\sigma) \,.
\ee
As discussed in \cite{Babich:1992mc,AM}, this condition tells us that $\partial \hat{\gamma}_A$ is un umbilic line, i.e. for any of its points the two principal curvatures coincide and therefore, locally, the surface looks like a sphere.

\subsection{Black holes and domain wall geometries}
\label{app:unit vectors backgrounds}

In this subsection we give explicit expressions for the unit vectors and for the area elements that are needed in the computation of $F_A$ for specific domains. 

For static backgrounds, let us consider surfaces parameterised either by $z=z(x,y)$ if cartesian coordinates $\{x,y\}$ have been chosen for the $z=0$ plane or by $z=z(\rho,\theta)$ for polar coordinates $\{\rho,\theta\}$ in the $z=0$ plane.
\\

{\bf Black holes.}
Let us consider first the black hole metric (\ref{bh 4dim}), which includes the special case of AdS$_4$ when $f(z)=1$ identically.
Choosing the order $\{z,x,y\}$ or $\{z,\rho,\theta\}$, by employing (\ref{unit vector C}), for the unit normal vector we have
\be
\tilde{n}_\mu  
= \frac{1}{\sqrt{f(z) + z_x^2 + z_y^2}}  
\, \big(\, 1 \, , -z_x \, , \, -z_y \,  \big) \,,
\qquad
\tilde{n}_\mu  
= \frac{1}{\sqrt{f(z) + z_\rho^2 +  z_\theta^2/\rho^2}}  
\, \big(\, 1 \, , -z_\rho \, , \, -z_\theta \,  \big) \,,
\ee
and, raising the index, the corresponding vectors read
\be
\label{normal vectors up}
 \tilde{n}^\mu  = \big(\,   f(z) \,\tilde{n}_z \, , \, \tilde{n}_x \, , \, \tilde{n}_y  \, \big) \,,
\qquad
 \tilde{n}^\mu  = \big(\,   f(z) \,\tilde{n}_z \, , \, \tilde{n}_\rho \, , \, \tilde{n}_\theta/\rho^2  \, \big) \,.
\ee
As for the induced metric on $\Sigma$, it is given by
\bea
& &
d\tilde{s}^2\big|_\Sigma 
\,=\,
 \frac{1}{f(z)}
\Big[
\big( z_x^2 + f(z) \big) dx^2
+ \big( z_y^2 + f(z) \big) dy^2
+2\, z_x z_y \, dx dy\,
\Big]  \,,
\\
\rule{0pt}{.6cm}
& &
d\tilde{s}^2\big|_\Sigma 
\,=\,
\frac{1}{f(z)}
\Big[
\big( z_\rho^2 + f(z) \big) d\rho^2
+ \big( z_\theta^2 + \rho^2 f(z) \big) d\theta^2
+2\, z_\rho z_\theta \, d\rho d\theta\,
\Big] \,.
\eea
Computing the determinant coming from induced metric $ds^2 |_\Sigma$, one gets the area element
\be
\label{area tilde element}
d\tilde{\mathcal{A}}
=
\frac{\sqrt{ f(z) + z^2_x + z^2_y}}{\sqrt{f(z)}} 
\, dx dy  \,,
\qquad
d\tilde{\mathcal{A}} 
=
\frac{\sqrt{ f(z) + z^2_\rho + z^2_\theta/\rho^2}}{\sqrt{f(z)}} 
\, \rho\, d\rho d\theta
\,.
\ee
The above expressions for the unit vectors and the area elements have been employed in \S\ref{sec bh strip} and \S\ref{sec disk bh} to write $F_A$ for strips and disks from the general formulas given in \S\ref{sec:ads4static} and \S\ref{sec bh gen}.
However, they can be used for a much larger class of domains. 
\\

{\bf Domain wall geometries.}
A similar analysis can be performed when the background metric is (\ref{holog rg 4dim}).
From (\ref{unit vector C}), one finds
\be
\tilde{n}_\mu  
= \frac{1}{\sqrt{1 + p(z) \big[z_x^2 + z_y^2\big]}}  
\, \big(\, 1 \, , -z_x \, , \, -z_y \,  \big) \,,
\qquad
\tilde{n}_\mu  
= \frac{1}{\sqrt{1+ p(z)  \big[ z_\rho^2 +  z_\theta^2 / \rho^2\big] }}  
\, \big(\, 1 \, , -z_\rho \, , \, -z_\theta \,  \big) \,,
\ee
and we find it useful also to give the same unit vectors obtained by raising the index, namely
\be
\label{normal vectors up rg}
 \tilde{n}^\mu  = \big(\,   \tilde{n}_z \, , \, p(z) \,\tilde{n}_x \, , \, p(z) \,\tilde{n}_y  \, \big) \,,
\qquad
 \tilde{n}^\mu  = \big(\,   \tilde{n}_z \, , \, p(z) \,\tilde{n}_\rho \, , \, p(z) \,\tilde{n}_\theta/\rho^2  \, \big) \,.
\ee
The two dimensional metric induced on the surface $\Sigma$ reads
\bea
& & 
d\tilde{s}^2\big|_\Sigma 
\,=\,
\left( z_x^2 + \frac{1}{p(z)} \right) dx^2
+ \left( z_y^2 + \frac{1}{p(z)}  \right) dy^2
+2\, z_x z_y \, dx dy \,,
\\
\rule{0pt}{.65cm}
& & 
d\tilde{s}^2\big|_\Sigma 
\,=\,
\left( z_\rho^2 + \frac{1}{p(z)} \,\right) d\rho^2
+ \left( z_\theta^2 + \frac{\rho^2}{p(z)} \, \right) d\theta^2
+2\, z_\rho z_\theta \, d\rho d\theta \,,
\eea
and the corresponding area elements are given respectively by
\be
\label{area element rg}
d\tilde{\mathcal{A}}
\,=\,
\frac{\sqrt{ 1 + p(z) \big[ z^2_x + z^2_y\big]}}{p(z)} 
\, dx dy  \,,
\qquad
d\tilde{\mathcal{A}}
\,=\,
\frac{\sqrt{ 1 + p(z) \big[ z^2_\rho + z^2_\theta/\rho^2 \big]}}{p(z)} 
\,  \rho\, d\rho d\theta \,.
\ee
The above expressions (\ref{normal vectors up rg}) and (\ref{area element rg}) have been used in \S\ref{sec rg strip} and \S\ref{sec disk rg} to specify $F_A$ for the strips and the disks starting from the general formulas given in \S\ref{sec static backs}; but they can be employed also for other domains.

\subsection{Vaidya-AdS backgrounds}
\label{app:unit vectors vaidya}

The analysis of \S\ref{app:unit vectors backgrounds} can be performed also for the Vaidya-AdS backgrounds (\ref{vaidya metric}).

Let us consider first the case where the spatial part of the  boundary at $z=0$ is parameterized by Cartesian coordinates.
Assuming that the surface $\gamma_A$ described by the functions $v=v(x,y)$ and $z=z(x,y)$, the vectors $m^{M}_{(1)} = ( v_x, z_x, 1, 0 ) $ and $m^{M}_{(2)} = ( v_y, z_y, 0, 1 )  $ are tangent to such surface (the order of the components is given by $(v,z,x,y)$).
From these vectors, one can construct two unit vectors $n^M_{(i)}$ normal to $\gamma_A$ such that $n^2_{(1)}=-1$, $n^2_{(2)}=1$, $n^M_{(1)} n_{M(2)}=0$ and $m^M_{(i)} n_{M(j)}=0$.
Moreover, we also require that $n^M_{(i)}$ reproduce the ones introduced in \S\ref{app:unit vectors backgrounds} for the static cases. 
By employing the notation $\boldsymbol{v}'= (v_x,v_y ) $ and $\boldsymbol{z}'= (z_x,z_y ) $, the unit vectors $n^M_{(i)}$ read
\bea
& & \hspace{-1.4cm}
n^M_{(1)}
=\,
-\,z\,
\frac{
\big(
1- \boldsymbol{v}' \cdot \boldsymbol{z}' \,, \,
\boldsymbol{z}'^2+f(v,z)  \boldsymbol{v}' \cdot \boldsymbol{z}' \,, \,
z_x +f(v,z) v_x + z_y  \boldsymbol{v}' \wedge \boldsymbol{z}'  \,, \,
z_y +f(v,z) v_y - z_x  \boldsymbol{v}' \wedge \boldsymbol{z}'
\big)
}{
\sqrt{[f(v,z)+\boldsymbol{z}'^2]\,
[1-2 \boldsymbol{v}' \cdot \boldsymbol{z}'-( \boldsymbol{v}' \wedge \boldsymbol{z}')^2 -f(v,z) \boldsymbol{v}'^2 ]}
} \,,
\\
& & \hspace{-1.4cm}
n^M_{(2)}
=\,
-\, z\,\frac{\big(
1 \,, \,
-f(v,z) \,, \,
z_x \,, \,
z_y 
\big)}{\sqrt{f(v,z)+\boldsymbol{z}'^2}} \,,
\eea
where $\boldsymbol{v}'\cdot \boldsymbol{z}'=v_x z_x +v_y z_y$, $\boldsymbol{v}' \wedge \boldsymbol{z}'=v_x z_y - v_y z_x$ and $\boldsymbol{z}'^2 = z_x^2+z_y^2$.
Then, it is not difficult to find that the two dimensional metric induced on the surface $\gamma_A$ reads
\be
\frac{1}{z^2}  
\left( 
\begin{array}{cc}
1-2 v_x z_x - f(v,z) v_x^2 & - f(v,z) v_x v_y-v_y z_x-v_x z_y \\
- f(v,z) v_x v_y-v_y z_x-v_x z_y & 1-2 v_y z_y - f(v,z) v_y^2 
\end{array}
\right) \,,
\ee
and its determinant provides the following area element
\be
d \mathcal{A} = \frac{\sqrt{1-2 \boldsymbol{v}' \cdot \boldsymbol{z}' -(\boldsymbol{v}' \wedge \boldsymbol{z}')^2 -f(v,z) \boldsymbol{v}'^2 }}{z^2} dx dy  \,.
\ee 
The previous expressions in the simpler case of $v=v(x)$ and $z=z(x)$ have been employed in \S\ref{sec vaidya strip}, where the strip has been considered.

For completeness, let us repeat the above analysis when polar coordinates are adopted for the spatial part of the boundary at $z=0$.
Ordering the coordinates as $(v,z,\rho,\theta)$ and restricting our attention to the surfaces $\gamma_A$ given by  $v=v(\rho,\theta)$ and $z=z(\rho,\theta)$, one first construct the tangent vectors $m^{M}_{(1)} =  ( v_\rho, z_\rho, 1, 0 ) $ and $m^{M}_{(2)} = ( v_\theta, z_\theta, 0, 1 )  $.
Then, by adopting the notation $\boldsymbol{v}'=\left(v_\rho,v_\theta/\rho\right) $ and $\boldsymbol{z}'=\left(z_\rho,z_\theta/\rho\right)$, we can construct the unit vectors $n^M_{(i)}$ such that $n^2_{(1)}=-1$, $n^2_{(2)}=1$ and $n^M_{(1)} n_{M(2)}=0$ as above. 
They read
\bea
& & \hspace{-1.7cm}
n^M_{(1)}
=\,
-\,z\,
\frac{
\big(
1- \boldsymbol{v}' \cdot \boldsymbol{z}' \,,\,
 \boldsymbol{z}'^2+f(v,z) \boldsymbol{v}' \cdot \boldsymbol{z}'  \,,\,
z_\rho +f(v,z) v_\rho + \tfrac{z_\theta}{\rho}\, \boldsymbol{v}' \wedge \boldsymbol{z}'  \,,\,
\tfrac{z_\theta}{\rho}\,
 + \tfrac{v_\theta}{\rho} f(v,z)  - z_\rho \boldsymbol{v}' \wedge \boldsymbol{z}'
\big)
}{
\sqrt{[f(v,z)+\boldsymbol{z}'^2] \,,
[1-2\boldsymbol{v}' \cdot \boldsymbol{z}'-(\boldsymbol{v}' \wedge \boldsymbol{z}')^2 -f(v,z) \boldsymbol{v}'^2 ]}
} \,,
\\
& & \hspace{-1.7cm}
n^M_{(2)}
=\,
- \, z\, \frac{\big(
1 \,,\,
-f(v,z) \,,\,
z_\rho \,,\,
z_\theta 
\big)}{\sqrt{f(v,z)+\boldsymbol{z}'^2}} \,.
\eea
The two dimensional metric induced on the surface $\gamma_A$ reads
\be
\frac{1}{z^2}  
\left( 
\begin{array}{cc}
1-2 v_\rho z_\rho - f(v,z) v_\rho^2 & - f(v,z) v_\theta v_\rho -v_\theta z_\rho-v_\rho z_\theta \\
- f(v,z) v_\theta v_\rho -v_\theta z_\rho-v_\rho z_\theta & \rho^2-2 v_\theta z_\theta - f(v,z) v_\theta^2 
\end{array}
\right) ,
\ee
and therefore for the corresponding area element we have
\be
d \mathcal{A} = \frac{\sqrt{1-2 \boldsymbol{v}' \cdot \boldsymbol{z}' -(\boldsymbol{v}' \wedge \boldsymbol{z}')^2 -f(v,z) \boldsymbol{v}'^2 }}{z^2} \rho \,d\rho d\theta \,.
\ee 
In \S\ref{sec vaidya disk}, where $A$ is a disk, we have employed these expressions in the special case of $v=v(\rho)$ and $z=z(\rho)$.


\section{On the higher dimensional cases}
\label{sec app higher dims}

In this appendix we briefly discuss the construction of the Weyl invariant expressions that occur in a natural way as one tries to generalize the construction of \S\ref{sec general case} to static backgrounds which are asymptotically AdS$_{d+1}$.

Given the $(d-1)$ dimensional spatial surface $\gamma$ embedded into a spatial time slice of the bulk spacetime, the induced metric and the extrinsic curvature are defined as in \S\ref{sec general case} but in this case the greek indices assume $d$ integer values.
The trace of the induced metric is $h_{\mu\nu} g^{\mu\nu} =  h_{\mu\nu} h^{\mu\nu} = d-1$ and the  traceless tensor to consider is
\be
\label{traceless Kmunu def d-dim}
\mathcal{K}_{\mu\nu}
= K_{\mu\nu}-\frac{\textrm{Tr}K}{d-1} \,h_{\mu\nu}  \,,
\ee
which becomes (\ref{traceless Kmunu def}) when $d=3$.

From (\ref{Kmunu law}) it is straightforward to find that, under Weyl transformations, $\textrm{Tr}K$ changes as follows 
\be
\label{extrinsic curv conf d-dim}
\textrm{Tr}K =  
e^{-\varphi} 
\Big( 
\textrm{Tr} \widetilde{K} + (d-1) \,\tilde{n}^\lambda \partial_\lambda \varphi\,
\Big) \,.
\ee
Combining this expression with (\ref{Kmunu law}), one finds the following simple transformation rule 
\be
\mathcal{K}_{\mu}{}^\nu
\,=\, e^{-\varphi} \,\widetilde{\mathcal{K}}_{\mu}{}^\nu \,.
\ee
Then, considering the determinants $h$ and $\tilde{h}$ of the induced metrics, they are related as $h=e^{2(d-1)\varphi} \tilde{h}$. 
This implies that for the area elements $d\mathcal{A}_{d-1}  = \sqrt{h} \, d\Sigma_{d-1}$ and $d\tilde{\mathcal{A}}_{d-1}  = \sqrt{\tilde{h}} \, d\Sigma_{d-1}$, where $d\Sigma_{d-1}=\prod_{i=1}^{d-1}d\sigma_i$, being  $\sigma_i$ some local coordinates, we have that $d\mathcal{A}_{d-1} = e^{(d-1)\varphi} d\tilde{\mathcal{A}}_{d-1} $.

Thus, from (\ref{extrinsic curv conf d-dim}) and the transformation rule of the area element, we can easily construct Weyl invariant expressions as follows
\be
\label{weyl invs traceless}
\prod_i  \big( \textrm{Tr} \mathcal{K}^{n_i} \big)^{a_i}  d\mathcal{A}_{d-1}  \,,
\qquad
\sum_i n_i a_i   = d-1  \,,
\qquad
n_i \geqslant  2 \,,
\hspace{.8cm}
a_i \geqslant  1 \,,
\ee
where the case $n_i =1$ is excluded because $\textrm{Tr} \mathcal{K}=0$.
Notice that (\ref{weyl invs traceless}) are defined only for $d \geqslant 3$.

When $d=3$ only the pair $(n,a) = (2,1)$ is allowed and, similarly, when $d=4$ one finds only the pair $(n,a) = (3,1)$.
Instead, for $d=5$ we can construct two terms of the form (\ref{weyl invs traceless}) with a single term in the product: one having $(n,a) = (4,1)$  and $(n,a) = (2,2)$.
Any linear combination of these two terms is Weyl invariant but let us mention that also other Weyl invariant terms different from (\ref{weyl invs traceless}) can be constructed \cite{Guven:2005abc}.

\section{A comment from the Helfrich energy}
\label{app helfrich}

The holographic entanglement entropy (\ref{RT formula intro}) for a two dimensional spatial domain $A$ is given by the area of the surface $\hat{\gamma}_A$ which minimises the area functional within the class of surfaces $\gamma_A$ such that $\partial \gamma_A =\partial A$, once the cutoff $z\geqslant \varepsilon >0$ has been introduced.
In \S\ref{sec static} it has been shown that, for smooth entangling curves and when the bulk spacetime is AdS$_4$, the $O(1)$ term in the $\varepsilon \to 0$ expansion is given by the Willmore energy of the surface $\hat{\gamma}_A$ embedded in $\mathbb{R}^3$ (see (\ref{FA ads4 sec})) \cite{Babich:1992mc, AM}.

Given an oriented, smooth and closed surface $\Sigma_g \subset \mathbb{R}^3$ of genus $g$, an interesting generalization of the Willmore functional is the Helfrich functional, which is defined as follows  \cite{Helfrich:1973abc}
\be
\label{helfrich def}
\mathcal{H}[\Sigma_g]
\equiv
\int_{\Sigma_g} 
\bigg[
\bigg( \frac{\textrm{Tr} \widetilde{K}}{2} - \widetilde{H}_0 \bigg)^2
+ \frac{\tilde{\lambda}}{2}\, \widetilde{\mathcal{R}} \,
\bigg] d\tilde{\mathcal{A}} \,,
\ee
where $\widetilde{H}_0$ and $\tilde{\lambda}$ are two constants. 
The functional (\ref{helfrich def}) plays a very important role in the study of the cell membranes \cite{cell-review}.
The last term in (\ref{helfrich def}) is topological and the Gauss-Bonnet theorem tells us that it is proportional to $(1-g)$.

In \S\ref{sec general case} it has been shown that, when the bulk geometry is AdS$_4$ and considering the surfaces $\gamma_A$ intersecting orthogonally  the boundary $z=0$, the area of  $\gamma_A$ restricted to $z\geqslant \varepsilon$ is (\ref{area generic J2 bis}), where the $O(1)$ term is given by (\ref{FA generic surface AdS4}). 
A natural question to ask is whether exists a surface $\hat{\gamma}_\varepsilon^{\textrm{\tiny (H)}}$ within this class of surfaces whose part having $z\geqslant \varepsilon$ (denoted by $\hat{\gamma}_\varepsilon^{\textrm{\tiny (H)}}$) has an area given by (\ref{area generic J2 bis}) with the $O(1)$ term given by the Helfrich energy of $\hat{\gamma}_A^{\textrm{\tiny (H)}}$ embedded in $\mathbb{R}^3$.
Thus, for the surface $\hat{\gamma}_\varepsilon^{\textrm{\tiny (H)}}$ we have 
\be
\label{area helfrich eps expansion}
\mathcal{A}[\hat{\gamma}_\varepsilon^{\textrm{\tiny (H)}}]
= 
\frac{P_A}{\varepsilon} 
- F_A^{\textrm{\tiny (H)}}
+ o(1) \,,
\ee
where
\be
F_A^{\textrm{\tiny (H)}} = \,\frac{1}{2}\, \mathcal{H}\big[\hat{\gamma}_A^{\textrm{\tiny (H,d)}}\big] \,,
\ee
being $\hat{\gamma}_A^{\textrm{\tiny (H,d)}} \equiv \hat{\gamma}_A^{\textrm{\tiny (H)}} \cup \hat{\gamma}_A^{\textrm{\tiny (H,r)}}$ the closed smooth surface in $\mathbb{R}^3$ obtained by introducing the reflected surface $\hat{\gamma}_A^{\textrm{\tiny (H,r)}}$ in the half space $z\leqslant 0$, as explained in \S\ref{sec:ads4static} (see Fig.\,\ref{3Ddoubling} an example of this construction involving the minimal area surface $\hat{\gamma}_A$). 

By employing the transformation properties of the extrinsic curvature and of the Ricci scalar introduced in \S\ref{sec time-dep general}, from the integrands in (\ref{helfrich def}) and (\ref{FA generic surface AdS4}) we find that $\hat{\gamma}_A^{\textrm{\tiny (H)}}$ is defined by the following equation 
\be
\frac{1}{4}
\big(\textrm{Tr} \widetilde{K}\big)^2 
+ \big (    \tilde{n}^\mu \partial_\mu \varphi  -\widetilde{H}_0 \big) \textrm{Tr} \widetilde{K}
+ ( \tilde{n}^\mu \partial_\mu \varphi)^2
+\frac{\tilde{\lambda}}{2} \,\widetilde{\mathcal{R}}
+ \widetilde{H}_0^2
\,=\,0 \,,
\ee
which is written through the curvature of $\hat{\gamma}_A^{\textrm{\tiny (H)}}$ embedded in $\mathbb{R}^3$.
In terms of the curvature of $\hat{\gamma}_A^{\textrm{\tiny (H)}}$ as  surface in $\mathbb{H}_3$, it reads 
\be
\label{eq helfrich surf H3}
\frac{1}{4}
\big(\textrm{Tr} K\big)^2 
\,=\,
\widetilde{H}_0 \, e^{-\varphi}  \big[
(\textrm{Tr} K -2 \,\tilde{n}^\mu \partial_\mu \varphi )
- \widetilde{H}_0 \, e^{-\varphi} 
\,\big]
-\frac{\tilde{\lambda}}{2} \big(
\mathcal{R} + 2 \,\mathcal{D}^2 \varphi
\big) \,.
\ee
As a simple consistency check, one observes that, by setting $\widetilde{H}_0 = 0 $ and $\tilde{\lambda} =0$ in (\ref{eq helfrich surf H3}), the minimal area condition $\textrm{Tr} K=0$ is recovered.
Thus, the surface $\hat{\gamma}_A^{\textrm{\tiny (H)}} \subset \mathbb{H}_3$, which is characterised by the parameters $\widetilde{H}_0 $ and $\tilde{\lambda}$, reduces to the minimal area surface $\hat{\gamma}_A$ occurring in the holographic entanglement entropy formula when $\widetilde{H}_0 = \tilde{\lambda} =0$.

It would be interesting to find a CFT quantity related in some way to the surface $\hat{\gamma}_A^{\textrm{\tiny (H)}}$. Such quantity should depend on the parameters $\widetilde{H}_0 $ and $\tilde{\lambda} $, and reduce to the entanglement entropy when they both vanish. Moreover, it should have the same leading divergence of the entanglement entropy as $\varepsilon \to 0$, as it can be seen from (\ref{area helfrich eps expansion}). Thus,  the R\'enyi entropies are excluded.

\section{Some technical details for the infinite wedge}
\label{sec app wedge}

In this appendix we discuss the computations leading to the results presented in \S\ref{sec wedge} for the holographic entanglement entropy of the infinite wedge when the bulk geometry is AdS$_4$.

For the surfaces $\gamma_A$ characterised by the ansatz (\ref{ansatz wedge}), the area of the part having $z \geqslant \varepsilon$ reads \cite{Drukker:1999zq}
\be
\mathcal{A}[\gamma_\varepsilon] =   
\int_{\gamma_\varepsilon} \frac{\rho}{q^2} \,\sqrt{q'^2+q^2+q^4} \,d\theta \,d \rho  \,.
\ee
Since the integrand does not depend explicitly on $\theta$, we have that $(q^4+q^2)/\sqrt{(q')^2+q^4+q^2}$ is independent of $\theta$.
Then, since for $\theta=0$ we have $q(0)=q_0$ and $q'(0)=0$ (see (\ref{rho minimax})), the first order differential equation providing $q(\theta)$ reads
\be
\label{cusp conservation law}
(q')^2 
\,=\, 
(q^2+q^4)\left(\frac{q^2+q^4}{q_0^2+q_0^4} -1 \right) .
\ee
By employing (\ref{unit vector C}) for AdS$_4$ and the ansatz (\ref{ansatz wedge}), one finds that the unit normal vector
\be
\label{normal vector cusp}
\tilde{n}^\mu = \frac{1}{\sqrt{q^4+q^2+ (q')^2}}
\big( \,q^2\, , \, -\, q\,,\,  q'/\rho \,\big)  \,.
\ee
Taking the component $\tilde{n}^z $ of this vector, the integrand of (\ref{FA ads4 sec}) is given by
\be
\label{cuspWillmore}
\frac{(\tilde{n}^z)^2}{z^2} = 
\frac{q^6}{\rho^2 \big[ q^4+q^2+ (q')^2\big]}
=
\frac{(q_0^2+q_0^4) \, q^2}{\rho^2\,  (q^2 +1)^2 } \,,
\ee
while the area element can be easily computed from (\ref{ansatz wedge}), finding
\be
\label{area element wedge}
d\tilde{\mathcal{A}} 
\,=\, 
\sqrt{\tilde{h}}\, d\rho  \,d\theta
\,=\,
 \frac{\sqrt{(q')^2+q^4+q^2}}{q^2}\,\rho\, d\rho\, d\theta
   \,=\,
   \frac{q^2+1}{\sqrt{q_0^4+q_0^2}}\,\rho\,d\rho \, d\theta \,,
\ee
Putting (\ref{cuspWillmore}) and (\ref{area element wedge}) together into the expression (\ref{FA ads4 sec}) for $F_A$ and changing the angular integration variable from $\theta$ to $q$, we get
\be
F_A
\,=\, 
2\int_{\rho_{\textrm{\tiny min}}}^{\rho_{\textrm{\tiny max}}}
\frac{d\rho}{\rho} 
 \int_{q_0}^{\rho/\varepsilon} 
dq\,
\frac{\sqrt{q_0^2+q_0^4} \,q^2}{(q^2 +1) q'}
\,=\, 
2\int_{\rho_{\textrm{\tiny min}}/\varepsilon}^{\rho_{\textrm{\tiny max}}/\varepsilon}
\frac{d\tilde{\rho}}{\tilde{\rho}} 
 \int_{\rho_{\textrm{\tiny min}}/\varepsilon}^{\tilde{\rho}} 
dq\,
\frac{\sqrt{q_0^2+q_0^4} \,q^2}{(q^2 +1) q'} \,,
\qquad
\tilde{\rho} =\frac{\rho}{\varepsilon} \,,
\ee
where $\rho_{\textrm{\tiny min}}$ and $\rho_{\textrm{\tiny max}}$ have been defined in (\ref{rho minimax}).
Now, by exchanging the order of integration first and then performing the integration over $\rho$, we find
\begin{subequations}
\bea
F_A
&=&
2 \, \sqrt{q_0^2+q_0^4} 
\int_{\rho_{\textrm{\tiny min}}/\varepsilon}^{\rho_{\textrm{\tiny max}}/\varepsilon}
dq\,
\frac{q^2}{(q^2 +1) q'}
 \int_{q}^{\rho_{\textrm{\tiny max}}/\varepsilon}
\frac{d\tilde{\rho}}{\tilde{\rho}} 
\\
\rule{0pt}{.7cm}
\label{cusp log term birth}
&=&
2 \, \sqrt{q_0^2+q_0^4} 
\left(\log\frac{\rho_{\textrm{\tiny max}}}{\varepsilon} 
\int_{\rho_{\textrm{\tiny min}}/\varepsilon}^{\rho_{\textrm{\tiny max}}/\varepsilon}
\frac{q^2}{(q^2 +1) q'}\, dq
-
\int_{\rho_{\textrm{\tiny min}}/\varepsilon}^{\rho_{\textrm{\tiny max}}/\varepsilon}
\frac{q^2 \log q}{(q^2 +1) q'}\, dq
\right) .
\eea
\end{subequations}
Thus, a logarithmic divergence when $\varepsilon \to 0$ is obtained from the first term in (\ref{cusp log term birth}), namely
\be
\label{FA wedge app}
F_A \,=\,
2\, b(\Omega) \log(\rho_{\textrm{\tiny max}}/\varepsilon)  + O(1) \,,
\ee
where
\be
\label{cusp v1}
b(\Omega) =
\sqrt{q_0^2+q_0^4} \int_{q_0}^{\infty}\frac{q^2}{(q^2 +1) q'}\, dq \,.
\ee
From (\ref{rho minimax}), it is straightforward to observe that the (\ref{FA wedge app}) can be written in the form (\ref{F_A wedge willmore}), but the two expressions (\ref{cusp exact result}) and (\ref{cusp v1}) for $b(\Omega)$ look quite different.
Nevertheless, one can show that they coincide through some manipulations. 
Starting from (\ref{cusp exact result}) and perform an integration by parts, we get
\begin{subequations}
\bea
\label{balpha}
& & \hspace{-.9cm}
b(\Omega) \;=\;
-\int_0^\infty   \zeta\; \frac{d}{d\zeta} \Bigg(1-\sqrt{\frac{\zeta ^2+{q_0}^2+1}{\zeta ^2+2 {q_0}^2+1}}\,\Bigg) d\zeta
\,=\,
\int_0^\infty \frac{q_0^2\, \zeta ^2 }{\sqrt{\zeta ^2+q_0^2+1} \,\big(\zeta ^2+2 q_0^2+1\big)^{3/2}}\, d\zeta 
   \\
   \label{cusp step integ 1}
 \rule{0pt}{.6cm}
& & \hspace{-.9cm}
\phantom{b(\Omega)} \;=\;
  -  \int_0^\infty \frac{q_0^2\, \zeta}{\sqrt{\zeta ^2+q_0^2+1} }
   \; \frac{d}{d\zeta}\bigg(\frac{1}{\sqrt{\zeta ^2+2 q_0^2+1}}\bigg)d\zeta 
\,=\,
   \int_0^\infty \frac{q_0^4+q_0^2}{\sqrt{\zeta ^2+2 q_0^2+1}\,\big(\zeta ^2+q_0^2+1\big)^{3/2}}\, d\zeta  \,, 
   \hspace{1cm}
\eea
\end{subequations}
where in (\ref{cusp step integ 1}) another integration by parts has been performed.
Now, by first introducing the variable $q$ as $\zeta=\sqrt{q^2-q_0^2}$ and then employing (\ref{cusp conservation law}),
the expression (\ref{cusp v1}) is recovered.

As for the contribution of the boundary $\partial \hat{\gamma}_\varepsilon$, in \S\ref{sec wedge} we discussed that it is given by two terms, according to the decomposition of $\partial \hat{\gamma}_\varepsilon = \partial \hat{\gamma}_\varepsilon^\parallel  \cup \partial \hat{\gamma}_\varepsilon^\perp $.
The curve of $\partial \hat{\gamma}_\varepsilon^\parallel$ is (see also (\ref{rho minimax}))
\be
\label{P_eps}
\partial \hat{\gamma}_\varepsilon^\parallel \, :
\qquad
\{ z, \rho, \theta \} = 
\{ \varepsilon , \varepsilon\, q(\theta) , \theta \}   \,, 
\qquad
|\theta| \leqslant \Omega/2-\omega   \,.  
\ee
The unit vector $\tilde{u}^\mu$ tangent to $\partial \hat{\gamma}_\varepsilon^\parallel$ can be easily found from (\ref{P_eps}), while (\ref{normal vector cusp}) provides the unit normal vector $\tilde{n}^\mu$. They read respectively
\be
\label{unit vectors bdy cusp}
\tilde{u}^\mu = \frac{1}{\varepsilon  \, \sqrt{q^2+   (q')^2}}\,
\big( \,0\, , \, \varepsilon  \,q'\,,\, 1 \,\big)  \,, 
\qquad
\tilde{n}^\mu = \frac{1}{\sqrt{q^4+q^2+ (q')^2}}\,
\big( \,q^2 , \, -q\,,\,  q'/(\varepsilon  \, q) \,\big) \,.
\ee
The unit vector  $\tilde{b}_\mu $ which is normal to the boundary curve $\partial \hat{\gamma}_\varepsilon$ and also tangent to the minimal surface is obtained by taking the wedge product of the two vectors in (\ref{unit vectors bdy cusp}), i.e. $\tilde{b}_\mu = \rho \, \varepsilon_{\mu\nu\lambda} \tilde{n}^\nu \tilde{u}^\lambda$ (we recall that, since we are using cylindrical coordinates for $\mathbb{R}^3$, a factor $\sqrt{\tilde{g}} = \rho$ occurs).
The result is
\be
\tilde{b}_\mu  \,=\, 
-\,\frac{1}{\sqrt{\big(q^2+   (q')^2\big)\big(q^2+q^4+(q')^2\big)}}\,
\big( \,q^2+(q')^2\, , \, q^3\,,\, -\, \varepsilon \, q^3 \,q' \,\big) \,.
\ee
Then, since $\tilde b^z=\tilde b_z$ and by employing the line element along $\partial \hat{\gamma}_\varepsilon$, i.e. $ d\tilde s = \sqrt{(\rho')^2+\rho^2} \,d\theta $ with $\rho = \varepsilon \, q$, for the boundary integral along $\partial \hat{\gamma}_\varepsilon^\parallel$ we get
\be
\label{bdy wedge parallel 1}
\int_{\partial \hat{\gamma}_\varepsilon^\parallel} \frac{\tilde b^z}{z} \, d\tilde{s}
\,=\,
- \,2 \int_0^{\Omega/2-\omega}
\frac{q^2+ (q')^2}{ \sqrt{q^4+q^2+ (q')^2}}  \,d\theta
\,=\,
-\, 2\int_{q_0}^{L/\varepsilon}  \frac{q\big[(1+q^2)^2-q_0^2-q_0^4\big]}{\sqrt{(1+q^2)^3(q^2-q_0^2)(1+q_0^2+q^2)}}\, dq \,,
\ee
where we have changed the integration variable to $q$ first and then used (\ref{cusp conservation law}).
Since the integrand converges to $1$ for $q\to\infty$, the integral is linearly divergent for $L/\varepsilon\to\infty$.
By adding $+1$ and $-1$ to the integrand of the last expression in (\ref{bdy wedge parallel 1}), we get
\be
\label{bdy wedge parallel fin}
\int_{\partial \hat{\gamma}_\varepsilon^\parallel} \frac{\tilde b^z}{z} \, d\tilde{s}
\,=\,
-\,\frac{2L}{\varepsilon} - 4\,\frac{\mathbb{E}(\tilde{q}_0^2)-(1-\tilde{q}_0^2)\mathbb{K}(\tilde{q}_0^2)}{\sqrt{1-2\tilde{q}_0^2}} + o(1) \,,
\qquad
\tilde{q}_0 = \frac{q_0}{\sqrt{1+2q_0^2}} \,.
\ee
Notice that this boundary contribution does not provide any $\log \varepsilon$ divergence.

The line integral along the boundary $\partial \hat{\gamma}_\varepsilon^\perp $ can be addressed in the same way.
The curve $\partial \hat{\gamma}_\varepsilon^\perp $ is given by 
\be
\label{P_eps vert}
\partial \hat{\gamma}_\varepsilon^\perp \, :
\qquad
\{ z, \rho, \theta \} = 
\{ \rho_{\textrm{\tiny max}}/ q(\theta) , \rho_{\textrm{\tiny max}} , \theta \}  \,,
\qquad
|\theta| \leqslant \Omega/2-\omega \,.
\ee
Then, the unit vector which are tangent and normal to $\partial \hat{\gamma}_\varepsilon^\perp $ are respectively
\be
\tilde{u}^\mu = \frac{1}{ \sqrt{q^4+   (q')^2}}\,
\big( \,-q'\, , \, 0\,,\, q^2 /\rho_{\textrm{\tiny max}}  \,\big) \,,
\qquad
\tilde{n}^\mu = \frac{1}{\sqrt{q^4+q^2+ (q')^2}}\,
\big( \,q^2\, , \, -q\,,\,  q'/\rho_{\textrm{\tiny max}} \,\big) \,.
\ee
The wedge product of these vector provides the unit vector $\tilde{b}_\mu $ normal to $\partial \hat{\gamma}_\varepsilon^\perp $ and tangent to the minimal surface at $\partial \hat{\gamma}_\varepsilon^\perp $:
\be
\tilde{b}_\mu = \frac{1}{\sqrt{[ q^4+   (q')^2 ] \, [q^2+q^4+(q')^2]}}\,
\big( \,q^3\, , \, q^4+   (q')^2\,,\, \rho_{\textrm{\tiny max}}  \,q  \,q' \,\big)  \,.
\ee
In this case the invariant measure is $ d\tilde s=\sqrt{(z')^2+\rho^2} \,d\theta$ specified to (\ref{P_eps vert}).
Thus, for the contribution of the boundary integral along $\partial \hat{\gamma}_\varepsilon^\perp $ to the holographic entanglement entropy of the wedge we obtain
\be
\int_{\partial \hat{\gamma}_\varepsilon^\perp} \frac{\tilde b^z}{z}\, d\tilde s 
\,=\, 
 \int_{\partial \hat{\gamma}_\varepsilon^\perp} 
\frac{q^2}{ \sqrt{q^4+q^2+ (q')^2}}  \, d\theta
\,=\, 
2\int_{q_0}^{\rho_{\textrm{\tiny max}}/\varepsilon}  \frac{(1+q_0^2)^2 q_0^2}{ q \,\sqrt{(1+q^2)^3(q^2-q_0^2)(1+q_0^2+q^2)}}\, dq  \,,
\ee
where (\ref{cusp conservation law}) has been employed. 
Notice that for $\rho_{\textrm{\tiny max}}/\varepsilon \to \infty$ the integral is convergent. Thus, we get
\be 
\label{wedge bdy int perp}
\int_{\partial \hat{\gamma}_\varepsilon^\perp} \frac{\tilde b^z}{z}\, d\tilde s 
\,=\, 
2\,
\frac{\mathbb{E}(\tilde{q}_0^2)-(1-2 \tilde{q}_0^2)\mathbb{K}(\tilde{q}_0^2)-\tilde{q}_0^2\, \Pi (1-\tilde{q}_0^2,\tilde{q}_0^2)}{\sqrt{1-2\tilde{q}_0^2}}  \,,
\ee
where $\tilde{q}_0$ has been defined in (\ref{bdy wedge parallel fin}).

Thus, in this case the boundary integral along $\partial \hat{\gamma}_\varepsilon$ occurring in (\ref{area generic J2 bdy})  is the sum of (\ref{bdy wedge parallel fin}) and (\ref{wedge bdy int perp}). The result is given in (\ref{wedge bdy int total}): it contains the expected area law divergence but a logarithmic divergence does not occur.

\section{Annulus}
\label{sec app annulus}

In this appendix we apply (\ref{FA ads4 sec}) for the annulus, recovering the minimal surface $\hat{\gamma}_A$ discussed in \cite{Drukker:2005cu, Hirata:2006jx, Dekel:2013kwa}.

When $A$ is an annulus delimited by two concentric circumferences with radii $R_- < R_+$ and the gravitational background is AdS$_4$, 
the global minimum of the area functional among the surfaces $\gamma_A$ which provides the holographic entanglement entropy depends on the ratio $\eta \equiv R_- / R_+ \in (0,1)$.

In particular, for $\eta \geqslant 0.367$ there are two topologically different local minima of the area functional: one is the union of the two disjoint hemispheres $\hat{\gamma}_{A_1} \cup \hat{\gamma}_{A_2}$, while the other one is a surface $\hat{\gamma}_{A_1, A_2}^{\textrm{\tiny  \,con}}$ connecting 
the two boundaries of the annulus through the bulk (there are two of them having the same $\eta$, but we consider only the one having minimal area). 
For a thin annulus $\eta \sim 1$ and  $\hat{\gamma}_{A_1, A_2}^{\textrm{\tiny  \,con}}$ is the global minimum.
At $\eta_c = 0.419$ the transition occurs and for $\eta < \eta_c$ the global minimum is given by the two disjoint hemispheres. 
For $\eta < \eta_\ast$ the solution $\hat{\gamma}_{A_1, A_2}^{\textrm{\tiny  \,con}}$ does not exist and only $\hat{\gamma}_{A_1} \cup \hat{\gamma}_{A_2}$ remains as extremal area surface.

Choosing polar coordinates $(\rho,\theta)$ in the $z=0$ plane centered in the origin, the expression for $\hat{\gamma}_{A_1, A_2}^{\textrm{\tiny  \,con}}$ can be written in a parametric form as the union of two branches
\be
\label{eq:annulsol}
z_\pm (t) \,=\, R_\pm\,t\, e^{-f_\pm (t)}  \,,
\qquad
\rho_\pm (t) \,=\, R_\pm \,e^{-f_\pm(t)}  \,,
\qquad
t\in\big[0,t_{\textrm{\tiny max}}\big] \,,
\ee
where $t_{\textrm{\tiny max}}$ is a function of $\eta$ coming from the matching condition of the two branches and  the functions $f_\pm(t)$ are given in terms of the incomplete elliptic functions of the first kind $\mathbb{F}$ and of the third kind $\Pi$ as follows
\be
\label{eq:fpm}
f_\pm (t) 
\,=\,
 \frac{1}{2}\log(1+t^2) \pm \kappa \,t_{\textrm{\tiny max}} 
\big[\mathbb{F}(\omega|\kappa^2) - \Pi(1-\kappa^2,\omega| \kappa^2)\big] \,,
\qquad
\sin{\omega} \equiv \frac{t}{t_{\textrm{\tiny max}} \sqrt{1+\kappa^2(t/t_{\textrm{\tiny max}}-1)}} \,,
\ee
being $\kappa \equiv \sqrt{(1+t_{\textrm{\tiny max}}^2)/(2+t_{\textrm{\tiny max}}^2)}$.
The boundary condition at $t=t_{\textrm{\tiny max}}$ provides a relation between $\kappa$ and $\eta$. 
Indeed, by imposing the joining of the two branches, i.e. $z_+(t_{\textrm{\tiny max}}) = z_{-}(t_{\textrm{\tiny max}}) $, one finds
\be
\log \eta 
=  f_-(t_{\textrm{\tiny max}})-f_+(t_{\textrm{\tiny max}})  
= 2 \kappa \,t_{\textrm{\tiny max}} \big[\mathbb{K}(\kappa^2) - \Pi(1-\kappa^2, \kappa^2)\big] \,.
\ee

The Willmore energy (\ref{FA ads4 sec}) of $\hat{\gamma}_{A_1, A_2}^{\textrm{\tiny  \,con}}$ can be found by summing the contributions of the two branches
\be
\label{willmore energy annulus}
F_A
\,=\,
\frac{\pi}{2}  \int_0^{t_{\textrm{\tiny max}}}  \left(\sqrt{\det h_+} \big(\textrm{Tr}\widetilde{K}_+\big)^2 
+\sqrt{\det h_-} \big(\textrm{Tr}\widetilde{K}_-\big)^2 \right) dt \,,
\ee
where the determinants of the induced metric are given by
\be 
\label{det h annulus}
\det h_\pm= \rho_\pm(t)^4 \Big(1-2t f_\pm (t) + (1+t^2)f_\pm (t)^2 \Big) \,,
\ee
and $\textrm{Tr}\widetilde{K}$ for a surface with cylindrical symmetry given by $z=z(t)$ and $\rho=\rho(t)$ reads
\be
\label{trK cylindrical}
\textrm{Tr}\widetilde{K}
 = \frac{z'}{\rho \big[(\rho')^2+(z')^2\big]^{1/2}} - \frac{z' \rho'' - \rho' z'' }{\big[(\rho')^2+(z')^2\big]^{3/2}} \,.
\ee
Plugging the solution (\ref{eq:annulsol}) into (\ref{trK cylindrical}), for the two branches we find that
\be
\label{trK pm annulus}
\textrm{Tr}\widetilde{K}_\pm \,=\, 
\frac{f_\pm (t)-(1+2t)f_\pm (t)^2+(1+t^2)f_\pm (t)^3-f'_\pm (t)}{
\rho_\pm(t)\big[1-2t f_\pm (t) + (1+t^2)f_\pm (t)^2 \big]^{3/2}} \,.
\ee
Thus, from (\ref{det h annulus}) and (\ref{trK pm annulus}), the Willmore energy (\ref{willmore energy annulus}) of $\hat{\gamma}_{A_1, A_2}^{\textrm{\tiny  \,con}}$ becomes
\begin{subequations}
\bea
F_A
&=&
4 \pi  \int_0^{t_{\textrm{\tiny max}}} 
\frac{(t^2- t_{\textrm{\tiny max}}^2	\tau_{\textrm{\tiny max}}^+)(t^2-t_{\textrm{\tiny max}}^2 \tau_{\textrm{\tiny max}}^-)}{(1+t^2)^2 \,t_{\textrm{\tiny max}}^2\sqrt{(t^2-t_{\textrm{\tiny max}}^2)(t^2+ t_{\textrm{\tiny max}}^2/[1+t_{\textrm{\tiny max}}^2])(1+t_{\textrm{\tiny max}}^2)}} \, dt
\\
\rule{0pt}{.7cm}
\label{fin}
&=&
2 \pi \sqrt{y_{\textrm{\tiny max}}(1+y_{\textrm{\tiny max}})}\int_0^1 
\frac{(	  \alpha-\tau_{\textrm{\tiny max}}^-)(\alpha-\tau_{\textrm{\tiny max}}^+)}{(1+  y_{\textrm{\tiny max}} \alpha)^2
\sqrt{ \alpha (1-\alpha)(\alpha + 1/[1+y_{\textrm{\tiny max}}])}}\, d\alpha 
\\
\label{fin fin}
\rule{0pt}{.6cm}
&=&
4 \pi \, \frac{\mathbb{E}(\kappa^2)-(1-\kappa^2)\mathbb{K}(\kappa^2)}{\sqrt{2\kappa^2-1}} \,,
\eea
\end{subequations}
where for the integration variables we employed $y=t^2$ and $\alpha= y/y_{\textrm{\tiny max}}$, introducing also the following notation
\be
\tau_{\textrm{\tiny max}}^{\pm}
\equiv
\frac{y_{\textrm{\tiny max}}^2 +y_{\textrm{\tiny max}}+1 \pm \sqrt{y_{\textrm{\tiny max}}^4+6 y_{\textrm{\tiny max}}^3+7 y_{\textrm{\tiny max}}^2+2 y_{\textrm{\tiny max}}+1}}{2 y_{\textrm{\tiny max}} \left(1+y_{\textrm{\tiny max}}\right)}  \,.
\ee
Thus, the general expression (\ref{FA ads4 sec}) for $F_A$ written through the Willmore energy reproduces the result for the annulus already found by computing the explicitly the area of $\hat{\gamma}_{A_1, A_2}^{\textrm{\tiny  \,con}}$.


\end{document}